\documentclass[longauth]{aa}
\usepackage{graphicx}

\usepackage[varg]{txfonts}
\usepackage{natbib}
\bibpunct{(}{)}{;}{a}{}{,}
\usepackage{array}
\usepackage{xspace}
\usepackage{lscape}
\usepackage{url}
\usepackage{arydshln}
\usepackage[hyperfootnotes=false, linktocpage=true, breaklinks=true, colorlinks=true, linkcolor=blue, citecolor=blue, urlcolor=blue]{hyperref}
\usepackage[all]{hypcap}
\usepackage{longtable}
\usepackage{afterpage}

\usepackage[dvipsnames]{xcolor}
\usepackage{xspace}

\newcommand{\kms}{\ensuremath{\mathrm{km\ s^{-1}}}\xspace}

\newcommand{\cf}{\ensuremath{C_f}\xspace}

\newcommand{\xmm}{{\it XMM-Newton}\xspace}
\newcommand{\chandra}{{\it Chandra}\xspace}

\newcommand{\iras}{{IRAS~17020+4544}\xspace}

\newcommand{\nustar}{{\it NuSTAR}\xspace}

\newcommand{\ergflux}{{\ensuremath{\rm{erg\ cm}^{-2}\ \rm{s}^{-1}}}\xspace}
\newcommand{\ergs}{{\ensuremath{\rm{erg\ s}^{-1}}}\xspace}

\newcommand{\lya}{Ly\ensuremath{\alpha}\xspace}
\newcommand{\lyb}{Ly\ensuremath{\beta}\xspace}
\newcommand{\civ}{\ion{C}{iv}\xspace}
\newcommand{\nv}{\ion{N}{v}\xspace}
\newcommand{\ovi}{\ion{O}{vi}\xspace}
\newcommand{\siiv}{\ion{Si}{iv}\xspace}

\newcommand{\sub}{SUBWAYS\xspace}

\newcommand{\lbol}{\xspace{$L_{\rm bol}$}\xspace}

\usepackage{xcolor}

\usepackage{xfrac}

\mathchardef\mhyphen="2D

\defcitealias{Matz22}{Paper~I}

\newcommand{\paperI}{{\citetalias{Matz22}}\xspace}

\graphicspath{{./}{figures_final/}}

\begin{document}

\title{Supermassive Black Hole Winds in X-rays: SUBWAYS}
\subtitle{II. HST UV spectroscopy of winds at intermediate redshifts}
\author{
M. Mehdipour \inst{1}
\and
G. A. Kriss \inst{1}
\and
M. Brusa \inst{2,3}
\and
G. A. Matzeu \inst{2,3,4}
\and
M. Gaspari \inst{5}
\and
S. B. Kraemer \inst{6}
\and
S. Mathur \inst{7,8}
\and
E. Behar \inst{9}
\and
S. Bianchi \inst{10}
\and
M. Cappi \inst{3}
\and
G. Chartas \inst{11}
\and
E. Costantini \inst{12,13}
\and
G. Cresci \inst{14}
\and
M. Dadina \inst{3}
\and
B. De Marco \inst{15}
\and
A. De Rosa \inst{16}
\and\newline
J. P. Dunn \inst{17}
\and
V. E. Gianolli \inst{18,10}
\and
M. Giustini \inst{19}
\and
J. S. Kaastra \inst{12,20}
\and
A. R. King \inst{21,13,20}
\and
Y. Krongold \inst{22}
\and
F. La Franca \inst{10}
\and\newline
G. Lanzuisi \inst{3,2}
\and
A. L. Longinotti \inst{22}
\and
A. Luminari \inst{16,23}
\and
R. Middei \inst{24,23}
\and
G. Miniutti \inst{19}
\and
E. Nardini \inst{14}
\and
M. Perna \inst{25,14}
\and\newline
P.-O. Petrucci \inst{18}
\and
E. Piconcelli \inst{23}
\and
G. Ponti \inst{26,27}
\and
F. Ricci \inst{10,3}
\and
F. Tombesi \inst{28,23,29,30,31}
\and
F. Ursini \inst{10}
\and
C. Vignali \inst{2,3}
\and\newline
L. Zappacosta \inst{23}
}
\institute{
% 1
Space Telescope Science Institute, 3700 San Martin Drive, Baltimore, MD 21218, USA \\ \email{mmehdipour@stsci.edu}
\and
% 2
Department of Physics and Astronomy ``Augusto Righi'' (DIFA), University of Bologna, Via Gobetti, 93/2, I-40129 Bologna, Italy
\and
% 3
INAF-Osservatorio di Astrofisica e Scienza dello Spazio di Bologna, Via Gobetti, 93/3, I-40129 Bologna, Italy
\and
% 4
European Space Agency (ESA), European Space Astronomy Centre (ESAC), E-28691 Villanueva de la Cañada, Madrid, Spain
\and
% 5
Department of Astrophysical Sciences, Princeton University, 4 Ivy Lane, Princeton, NJ 08544-1001, USA
\and
% 6
Department of Physics, Institute for Astrophysics and Computational Sciences, The Catholic University of America, Washington, DC 20064, USA
\and
% 7
Department of Astronomy, The Ohio State University, 140 West 18th Avenue, Columbus, OH 43210, USA
\and
% 8
Center for Cosmology and Astroparticle Physics, 191 West Woodruff Avenue, Columbus, OH 43210, USA
\and
% 9
Department of Physics, Technion-Israel Institute of Technology, 32000 Haifa, Israel
\and
% 10
Dipartimento di Matematica e Fisica, Universit\`a degli Studi Roma Tre, via della Vasca Navale 84, 00146 Roma, Italy
\and
% 11
Department of Physics and Astronomy, College of Charleston, Charleston, SC, 29424, USA
\and
% 12 
SRON Netherlands Institute for Space Research, Niels Bohrweg 4, 2333 CA Leiden, the Netherlands
\and
% 13
Anton Pannekoek Institute, University of Amsterdam, Postbus 94249, 1090 GE Amsterdam, The Netherlands
\and
% 14
INAF – Osservatorio Astrofisico di Arcetri, Largo Enrico Fermi 5, I-50125 Firenze, Italy
\and
% 15
Departament de F\'isica, EEBE, Universitat Polit\`ecnica de Catalunya, Av. Eduard Maristany 16, E-08019 Barcelona, Spain 
\and
% 16
INAF - Istituto di Astrofisica e Planetologia Spaziali (IAPS), via Fosso del Cavaliere, Roma, I-00133, Italy
\and
% 17
Department of Physics and Astronomy, Georgia State University, Atlanta, GA 30303, USA
\and
% 18
Universit\'e Grenoble Alpes, CNRS, IPAG, 38000 Grenoble, France
\and
% 19
Centro de Astrobiologia (CAB), CSIC-INTA, Camino Bajo del Castillo s/n, 28692, Villanueva de la Cañada, Madrid, Spain
\and
% 20
Leiden Observatory, Leiden University, PO Box 9513, 2300 RA Leiden, the Netherlands
\and
% 21
School of Physics and Astronomy, University of Leicester, Leicester, LE1 7RH, UK
\and
% 22
Instituto de Astronom\'{i}a, Universidad Nacional Aut\'{o}noma de M\'{e}xico, Circuito Exterior, Ciudad Universitaria, Ciudad de M\'{e}xico 04510, M\'{e}xico
\and
% 23
INAF - Osservatorio Astronomico di Roma, Via Frascati 33, I-00040 Monte Porzio Catone, Italy
\and
% 24
Space Science Data Center, Agenzia Spaziale Italiana, Via del Politecnico snc, 00133 Roma, Italy
\and
% 25
Centro de Astrobiología, (CAB, CSIC–INTA), Departamento de Astrofísica, Cra. de Ajalvir Km. 4, 28850 – Torrejón de Ardoz, Madrid, Spain
\and
% 26
INAF-Osservatorio Astronomico di Brera, Via E. Bianchi 46, I-23807 Merate (LC), Italy
\and
% 27
Max-Planck-Institut f{\"u}r Extraterrestrische Physik, Giessenbachstrasse, 85748 Garching, Germany
\and
% 28
Physics Department, Tor Vergata University of Rome, Via della Ricerca Scientifica 1, 00133 Rome, Italy
\and
% 29
Department of Astronomy, University of Maryland, College Park, MD 20742, USA
\and
% 30
NASA - Goddard Space Flight Center, Code 662, Greenbelt, MD 20771, USA
\and
% 31
INFN - Roma Tor Vergata, Via della Ricerca Scientifica 1, 00133 Rome, Italy
}
%%%%%%%%%%%%%%%%%%%%%%%%%%%%%%%%%%%%%%%%%%%%%%%%%%%%%%%%%%%%%%%%%%%%%%%%%%%%%%%%%%%%%%%%%%%%%%%%%%%%%%%
\date{Received 23 September 2022 / Accepted 2 November 2022}
%%%%%%%%%%%%%%%%%%%%%%%%%%%%%%%%%%%%%%%%%%%%%%%%%%%%%%%%%%%%%%%%%%%%%%%%%%%%%%%%%%%%%%%%%%%%%%%%%%%%%%%
\abstract{
We present a UV spectroscopic study of ionized outflows in 21 active galactic nuclei (AGN), observed with the {\it Hubble Space Telescope} (HST). The targets of the Supermassive Black Hole Winds in X-rays (\sub) sample were selected with the aim to probe the parameter space of the underexplored AGN between the local Seyfert galaxies and the luminous quasars at high redshifts. Our targets, spanning redshifts of $0.1$--$0.4$ and bolometric luminosities ($L_{\rm bol}$) of $10^{45}$--$10^{46}$ erg~s$^{-1}$, have been observed with a large multi-wavelength campaign using \xmm, \nustar, and HST. Here, we model the UV spectra and look for different types of AGN outflows that may produce either narrow or broad UV absorption features. We examine the relations between the observed UV outflows and other properties of the AGN. We find that 60\% of our targets show a presence of outflowing \ion{H}{i} absorption, while 40\% exhibit ionized outflows seen as absorption by either \ion{C}{iv}, \ion{N}{v}, or \ion{O}{vi}. This is comparable to the occurrence of ionized outflows seen in the local Seyfert galaxies. All UV absorption lines in the sample are relatively narrow, with outflow velocities reaching up to $-3300$~\kms. We did not detect any UV counterparts to the X-ray ultra-fast outflows (UFOs), most likely due to their being too highly ionized to produce significant UV absorption. However, all \sub targets with an X-ray UFO that have HST data demonstrate the presence of UV outflows at lower velocities. We find significant correlations between the column density ($N$) of the UV ions and $L_{\rm bol}$ of the AGN, with $N_{\rm H\,I}$ decreasing with $L_{\rm bol}$, while $N_{\rm O\,VI}$ is increasing with $L_{\rm bol}$. This is likely to be a photoionization effect, where toward higher AGN luminosities, the wind becomes more ionized, resulting in less absorption by neutral or low-ionization ions and more absorption by high-ionization ions. In addition, we find that $N$ of the UV ions decreases as their outflow velocity increases. This may be explained by a mechanical power that is evacuating the UV-absorbing medium. Our observed relations are consistent with multiphase AGN feeding and feedback simulations indicating that a combination of both radiative and mechanical processes are in play.
}
\authorrunning{M. Mehdipour et al.}
\titlerunning{SUBWAYS. II.}
\maketitle
%%%%%%%%%%%%%%%%%%%%%%%%%%%%%%%%%%%%%%%%%%%%%%%%%%%%%%%%%%%%%%%%%%%%%%%%%%%%%%%%%%%%%%%%%%%%%%%%%%%%%%%
%%%%%%%%%%%%%%%%%%%%%%%%%%%%%%%%%%%%%%%%%%%%%%%%%%%%%%%%%%%%%%%%%%%%%%%%%%%%%%%%%%%%%%%%%%%%%%%%%%%%%%%
%%%%%%%%%%%%%%%%%%%%%%%%%%%%%%%%%%%%%%%%%%%%%%%%%%%%%%%%%%%%%%%%%%%%%%%%%%%%%%%%%%%%%%%%%%%%%%%%%%%%%%%
\section{Introduction} 
\label{sect_intro}

A major finding in modern astronomy has been the discovery that the growth of supermassive black holes (SMBHs) may be linked to the growth of their host galaxies. The observational relations between SMBHs and their host galaxies \citep{Korm13,Gasp19} have suggested that they undergo a process of co-evolution; however, the underlying mechanism  is not yet fully understood. In active galactic nuclei (AGN), accretion onto the SMBH leads to the release of an enormous amount of power, but how this power is transferred from the small scale of the SMBH to the larger scale of the galaxy is uncertain (see \citealt{Gasp20} for a recent review). Importantly, accretion in AGN is accompanied by outflowing winds, which can transfer matter and energy away from the nucleus and into the interstellar medium (ISM) of the host galaxy. Thus, the AGN outflows may couple the SMBHs to their host galaxies \citep{King15,Gasp17,Harr18}. The resulting feedback mechanism between the AGN and star formation may have important implications for understanding how the observed galaxy population in the universe is formed and how macro-scale cooling flows are quenched \citep{Silk98,King15,Gasp13}.

Case studies, alongside statistical surveys, both play an important role in advancing our understanding of winds in AGN. Bright Seyfert-1 galaxies in the local universe have been ideal laboratories for carrying out the most detailed high-resolution UV/X-ray spectroscopic analyses of outflows (e.g., the multi-wavelength campaign on Mrk~509, \citealt{Kaa11a}). Such case-studies have uncovered and mapped the multi-component ionization and kinematic structure of winds at the core of AGN (e.g., in NGC~5548, \citealt{Kaas14,Arav15}). Statistical surveys of the parameters of both the ionized disk winds (e.g., \citealt{Blu05,Kris06,Tomb10,Laha14,Mehd19}) and the large galactic-scale molecular and ionized outflows (\citealt{Cico14,Fior17,Veil17,Brus18,Huse19,Robe20,Flue21}) have established important relations between winds and the parameters characterizing the AGN and the galaxy (e.g., \citealt{Carn15,Feru15,Tomb15,Fior17,Smit19,Mara20,Tozz21}). These scaling relations provide us with key insights into the launch and driving mechanisms of AGN winds and understanding how their energy and momentum are propagated throughout the AGN and its host galaxy. Such wind surveys, conducted over wide ranges of luminosities and redshifts, are needed for tracing the role and impact of AGN outflows in the evolution of galaxies. Such observational results would provide useful constraints and diagnostic information for theoretical campaigns and hydrodynamical simulations that are aimed at advancing our overall understanding of the SMBH growth and the AGN feedback cycle together on all scales. One example is  the chaotic cold accretion (CCA) scenario \citep{Gasp13b,Gasp17b,Gasp18}, which is suggested to trigger powerful multi-phase AGN outflows that are capable of quenching star formation and macro-scale cooling flows \citep{Gasp17}.

Different types or forms of ionized outflows, with their distinct characteristics, have been discovered in AGN. High-resolution UV spectroscopy, facilitated by HST, has played a crucial role in studies of these AGN outflows by constraining their kinematics and physical structure. The HST spectroscopy, alongside studies in other wavelengths, allows for a more complete picture of AGN ionized outflows to be obtained. The UV and X-ray observations of bright AGN have revealed a complex and multifarious landscape of ionized outflows and winds in AGN, seen as the broad absorption line (BAL) outflows (e.g., \citealt{Arav01,Xu20}),  warm absorber (WA) outflows (e.g., \citealt{Math94,Math95,Kaa00,Crens12,Laha21}), ultra-fast outflows (UFOs, e.g., \citealt{Char02,Tomb10,Reev18a}), and transient obscuring winds (e.g., \citealt{Kaas14,Mehd17}). In addition, each type of outflow is commonly seen with multiple ionization and velocity components. Currently, the relations between these different types of AGN outflows are not fully understood and their origin and driving mechanisms -- either thermal \citep{Krol01,Mizo19}, radiative \citep{Prog04,Gius19}, or magnetic \citep{Fuku10,Sado17} -- are still open questions. It has not yet been determined whether different types of outflows are manifestation of the same primary outflow \citep{Tomb12} or if they are outflows with different origins and launching mechanisms. Moreover, their association with molecular and galactic-scale outflows has not been fully explored as studying the connection between different spatial and time scales remains challenging \citep{Cico18}. The above types of outflows are found to produce spectral signatures in the UV band either as narrow and/or broad absorption lines. Thus, their UV spectroscopy complements the X-ray studies of the outflows and provides additional information, such as details of their kinematic and ionization structure, as well as their metalicities, which cannot currently be fully ascertained from the X-rays alone.

Models of AGN feedback suggest that the kinetic luminosity of AGN outflows needs to be at least 0.5--5\% of the Eddington luminosity to exert a significant impact on the galaxy evolution \citep{DiMat05,Hopk10}. The energetic UFOs, with relativistic outflow velocities, are a crucial component of AGN outflows as they have an adequate level of kinetic luminosity to play a key role in AGN feedback \citep{Tomb10,Tomb12,Tomb13,Goff15,Nard15}. UFOs usually consist of highly ionized outflows, which primarily imprint their absorption signatures in hard X-rays in the Fe-K band, namely through \ion{the Fe}{xxv} and \ion{Fe}{xxvi} lines; as seen, for instance, in the case studies of APM~08279+5255 \citep{Char02}, PG~1211+143 \citep{Poun03}, and PDS~456 \citep{Reev18a}. X-ray studies of nearby Seyfert-1 galaxies find that about 40\% of them have highly ionized UFOs, detected in dozens of targets \citep{Tomb10,Goff13}. The presence of UFOs alongside molecular outflows have been discovered, suggesting a physical connection between the small-scale winds from the accretion disk and large-scale galactic outflows \citep{Tomb15,Feru15,Long18,Char20,Mara20,Tozz21}. However, this connection is not straightforward and outstanding questions remain \citep{Nard18,Bisc19,Zubo20,Mara20,Tozz21}. There are still significant gaps in our understanding of how UFOs operate and how they relate to the less energetic warm-absorber outflows \citep{Blu05,Tomb13,Laha14} -- which are seen as narrow lines with moderate velocities in the UV and X-ray bands and which may also originate from the AGN torus \citep{Krol01}.

Observational results in recent years show that highly ionized UFOs may have lower ionization counterparts in the UV and soft X-rays (e.g., \citealt{Gupt13,Gupt15,Long15,Reev16,Park17,Kris18a,Sera19,Chart21,Kron21,Mehd22b}). The UV spectral signatures of X-ray UFOs have been discovered with HST as a broad and relativistically blueshifted H~I Ly$\alpha$ absorption feature in PG~1211+143 \citep{Kris18a}, and possibly a relativistic \ion{C}{iv} feature in PDS~456 \citep{Hama18}. In addition, a multi-component UV UFO has been found in quasar J1538+0855 \citep{Viet22}. Recently, the UV counterpart of a multi-component X-ray UFO was found in IRAS~17020+4544 as a narrow, relativistically blueshifted, Ly$\alpha$ absorption line \citep{Mehd22b}. Interestingly, the ionization parameter $\xi$ of the UV and soft X-ray UFOs is much lower than that of the hard X-ray UFOs, yet their outflows travel at relativistic velocities. The ionization parameter $\xi$ \citep{Kro81} is defined as ${\xi = {L}\, /\, {{n_{\rm{H}} r^2 }}}$, where $L$ is the luminosity of the ionizing source over 1--1000 Ryd, $n_{\rm{H}}$ the hydrogen density, and $r$ the distance between the gas and the ionizing source. One plausible explanation that has been proposed for the low-ionization UFOs is the entrained-UFO model \citep{Gasp17,Sanf18,Sera19,Long20}, where the inner primary UFO moving at relativistic speeds, entrains the surrounding gas, pushing it at velocities comparable to that of the UFO, while retaining its ionization state and column density. This results in the formation of multi-ionization phase UFOs, including lower ionization components, which would be detectable in the HST band mainly through Ly$\alpha$ absorption, as well as (possibly) other lines such as \ion{C}{iv} and \ion{N}{v}.

Apart from BAL outflows, warm absorbers, and UFOs, a new form of outflow has come into light in recent years: transient obscuring outflows. In contrast to the warm absorbers, which are located at pc-scale distances from the black hole with moderate column densities \citep{Kaas12,Krol01}, the obscuring winds that have been found in NGC~5548 \citep{Kaas14} and NGC~3783 \citep{Mehd17} are a different type of AGN wind. They are more massive and faster, and are located at only a few light days from the black hole. Also, they are transient, highly variable, and partially cover the central X-ray source. In particular, UV spectroscopy with HST \citep{Long13,Kaas14} has been crucial in understanding the nature of the X-ray transient obscuration events \citep{Mark14} and in deriving the parameters of the obscuring outflows. These obscuring winds, which heavily absorb the soft X-ray continuum, can appear with an associated blue-shifted and broad UV absorption component \citep{Kris19,Kris19b,Kara21}, and in some cases with a high ionization component in the Fe~K band \citep{Mehd17}. The studies suggest the location of these obscuring winds is the broad-line region (BLR) and they are likely to be accretion disk winds \citep{Kaas14,Mehd17,Long19}. This nuclear obscuration has a significant impact on the ionization state and our interpretation of both the warm absorbers and the BLR emission lines \citep{Dehg19,Dehg19b,Dehg20}. Such obscurations may also be a result of the CCA's condensation phase, which can produce large column density variations near the SMBH \citep{Gasp13b,Gasp17b}.

Our \sub program and its objectives are described in the following section. In Sect. \ref{sect_data}, we describe the HST observations and their data processing and preparation. The spectroscopic analysis and modeling of the HST spectra are presented in Sect. \ref{sect_model}. We discuss our findings in Sect. \ref{sect_discuss} and give our concluding remarks in Sect. \ref{sect_concl}.

%%%%%%%%%%%%%%%%%%%%%%%%%%%%%%%%%%%%%%%%%%%%%%%%%%%%%%%%%%%%%%%%%%%%%%%%%%%%%%%%%%%%%%%%%%%%%%%%%%%%%%%
%%%%%%%%%%%%%%%%%%%%%%%%%%%%%%%%%%%%%%%%%%%%%%%%%%%%%%%%%%%%%%%%%%%%%%%%%%%%%%%%%%%%%%%%%%%%%%%%%%%%%%%
%%%%%%%%%%%%%%%%%%%%%%%%%%%%%%%%%%%%%%%%%%%%%%%%%%%%%%%%%%%%%%%%%%%%%%%%%%%%%%%%%%%%%%%%%%%%%%%%%%%%%%%
\section{The SUBWAYS campaign}
\label{sect_sample}

The \sub campaign, introduced in \cite{Matz22} (hereafter,  \paperI), is designed to probe the AGN wind properties of the previously underexplored region of the luminosity-and-redshift parameter space: between the low-luminosity Seyfert-1 galaxies in the local universe and the more luminous quasi-stellar objects (QSOs) at higher redshifts. This corresponds to AGN at intermediate redshifts of ${0.1 < z < 0.4}$ with bolometric luminosity $L_{\rm bol}$ of about $10^{45}$--$10^{46}$ \ergs. By deriving the parameters of all types of outflows that are seen in the UV and X-ray spectra of the \sub sample, their statistical properties and relations to each other and other AGN properties are established. A comparison of the wind parameters and energetics in the \sub sample with those in AGN populations at lower and higher redshifts and luminosities is useful for investigations of the AGN feedback models as a function of redshift and luminosity.

We selected targets for the \sub campaign from the \xmm serendipitous survey catalog (3XMM-DR7, \citealt{Rose16}), with confirmed matches in either the Sloan Digital Sky Survey catalog (SDSS-DR14, \citealt{Abol18}) or in the Palomar bright quasar survey catalog \citep{Schm83}. Thus, from their known X-ray and UV characteristics, the required exposures with \xmm, \nustar, and HTS were planned. At the end, 24 AGN satisfied the selection for the \sub sample with $z$ of $0.1$--$0.4$ and $L_{\rm bol}$ of $10^{44.6}$--$10^{46.3}$ erg~s$^{-1}$. These comprise 19 objects proposed in a dedicated \xmm large program (PI: M. Brusa) in 2019--2020 (with only 17 out of 19 actually observed), along with five AGN with high quality archival \xmm data that were already available. 

The \sub targets observed with HST (21 AGN) are listed in Table \ref{table_log}. This HST study covers the same targets observed in X-rays (\paperI), with the following exceptions. \paperI includes three additional targets (2MASS~J10514425+3539306, 2MASS~J16531505+2349427, and WISE~J053756.30-024513.1), which, due to their being heavily reddened, were not observed with HST. Furthermore, the HST program has two targets (2MASS~J14025120+2631175 and PG~1427+480) that are not included in \paperI due to a lack of adequate X-ray data. The HST targets with proposal ID 15890 in Table \ref{table_log} (16 targets) are those observed most recently through our proposal in HST Cycle 27 (PI: G. Kriss). For the remaining five targets, we made use of the archival HST UV spectra. 

In \paperI, the results of the X-ray spectroscopy of the \sub sample were presented. In seven of the targets, evidence of X-ray UFOs was found, where high-velocity Fe K absorption was detected at ${\gtrsim 95\%}$ confidence level. In this paper, we study the ionized outflows with HST and also look for any counterparts of the X-ray UFOs. We identify and parameterize all the intrinsic UV absorption lines in the sample. We also model the UV emission lines and examine their asymmetry. We look for relations between the parameters of the UV outflows and other properties of the AGN such as their bolometric luminosity.

%============================
% TABLE: data log
%
\begin{table*}[!tbp]
\begin{minipage}[t]{\hsize}
\setlength{\extrarowheight}{0.6pt}
\setlength{\tabcolsep}{5pt}
\caption{Log of the HST UV grating observations of the targets in the \sub sample.}
\centering
\footnotesize
\renewcommand{\footnoterule}{}
\begin{tabular}{c c c c c c c c c}
\hline \hline
Object & Redshift & $\log L_{\rm bol}$ & $\log M_{\rm BH}$    & Prop. & HST        & Grating\,/          & Obs. Date     & Exp.  \\
Name   & $z$      & (erg s$^{-1}$)     & (M$_{\odot}$)        & ID    & Inst.      & Central $\lambda$   & yyyy-mm-dd     & (ks)  \\
\hline
\object{2MASS J02201457-0728591}          & 0.21343 (1) & 46.33 & 8.42 & 15890 & COS  & G140L\,/\,1105 & 2020-01-03 &    4.4 \\
\object{2MASS J10514425+3539306}    & 0.15881 (1) & 44.88 & 8.40 & -     & -    & -              & -          &   -    \\
\object{2MASS J14025120+2631175}    & 0.18762 (1) & 45.44 & 8.55 & 15890 & COS   & G140L\,/\,1105 & 2019-12-03 &         4.4 \\
\object{2MASS J16531505+2349427}    & 0.10344 (1) & 45.37 & 6.98 & -     &      & -              & -          &   -    \\
\object{HB89 1257+286}              & 0.09117   (1) & 44.58 & 7.46 & 4952  & FOS  & G130H\,/\,1600 & 1993-07-15 &   0.8 \\
\object{HB89 1529+050}              & 0.21817 (1) & 45.17 & 8.75 & 15890 & COS  & G140L\,/\,1105 & 2019-08-30 &  7.0 \\
\object{LBQS 1338-0038}             & 0.23745 (1) & 45.27 & 7.77 & 15890 & COS  & G140L\,/\,1105 & 2020-01-28 &  4.3 \\   
\object{PG 0052+251}                  & 0.15445 (2) & 45.72 & 8.41 & 15890 & COS  & G140L\,/\,1105 & 2019-11-20 &  1.8 \\   
"                                         & "           & "     & "    & 14268 & COS  & G130M\,/\,1291,1327 & 2015-10-03 & 3.0    \\      
"                                         & "           & "     & "    & 14268 & COS  & G160M\,/\,1589,1623 & 2015-10-03 & 3.0    \\      
\object{PG 0804+761}                  & 0.10000 (3) & 45.27 & 8.31 & 11686 & COS  & G130M\,/\,1291,1300,1309,1318 & 2010-06-12 & 6.0        \\      
"                                       & "           & "     & "    & 11686 & COS  & G160M\,/\,1589,1600,1611,1623 & 2010-06-12 & 6.3        \\      
\object{PG 0947+396}                  & 0.20553 (1) & 45.89 & 8.68 & 15890 & COS  & G140L\,/\,1105 & 2020-04-25 &  0.3      \\
"                                       & "           & "     & "    & 15890 & COS  & G130M\,/\,1222 & 2020-04-25 &  0.7      \\
"                                       & "           & "     & "    & 15890 & COS  & G160M\,/\,1600 & 2020-04-25 &  0.6      \\
"                                       & "           & "     & "    & 15890 & COS  & G185M\,/\,1953 & 2020-04-26 &  1.0  \\
\object{PG 0953+414}                  & 0.23410 (4) & 46.33 & 8.24 & 15890 & COS  & G140L\,/\,1105 & 2020-04-04 &  1.5 \\
"                                       & "           & "     & "    & 12038 & COS  & G130M\,/\,1291,1300,1309,1318 & 2011-10-18 & 5.3        \\
"                                       & "           & "     & "    & 12038 & COS  & G160M\,/\,1589,1600,1611,1623 & 2011-10-18 & 5.6        \\
\object{PG 1114+445}                & 0.14373 (1) & 45.87 & 8.59 & 6484  & FOS  & G130H\,/\,1600  & 1996-11-23  & 12.8      \\
"                                   & "           & "     & "    & 6484  & FOS  & G190H\,/\,2300  & 1996-11-23  & 4.4      \\
"                                   & "           & "     & "    & 6781  & FOS  & G130H\,/\,1600  & 1996-05-13  & 9.3      \\
"                                   & "           & "     & "    & 6781  & FOS  & G190H\,/\,2300  & 1996-05-13  & 1.6      \\
"                                   & "           & "     & "    & 9871  & STIS & G140M\,/\,1173,1400  & 2004-05-28            & 5.3      \\
"                                   & "           & "     & "    & 9871  & STIS & G230M\,/\,1769       & 2004-05-28            & 2.9      \\
\object{PG 1202+281}                  & 0.16501 (5) & 46.30 & 8.61 & 15890 & COS  & G140L\,/\,1105 & 2019-12-14 &  0.9  \\
"                                       & "           & "     & "    & 15890 & COS  & G130M\,/\,1327 & 2019-12-14 &  2.9 \\
"                                         & "           & "     & "    & 12248 & COS  & G130M\,/\,1309,1327 & 2011-04-14 & 2.9  \\
"                                         & "           & "     & "    & 12248 & COS  & G160M\,/\,1577,1600 & 2011-04-14 & 4.8    \\
\object{PG 1216+069}                  & 0.33130 (6) & 45.84 & 9.20 & 15890 & COS  & G140L\,/\,1105 & 2019-12-26 &  1.8 \\           
"                                       & "           & "     & "    & 15890 & COS  & G185M\,/\,1953 & 2019-12-26 &  2.3 \\  
"                                       & "           & "     & "    & 12025 & COS  & G130M\,/\,1291,1300,1309,1318 & 2012-02-04 & 5.1        \\               
"                                       & "           & "     & "    & 12025 & COS  & G160M\,/\,1589,1600,1611,1623 & 2012-02-04 & 5.6        \\
\object{PG 1307+085}                  & 0.15384 (1) & 44.86 & 7.90 & 15890 & COS  & G140L\,/\,1105 & 2019-11-30 &  1.8 \\           
"                                       & "           & "     & "    & 12569 & COS  & G130M\,/\,1309,1327 & 2012-06-16 & 2.0  \\              
\object{PG 1352+183}                  & 0.15147 (1) & 46.26 & 8.42 & 15890 & COS  & G140L\,/\,1105 & 2020-01-12 &  1.8 \\
"                                       & "           & "     & "    & 13448 & COS  & G130M\,/\,1291 & 2014-07-31 & 8.2       \\
"                                       & "           & "     & "    & 13448 & COS  & G160M\,/\,1600 & 2014-07-31 & 8.7       \\
\object{PG 1402+261}                & 0.16430 (7) & 46.34 & 7.94 & 6781  & FOS  & G130H\,/\,1600 & 1996-08-25 & 1.6      \\
"                                   & "           & "     & "    & 6781  & FOS  & G190H\,/\,2300 & 1996-08-25 & 0.2      \\
\object{PG 1416-129}                  & 0.12894 (2) & 45.74 & 9.05 & 6528  & FOS  & G190H\,/\,2300 & 1996-08-23 & 1.0       \\
\object{PG 1425+267}                  & 0.36361 (1) & 46.06 & 9.22 & 15890 & COS  & G140L\,/\,1105 & 2020-01-25 &  3.6 \\   
"                                       & "           & "     & "    & 15890 & COS  & G230L\,/\,3360 & 2020-01-25 &  0.4  \\  
"                                         & "           & "     & "    & 12603 & COS  & G130M\,/\,1291,1327 & 2012-06-23 & 2.2    \\      
"                                       & "           & "     & "    & 14729 & COS  & G160M\,/\,1600 & 2017-05-29 & 5.4  \\
\object{PG 1427+480}                  & 0.22063 (1) & 45.82 & 8.09 & 15890 & COS   & G140L\,/\,1105 & 2019-10-28 &         0.9  \\  
"                                       & "           & "     & "    & 15890 & COS  & G185M\,/\,1864 & 2019-10-28 &  5.6 \\           
\object{PG 1435-067}                  & 0.12900 (8) & 45.51 & 7.77 & 15890 & COS  & G140L\,/\,1105 & 2020-01-19 &  1.8 \\
"                                       & "           & "     & "    & 12569 & COS  & G130M\,/\,1291,1309 & 2012-02-29 & 2.1  \\
\object{PG 1626+554}                  & 0.13170 (9) & 46.02 & 8.54 & 15890 & COS   & G140L\,/\,1105 & 2020-07-13 &         1.8 \\
"                                         & "           & "     & "    & 12029 & COS     & G130M\,/\,1291,1300,1309,1318 &       2011-06-15 & 3.9 \\
"                                       & "           & "     & "    & 12029 & COS   & G160M\,/\,1589,1600,1611,1623 &       2011-06-15 & 4.3  \\
\object{SDSS J144414.66+063306.7}         & 0.20768 (1) & 45.34 & 8.10 & 15890 & COS  & G140L\,/\,1105 & 2019-12-22 &    1.8 \\
\object{WISE J053756.30-024513.1}   & 0.10983 (1) & 44.86 & 7.73 & -     & -    & -              & -          &  -     \\        
\hline
\end{tabular}
\end{minipage}
\tablefoot{
The \sub targets observed with HST overlap with those observed in X-rays (\paperI) with the following exceptions: three of the \sub targets that are studied in X-rays in \paperI are too heavily reddened and thus are not observed with HST: 2MASS~J10514425+3539306, 2MASS~J16531505+2349427, and WISE~J053756.30-024513.1; two of the targets that are observed with HST are excluded in \paperI due to lack of sufficient X-ray data: 2MASS~J14025120+2631175 and PG~1427+480. The new HST data that we obtained in HST Cycle 27 correspond to the proposal ID number 15890 (16 targets) and the remaining targets have only archival HST data (5 targets). References for cosmological redshifts $z$ are: (1) \citet{Alba17}; (2) \citet{Ho09}; (3) \citet{Schm83}; (4) \citet{Marz96}; (5) \citet{Alba15}; (6) \citet{Gang13}; (7) \citet{Hu21}; (8) \citet{Monr16}; (9) \citet{Tang12}. For references of the bolometric luminosities, $L_{\rm bol}$, and the black hole masses, $M_{\rm BH}$, see \paperI and references therein. Targets are ordered alphabetically by their name for ease in looking them up.
}
\label{table_log}
\end{table*}
%============================

%============================
% FIG: Overview COS spectra (PART 1)
%
\begin{figure*}
\centering
\resizebox{0.99\hsize}{!}{
\includegraphics[angle=0]{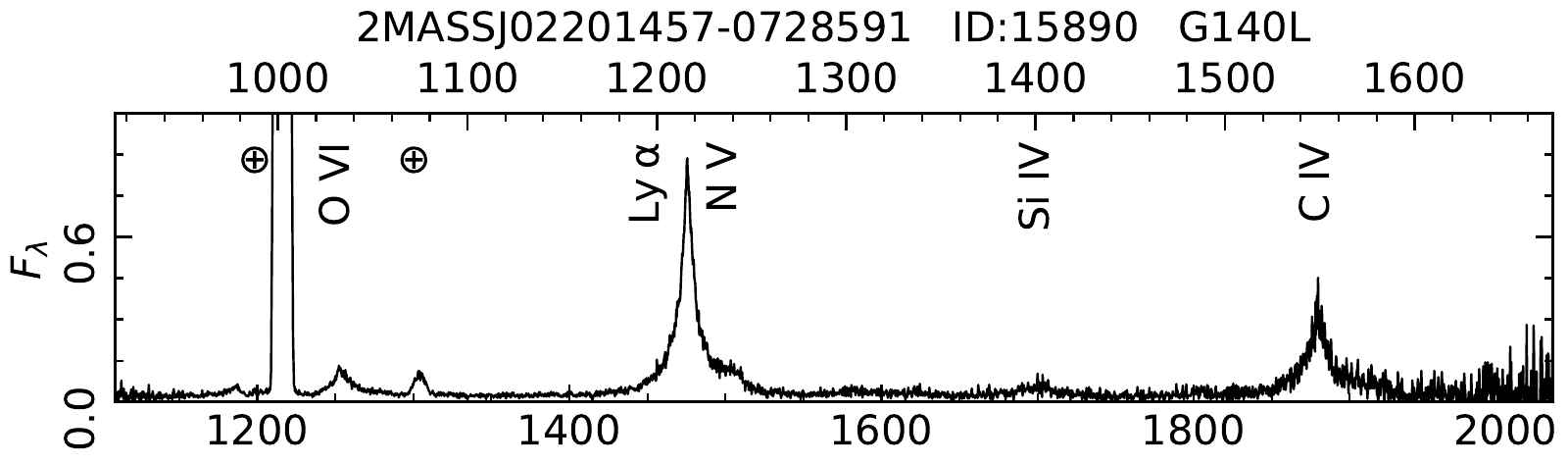}
\includegraphics[angle=0]{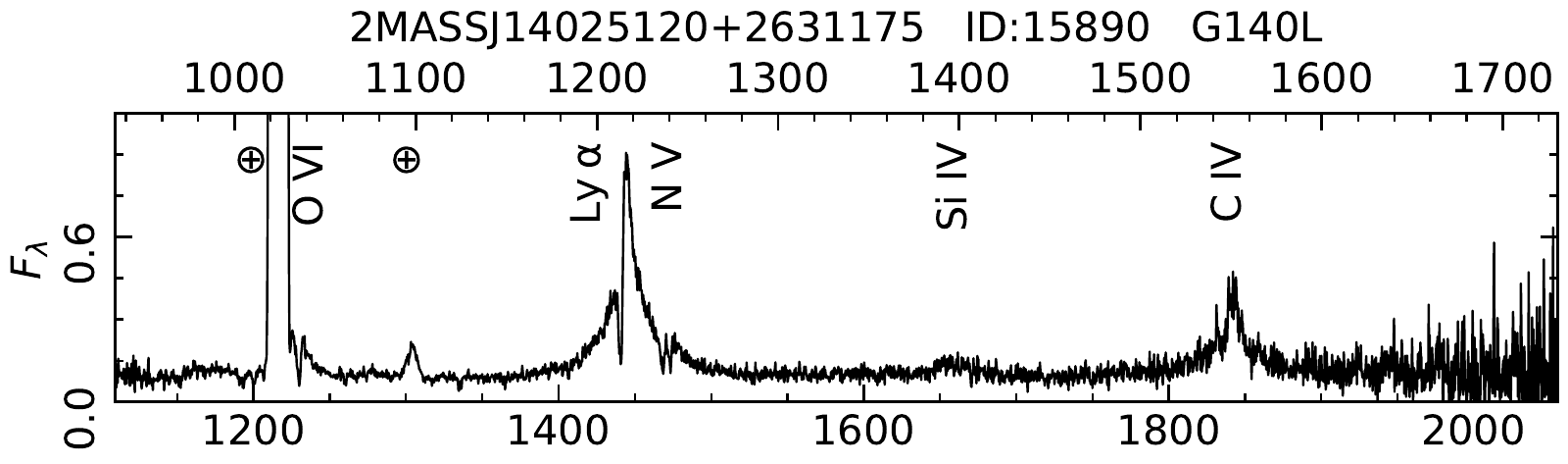}
}
\resizebox{0.99\hsize}{!}{
\includegraphics[angle=0]{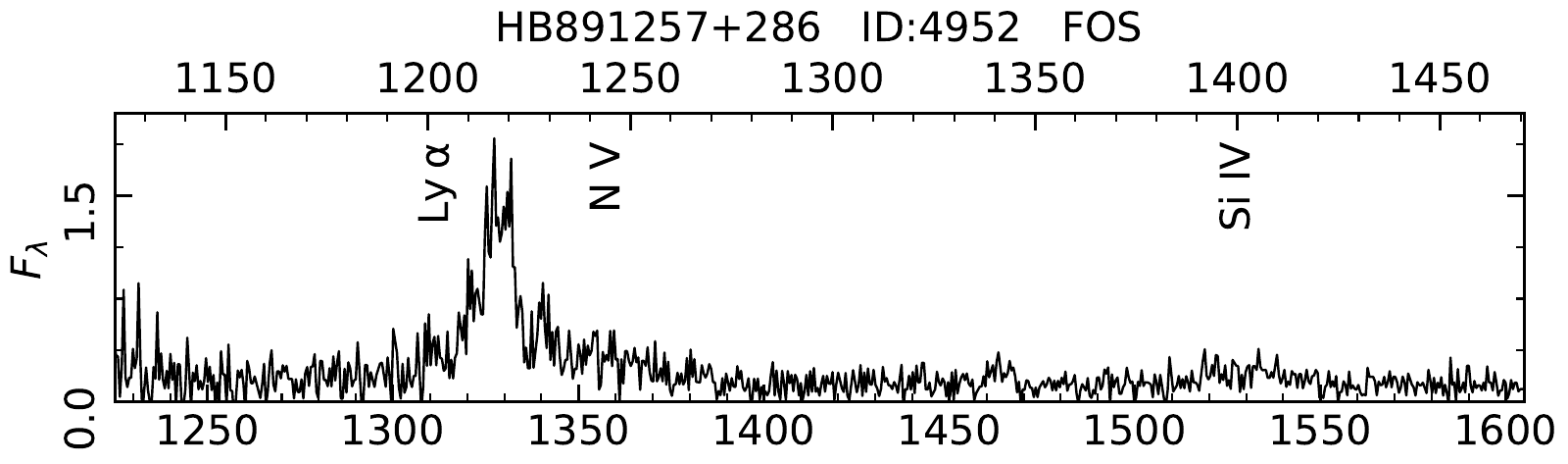}
\includegraphics[angle=0]{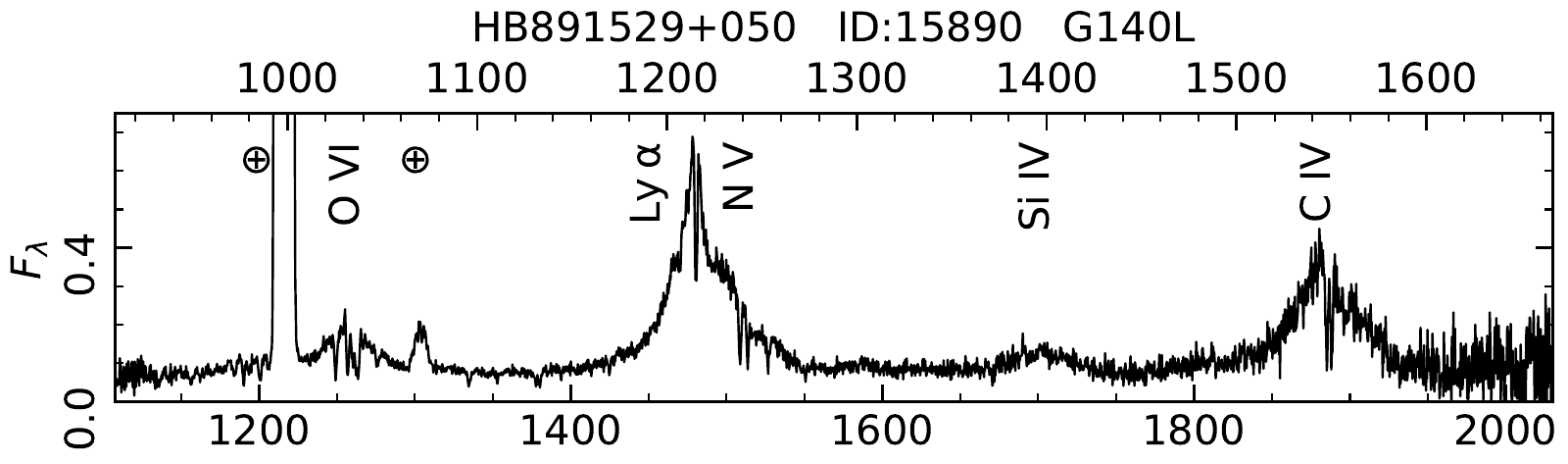}
}
\resizebox{0.99\hsize}{!}{
\includegraphics[angle=0]{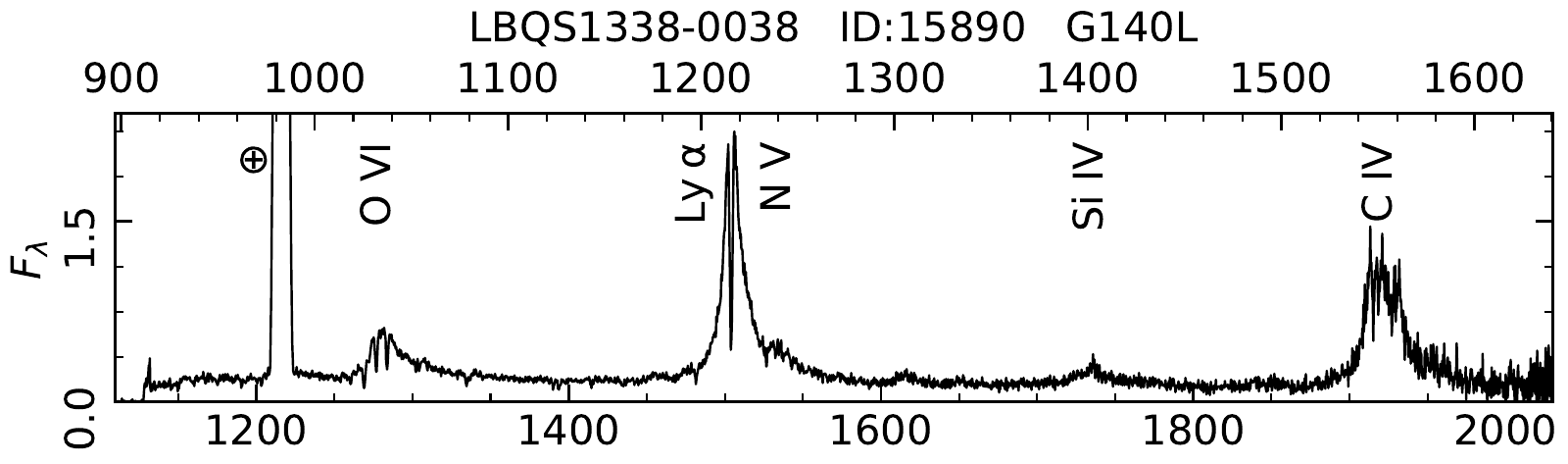}
\includegraphics[angle=0]{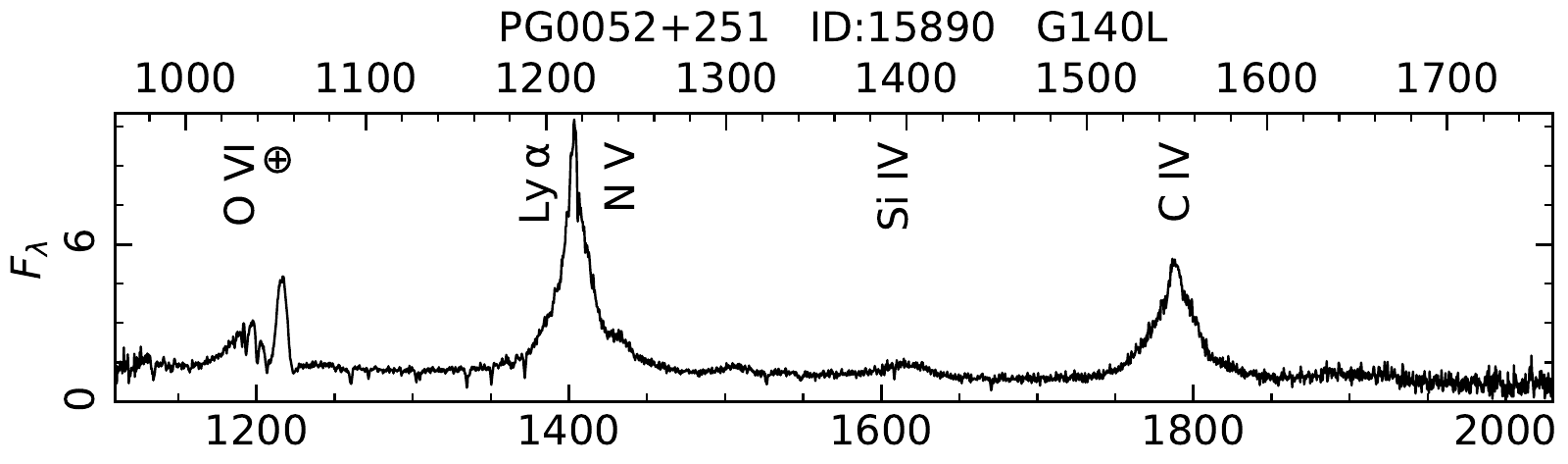}
}
\resizebox{0.99\hsize}{!}{
\includegraphics[angle=0]{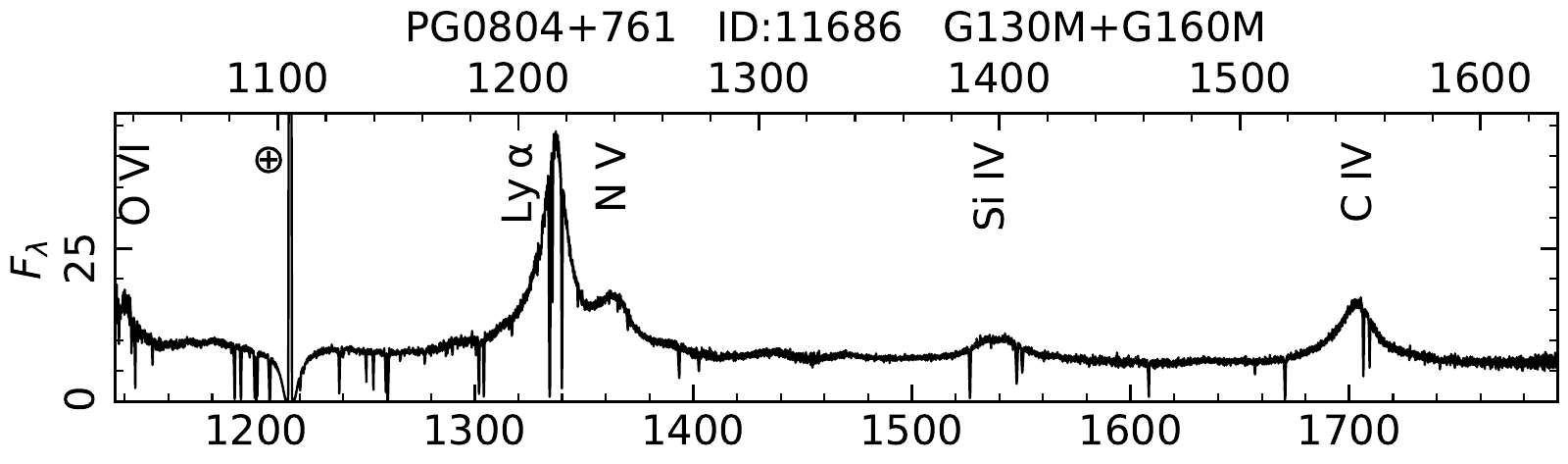}
\includegraphics[angle=0]{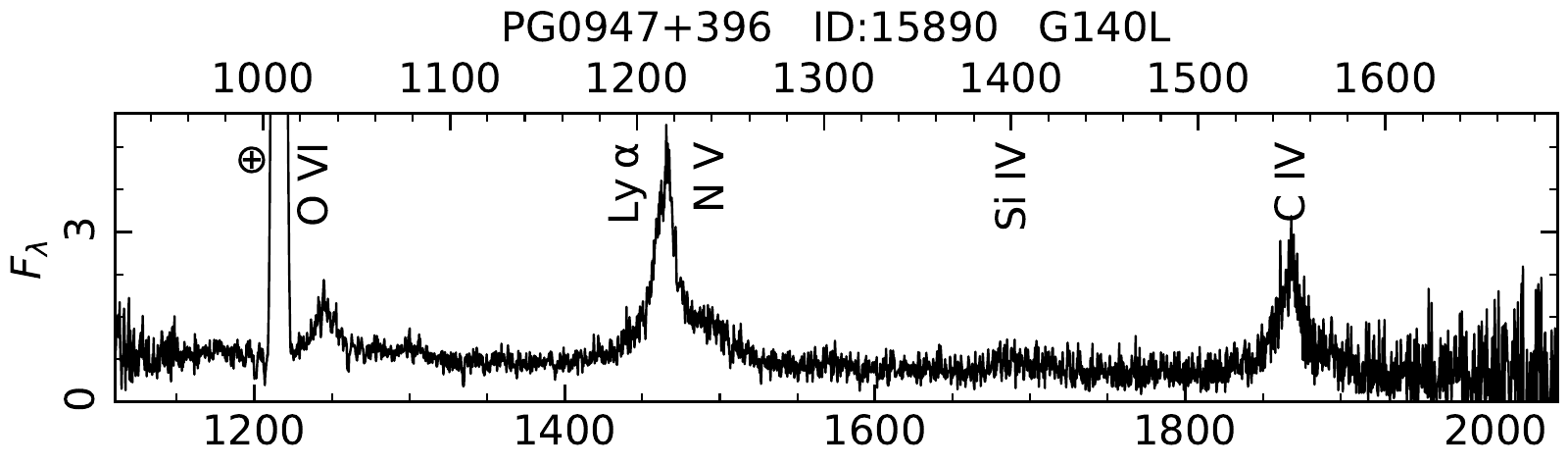}
}
\resizebox{0.99\hsize}{!}{
\includegraphics[angle=0]{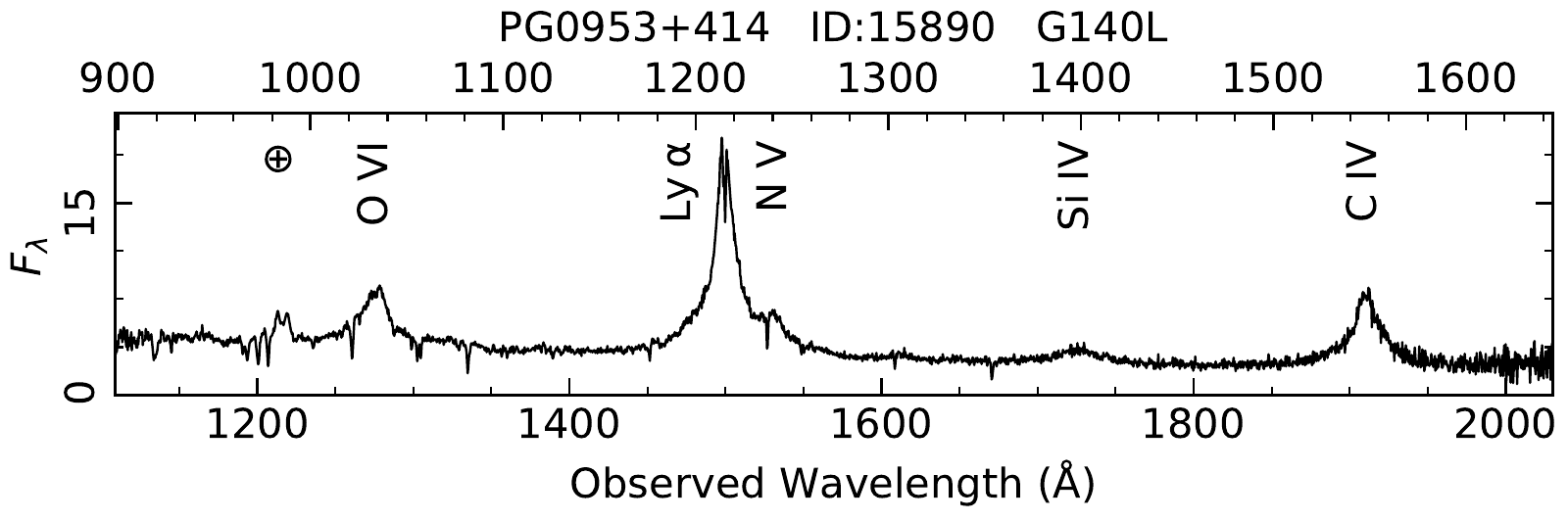}
\includegraphics[angle=0]{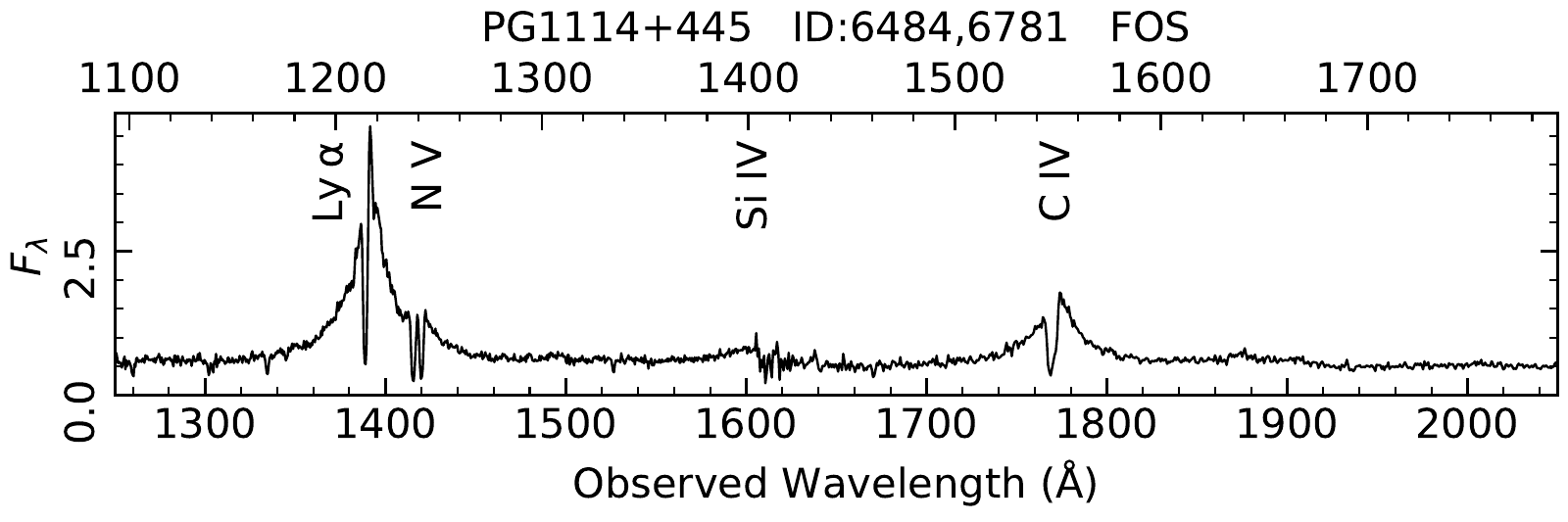}
}
\caption{Overview of the HST spectra of the AGN in the \sub sample. The observed flux $F_{\lambda}$ is in units of ${10^{-14}}$ erg~s$^{-1}$~cm$^{-2}$~\AA$^{-1}$. The observed (bottom axis) and rest-frame (top axis) wavelengths are shown in each panel. The most prominent AGN emission features are labeled. The geocoronal emission lines are indicated with the symbol $\oplus$. The "ID" label on the panels of this figure and the subsequent figures refers to the HST proposal ID number (Table \ref{table_log}). The figure is  continued on the next page.
\label{fig_overview}}
\end{figure*}
%============================

\setcounter{figure}{0}
%============================
% FIG: Overview COS spectra (PART 2)
%
\begin{figure*}
\centering
\resizebox{0.99\hsize}{!}{
\includegraphics[angle=0]{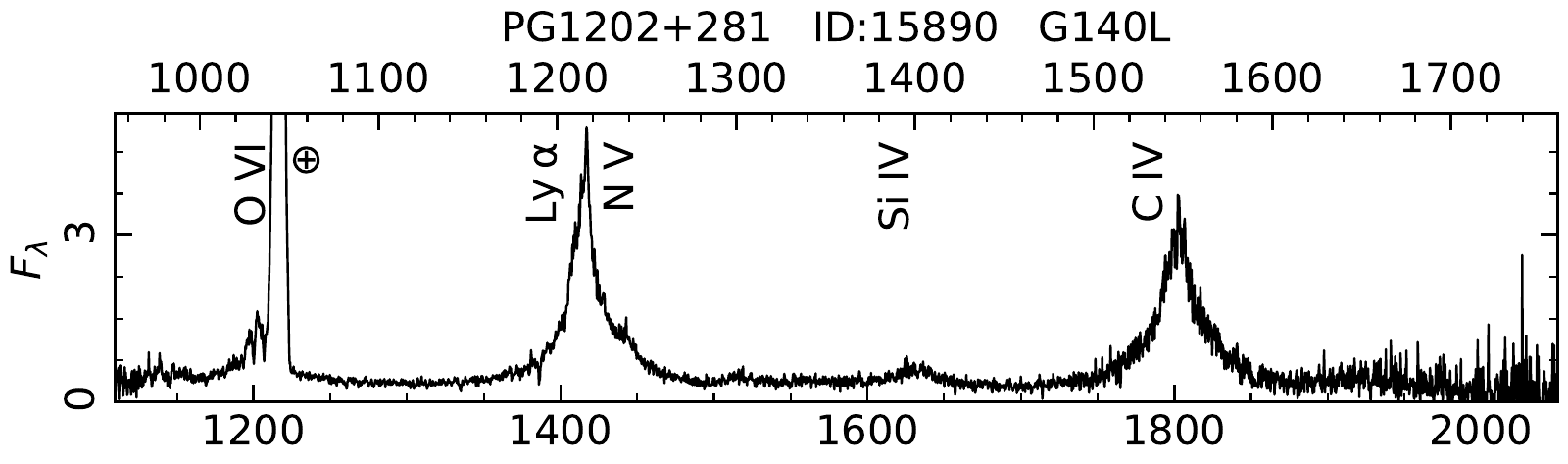}
\includegraphics[angle=0]{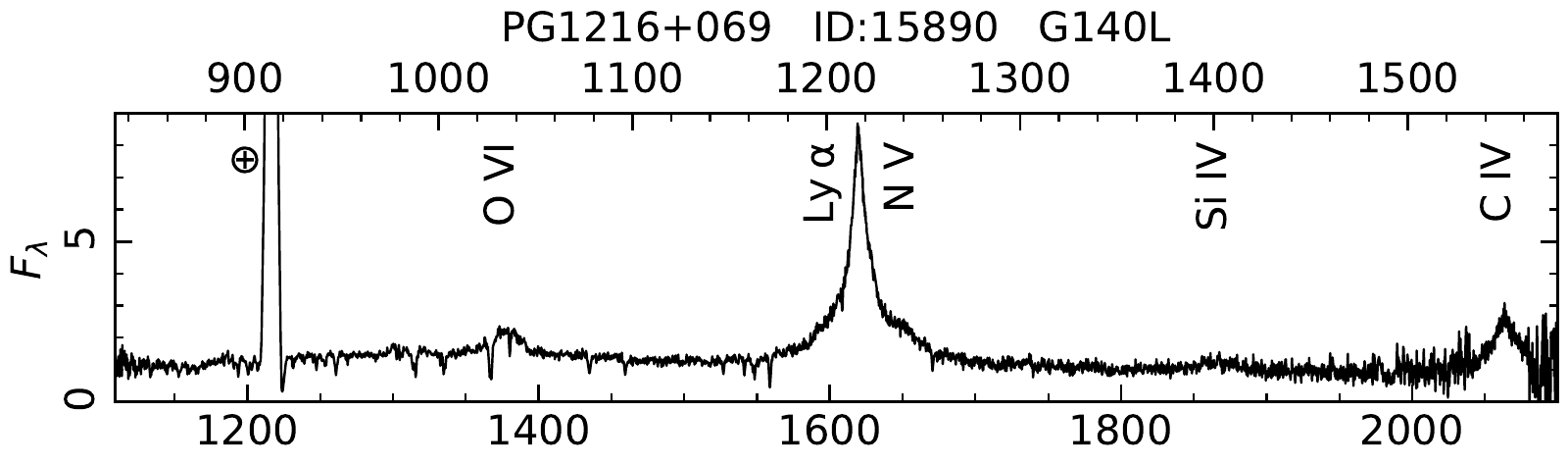}
}
\resizebox{0.99\hsize}{!}{
\includegraphics[angle=0]{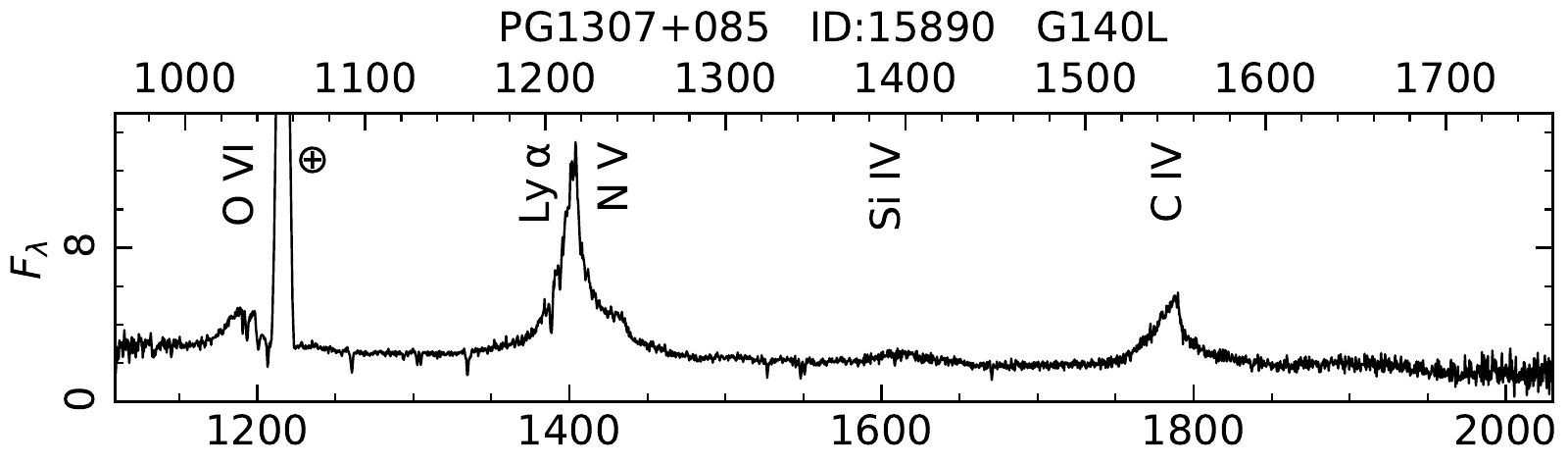}
\includegraphics[angle=0]{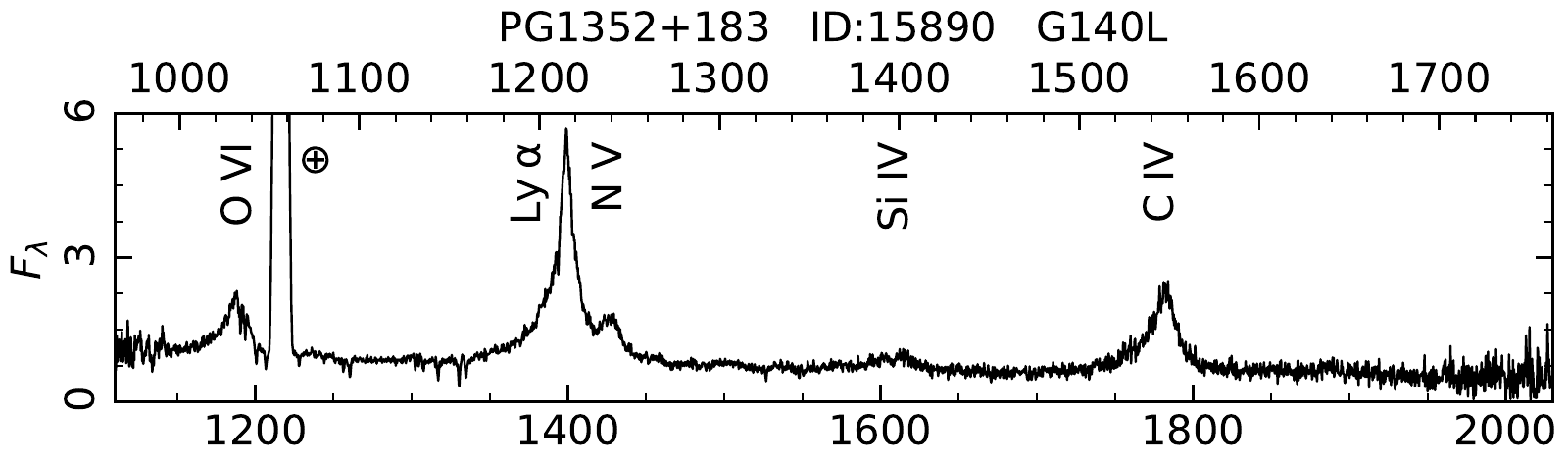}
}
\resizebox{0.99\hsize}{!}{
\includegraphics[angle=0]{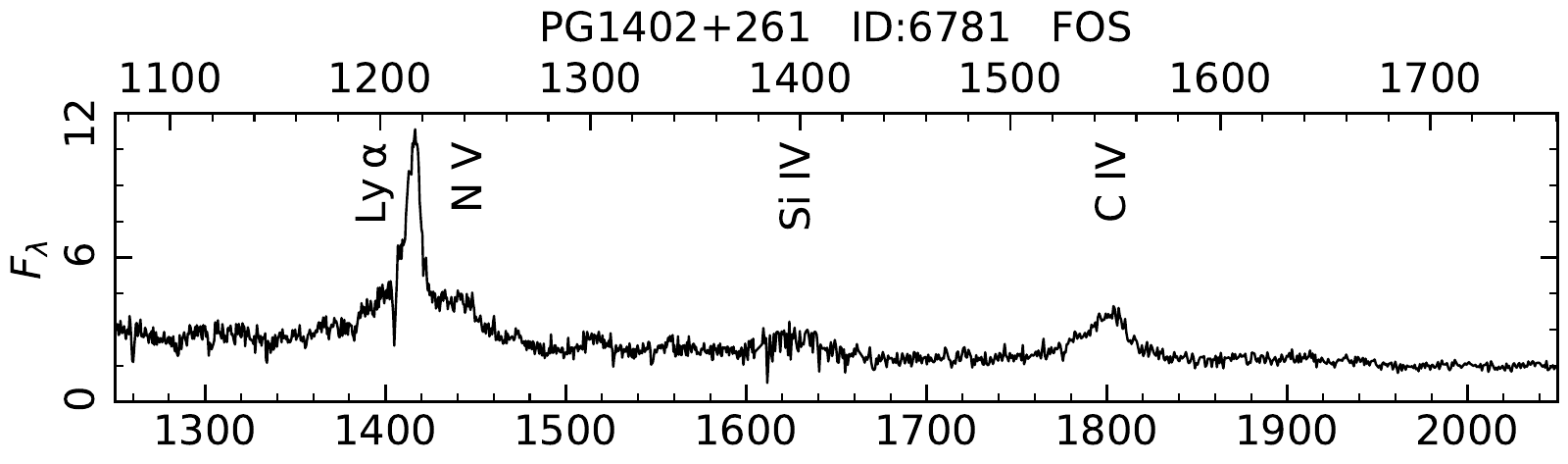}
\includegraphics[angle=0]{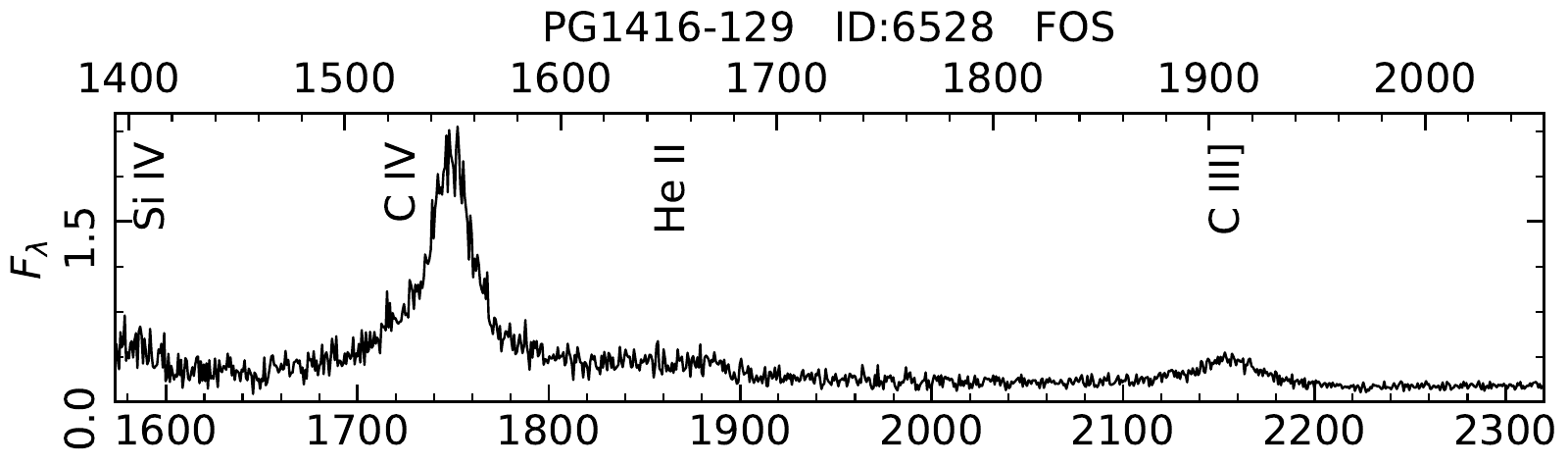}
}
\resizebox{0.99\hsize}{!}{
\includegraphics[angle=0]{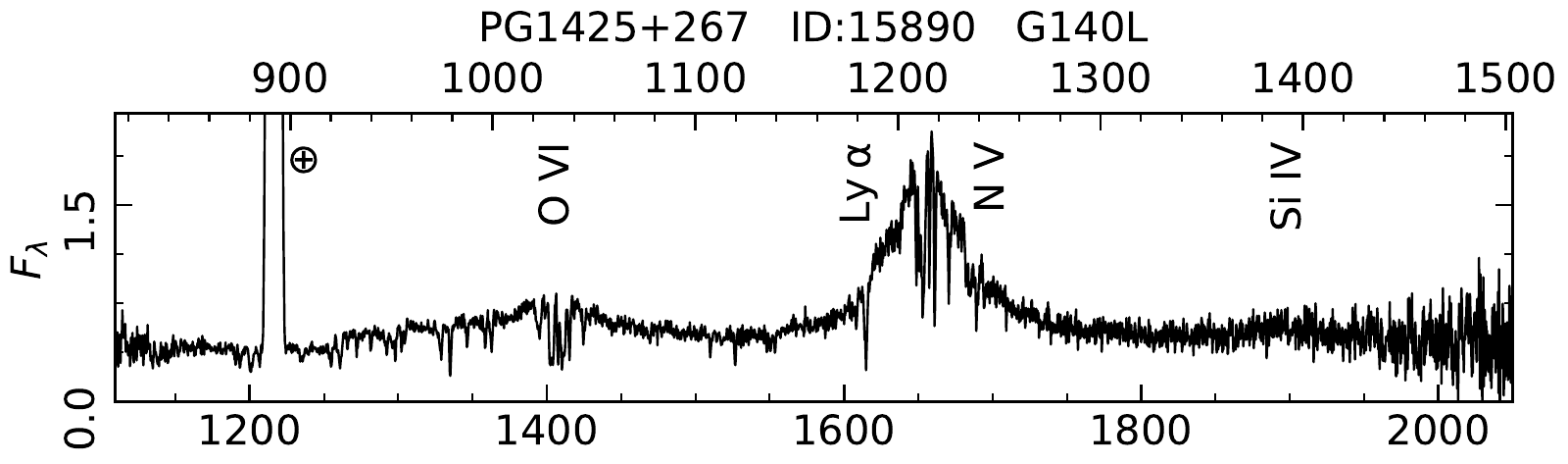}
\includegraphics[angle=0]{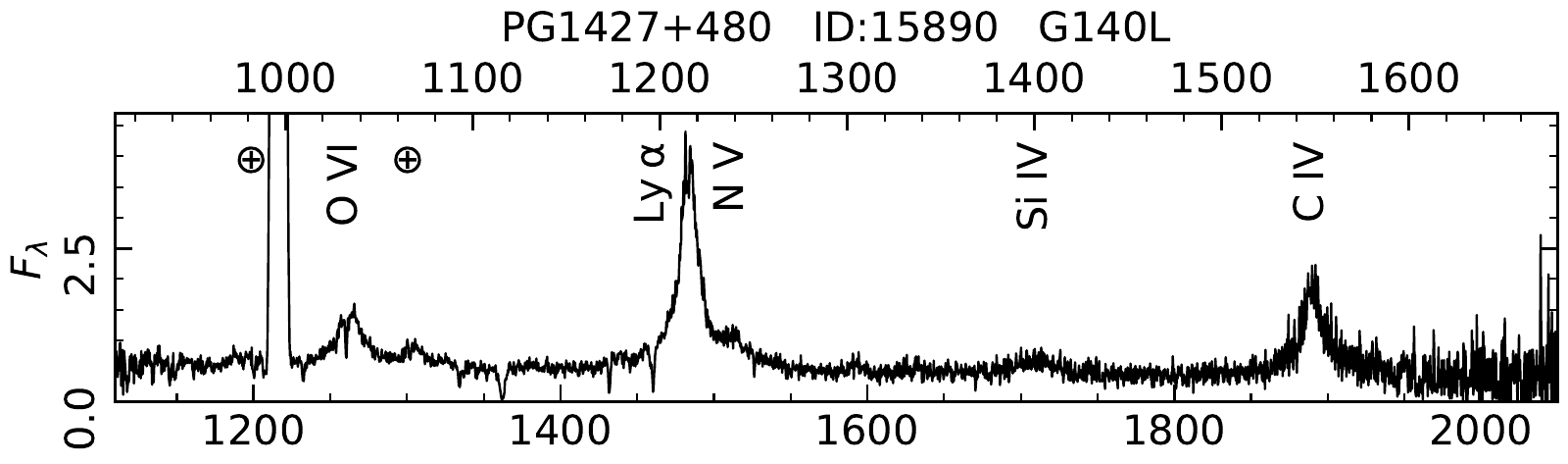}
}
\resizebox{0.99\hsize}{!}{
\includegraphics[angle=0]{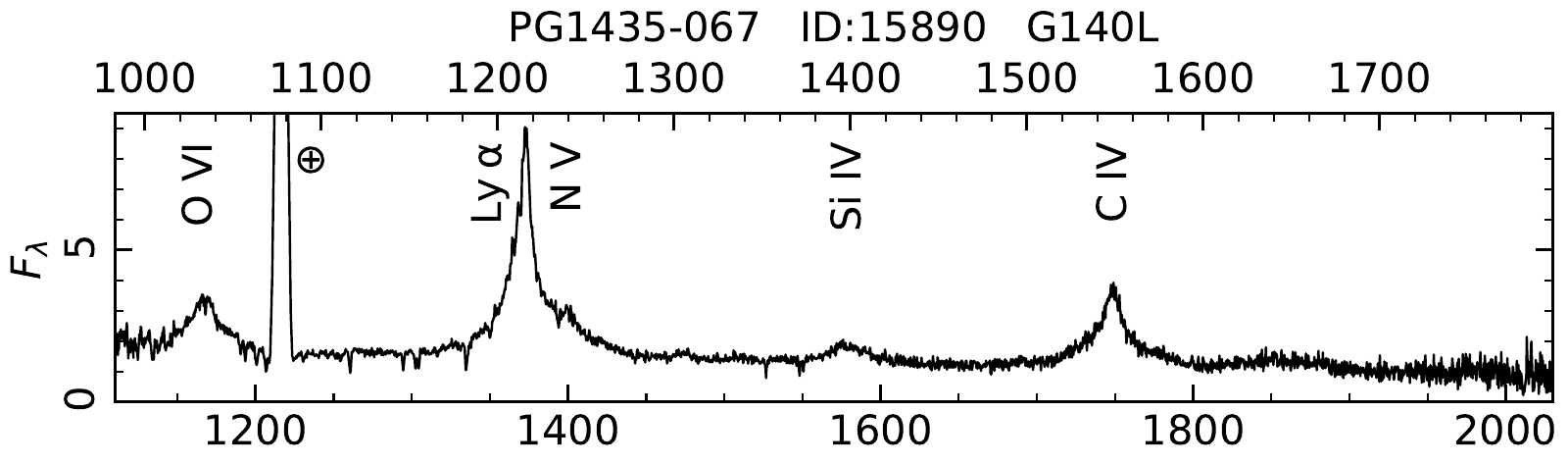}
\includegraphics[angle=0]{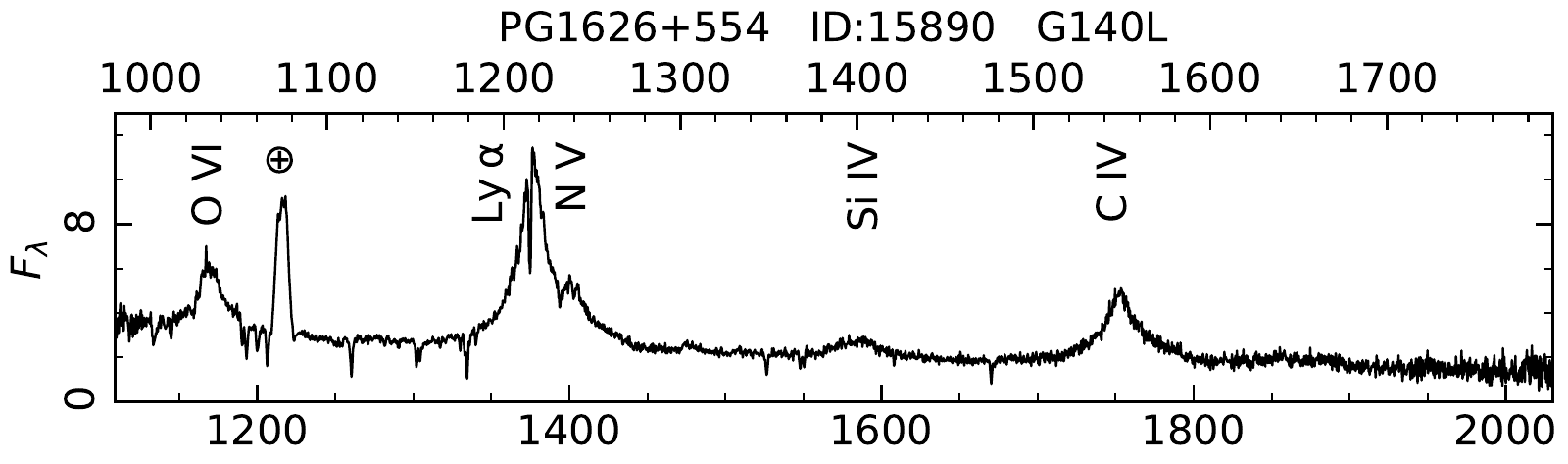}
}
\resizebox{0.496\hsize}{!}{
\includegraphics[angle=0]{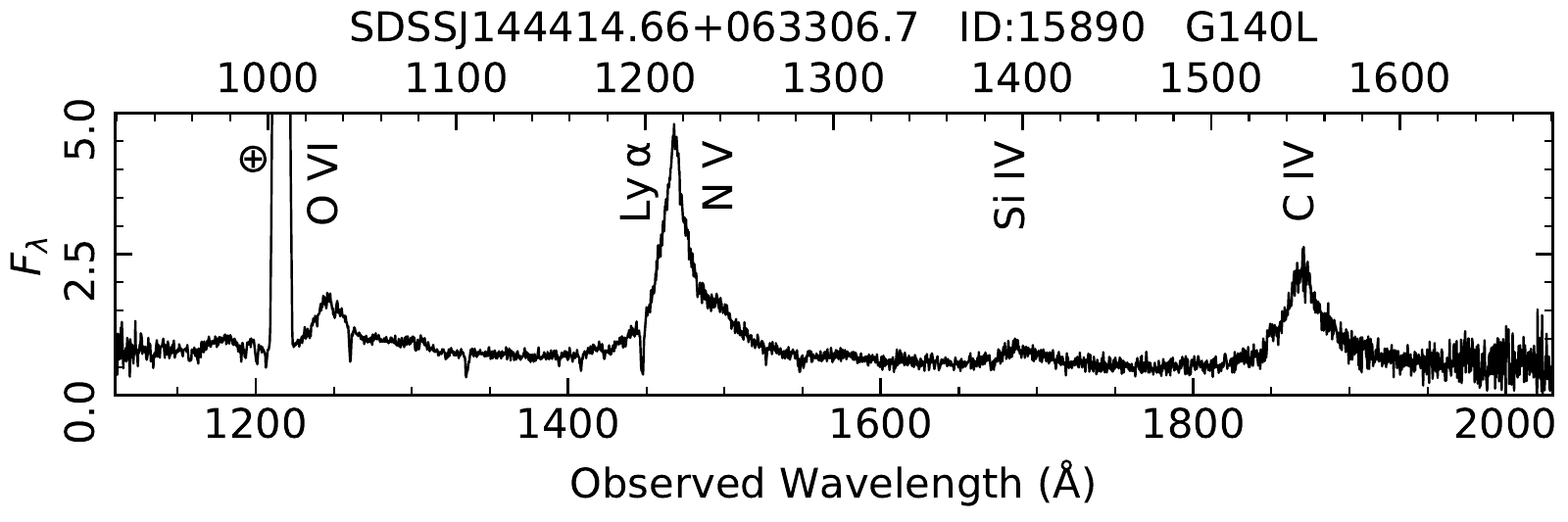}
}
\caption{Continued.
\label{fig_overview}}
%\vspace{0.3cm}
\end{figure*}
%============================

%%%%%%%%%%%%%%%%%%%%%%%%%%%%%%%%%%%%%%%%%%%%%%%%%%%%%%%%%%%%%%%%%%%%%%%%%%%%%%%%%%%%%%%%%%%%%%%%%%%%%%%
%%%%%%%%%%%%%%%%%%%%%%%%%%%%%%%%%%%%%%%%%%%%%%%%%%%%%%%%%%%%%%%%%%%%%%%%%%%%%%%%%%%%%%%%%%%%%%%%%%%%%%%
%%%%%%%%%%%%%%%%%%%%%%%%%%%%%%%%%%%%%%%%%%%%%%%%%%%%%%%%%%%%%%%%%%%%%%%%%%%%%%%%%%%%%%%%%%%%%%%%%%%%%%%
\section{HST observations and data processing} 
\label{sect_data}

The log of the HST observations of 21 targets in the \sub sample are provided in Table \ref{table_log}. For 17 out of 21 objects, we used HST spectra taken with the Cosmic Origins Spectrograph (COS, \citealt{Gree12}), 16 of which were observed by our proposal in HST Cycle 27. For the remaining four targets, which lack any COS data, we made use of archival HST data taken with the Space Telescope Imaging Spectrograph (STIS, \citealt{Wood98}) and the Faint Object Spectrograph (FOS, \citealt{Harm91}).

In planning our HST/COS observation of a \sub target, its redshift and UV brightness were taken into account, so that the selected COS gratings and settings would provide broad spectral coverage, spanning from the \lyb and \ovi regions to the \civ line, with an adequate signal-to-noise ratio (S/N). The exposure times were calculated to provide a S/N > 10 per resolution element in the continuum, thereby allowing the absorption lines to be significantly detected and their parameters to be constrained. Furthermore, the COS exposures were taken at all four grating offset positions (FP-POS), so that the spectrum falls on slightly different areas of the detector, thus eliminating the effects of detector artifacts such as the flat-field features \citep{Hirs21}. 

All fully calibrated HST data (COS, STIS, and FOS) were retrieved from the Mikulski Archive for Space Telescopes (MAST). The COS data have been processed with the latest COS calibration pipeline, CalCOS v3.3.10. Individual exposures of an observation were combined to produce one calibrated merged spectrum. The COS spectra were binned by four pixels to improve the S/N while still oversampling the resolution element on the detector \citep{Fox18}. The STIS and FOS spectra were binned by two pixels. An overview of the HST spectra of the \sub sample is shown in Fig. \ref{fig_overview}. We note that in all our tables and figures in this paper the targets are ordered alphabetically by their name for ease in looking them up.

We carried out an examination of the HST data to verify the wavelength calibration and accuracy of the final spectra. We selected suitable Galactic ISM absorption lines that have been detected and we measured their wavelengths. The lines we used for our checks are:
\ion{S}{ii} doublet $\lambda 1250.6$ and $\lambda 1253.8$,
\ion{Si}{iv} doublet $\lambda 1393.8$ and $\lambda 1402.8$,
\ion{Si}{ii} $\lambda 1526.7$, and
\ion{Al}{ii} $\lambda 1670.8$.
At least lines from two ions were examined in each spectrum. For each line, its corresponding velocity shift in the local standard of rest (LSR) was calculated and compared with the velocity of the \ion{H}{i} 21 cm line in the same line of sight \citep{Wakk11}. Thus, the difference ($\delta)$ in wavelength between the COS ISM lines and the 21 cm line were obtained. We found the mean and median $\delta$ for the COS spectra in our sample are $8$ and $10$~\kms, respectively. The $\delta$ measurements in our sample have a standard deviation of $18$~\kms. All $\delta$ measurements are less than 50~\kms. Also, for each target there is consistency between the COS wavelength measurements taken with different gratings and in different observations. These $\delta$ differences between the ionized ISM lines in the COS spectra and the 21 cm line are sufficiently low that they do not require wavelength re-adjustment. This is particularly appropriate as there are also cases of high velocity clouds (HVCs) toward some \sub targets that would contribute to the observed $\delta$ differences between the neutral 21 cm and the ionized ISM lines.

%%%%%%%%%%%%%%%%%%%%%%%%%%%%%%%%%%%%%%%%%%%%%%%%%%%%%%%%%%%%%%%%%%%%%%%%%%%%%%%%%%%%%%%%%%%%%%%%%%%%%%%
%%%%%%%%%%%%%%%%%%%%%%%%%%%%%%%%%%%%%%%%%%%%%%%%%%%%%%%%%%%%%%%%%%%%%%%%%%%%%%%%%%%%%%%%%%%%%%%%%%%%%%%
%%%%%%%%%%%%%%%%%%%%%%%%%%%%%%%%%%%%%%%%%%%%%%%%%%%%%%%%%%%%%%%%%%%%%%%%%%%%%%%%%%%%%%%%%%%%%%%%%%%%%%%
\section{Spectral analysis and modeling} 
\label{sect_model}

We began our analysis of the HST spectra by first selecting the most reliable cosmological redshifts ($z$) for the \sub sample. We assessed the different optically-determined redshift publications for each object, as provided on the NASA/IPAC Extragalactic Database (NED). We examined their reported uncertainties and gave preference to studies with the Sloan Digital Sky Survey (SDSS) for accurate measurements. Our selected redshift values and their references are provided in Table \ref{table_log}. 

In the following subsections, we describe our steps in modeling the HST spectra of the \sub sample. We note that in this paper, we do not explore any time variability of the objects and our best-fit parameters are given for one epoch. For targets that have multiple COS observations taken in different times and with different gratings, we gave preference to observations with COS G130M and G160M gratings over COS G140L as their higher spectral resolution is beneficial in fitting the narrow absorption lines. Nonetheless, the G140L spectra were also analyzed because their broad spectral range is useful for searching for UV counterparts to the X-ray UFOs. In our notation in this paper, negative $v$ means systematic blueshift (outflow) and positive $v$ means systematic redshift (inflow), with respect to the local rest frame of the object. The errors on parameters are given at the 90\% confidence level.

%%%%%%%%%%%%%%%%%%%%%%%%%%%%%%%%%%%%%%%%%%%%%%%%%%%%%%%%%%%%%%%%%%%%%%%%%%%%%%%%%%%%%%%%%%%%%%%%%%%%%%%
\subsection{Identification of the spectral features in the HST data}
\label{sect_iden}

We started by identifying all the significantly-detected ($\geq 5\sigma$) spectral features. Since we are interested in modeling only the AGN spectral features, we first identified all the non-AGN features and excluded them from our fitting. These non-AGN features consist of absorption lines from the Galactic ISM and the intergalactic medium (IGM), as well as the geocoronal airglow emission features. The wavelengths of the ISM and airglow lines are well-established and, hence, these lines were readily identified and excluded. On the other hand, any intervening IGM in our line of sight, would produce \ion{H}{i} lines at its intervening redshift with its strongest feature (i.e., the \lya line) appearing at wavelengths longer than 1215.7 \AA. Thus, such contaminating IGM features need to be excluded from our modeling of the intrinsic AGN absorption lines. To distinguish between an IGM and AGN \lya line we took the following steps.

First, we checked for the presence of any associated ionized lines at the same velocity as the \lya line: if there is either a \ion{C}{iv}, \ion{N}{v}, or \ion{O}{vi} line accompanying the \lya feature at the same velocity, then the feature is considered to be intrinsic to the AGN.

Second, we assessed the implied velocity shift if attributed to the AGN: if the \lya feature is at around the rest-frame wavelength of the AGN, then the line  is determined to likely be intrinsic to the AGN. However, if an isolated \lya line has an extremely shifted wavelength, with no ionized absorption counterpart at that velocity in either the UV or X-ray bands (including the X-ray UFO), then that line is not likely to be intrinsic to the AGN.

Third, we examined the velocity broadening: since narrow absorption lines from the AGN are still typically broader than IGM lines (FWHM of IGM lines is most often about 50~\kms, \citealt{Danf16}), if an isolated absorption line shows too little velocity broadening, with no other UV and X-ray counterparts, it is likely to be an IGM line.

Next, we checked for any previous IGM identifications in the literature: we verified our identification of the IGM lines by checking the literature for any previous reports of IGM lines toward our targets. 

\citet{Danf16}  carried out an IGM study of 82 UV-bright AGN at ${z < 0.85}$ using HST/COS observations. They identified IGM lines in the COS spectra and derived a cumulative column density distribution of IGM \ion{H}{i} systems. Six of our targets are in the \citet{Danf16} sample, namely: PG~0804+761, PG~0953+414, PG~1216+069, PG~1307+085, PG~1435-067, and PG~1626+554. The results of \citet{Danf16} are consistent with our identification of IGM lines in these targets. \citet{Shul17} also conducted a survey of \ion{H}{i} systems using HST/COS spectra of 102 AGN. Apart from the aforementioned targets, some of which are also included in the sample of \citet{Shul17}, PG~0052+251 and PG~1352+183 have IGM identification in \citet{Shul17}, which is consistent with ours. Furthermore, the distribution of IGM \ion{H}{i} absorbers as a function of redshift and column density \citep{Danf16,Shul17} is a useful statistical benchmark. These IGM distributions confirm that the high column density and the high detection frequency of our intrinsic AGN \lya lines (modeled in Sect. \ref{sect_fit}) cannot be explained by IGM absorbers in ${0.1 < z < 0.4}$.

Following our line identification procedure, we could see intrinsic absorption lines in 13 of the 21 targets in the \sub HST sample (${\sim 60\%}$). These intrinsic absorption lines belong to \lya, \lyb, \civ, \nv, and \ovi. In some targets intrinsic absorption by multiple outflow velocity components were found. We detected no significant intrinsic \ion{Si}{iv} absorption lines in our HST spectra. In Sect. \ref{sect_fit}, we describe our spectral modeling of the intrinsic absorption lines and we also present our search results for the UV counterparts to the X-ray UFOs.

%%%%%%%%%%%%%%%%%%%%%%%%%%%%%%%%%%%%%%%%%%%%%%%%%%%%%%%%%%%%%%%%%%%%%%%%%%%%%%%%%%%%%%%%%%%%%%%%%%%%%%%
\subsection{Modeling of the intrinsic spectral lines}
\label{sect_fit}

Following the identification of all spectral lines, we proceeded to model the intrinsic AGN emission and absorption lines (\ion{H}{i} \lya and \lyb, \ion{C}{iv}, \ion{N}{v}, and \ion{O}{vi}). Since we are not concerned with the study of the broadband continuum in this paper, we carried out local continuum modeling in each spectral region. This is adequate for our purpose of parameterizing the intrinsic emission and absorption lines. We thus fitted a linear function to the feature-free spectral bands surrounding each emission line. We then modeled the emission lines by applying Gaussian functions, while ignoring all data bins that contain absorption lines. We fit the flux, wavelength (line-centroid velocity shift $v$), and the full width at half maximum (FWHM) of the emission lines. For those BLR emission lines where their doublet lines are blended into one apparent line (such as \civ), we used a single Gaussian to take into account the doublet, as well  as the average of the doublet's rest wavelengths as reference for calculating $v$. Broad AGN emission lines are commonly composed of multiple velocity broadening components, so for lines that clearly display such a profile, we applied multiple Gaussian components until a good fit with reduced chi-squared approaching unity ($\chi^{2}_{\nu} \sim $ 1) was achieved. 

Our best-fit model to the UV emission lines and their parameters is provided in Appendix \ref{sect_append}. The \lya and \nv emission lines of the \sub sample are displayed in Fig. \ref{fig_emission_lya} and the \civ emission lines in Fig. \ref{fig_emission_civ}. The best-fit parameters of these emission lines are given in Table \ref{table_emission}. For the \ion{C}{iv} emission line, which is least contaminated by foreground absorption or blending with other emission features, we calculated its line asymmetry \citep{Netz90}. The nearest known emission lines, such as \ion{He}{ii} $\lambda$1640, are sufficiently far away from \ion{C}{iv}, and also we do not detect any other significant contaminating emission feature in modeling the \ion{C}{iv} emission line. The intrinsic asymmetry of emission lines is thought to be an indicator of disk wind activity (e.g., \citealt{Coat16}). We made use of two definitions for the line asymmetry ($a_f$ and $a_p$). The $a_f$ is the "flux asymmetry" parameter, defined as the ratio of flux on the blue over red side of the line centroid. The $a_p$ is the "profile asymmetry" parameter, defined as $({\lambda_{3/4} - \lambda_{1/4}) / {\rm FWHM}}$, where $\lambda_{3/4}$ and $\lambda_{\rm 1/4}$ are the line centroids at $3/4$ and $1/4$ of intensity, respectively (see e.g., \citealt{Bask05} for line asymmetry characterization of \ion{C}{iv}). Hence, in these definitions, a line with ${a_f = 1}$ and ${a_p = 0}$ would be fully symmetrical. The obtained asymmetry parameters of the \civ lines for the \sub HST sample are provided in Table \ref{table_asym}. In this table, we also give the flux-weighted average velocities ($v_{\rm mean}$ and FWHM$_{\rm mean}$) of the \civ emission line, calculated using the individual model components that comprise the line.

\begin{figure*}
\centering
\resizebox{0.94\hsize}{!}{
\includegraphics[angle=0]{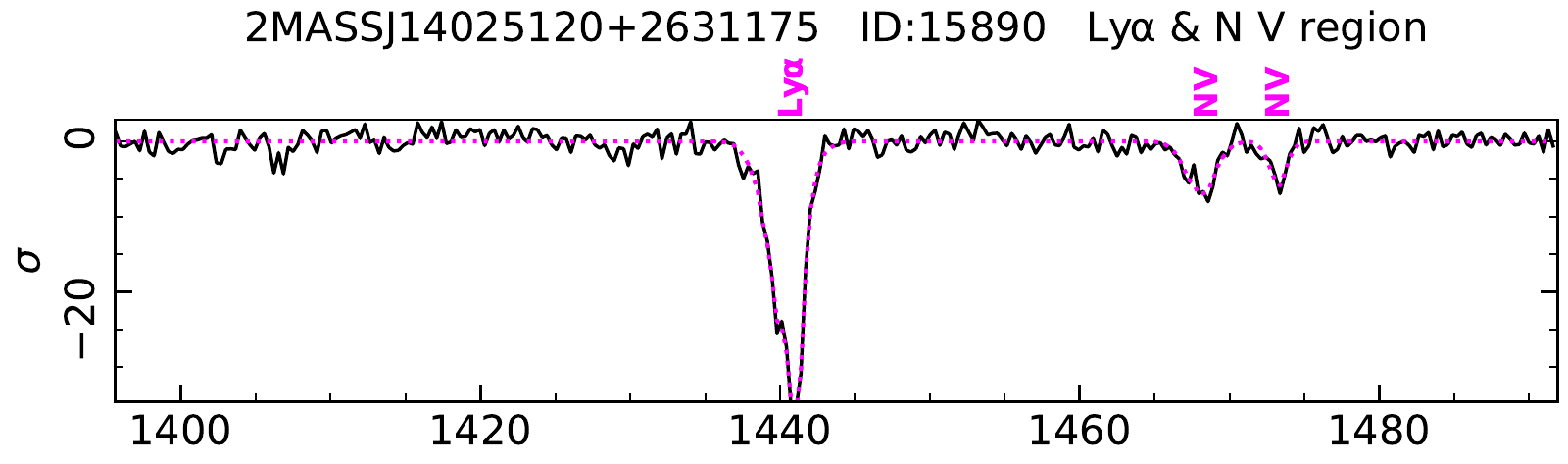}
\includegraphics[angle=0]{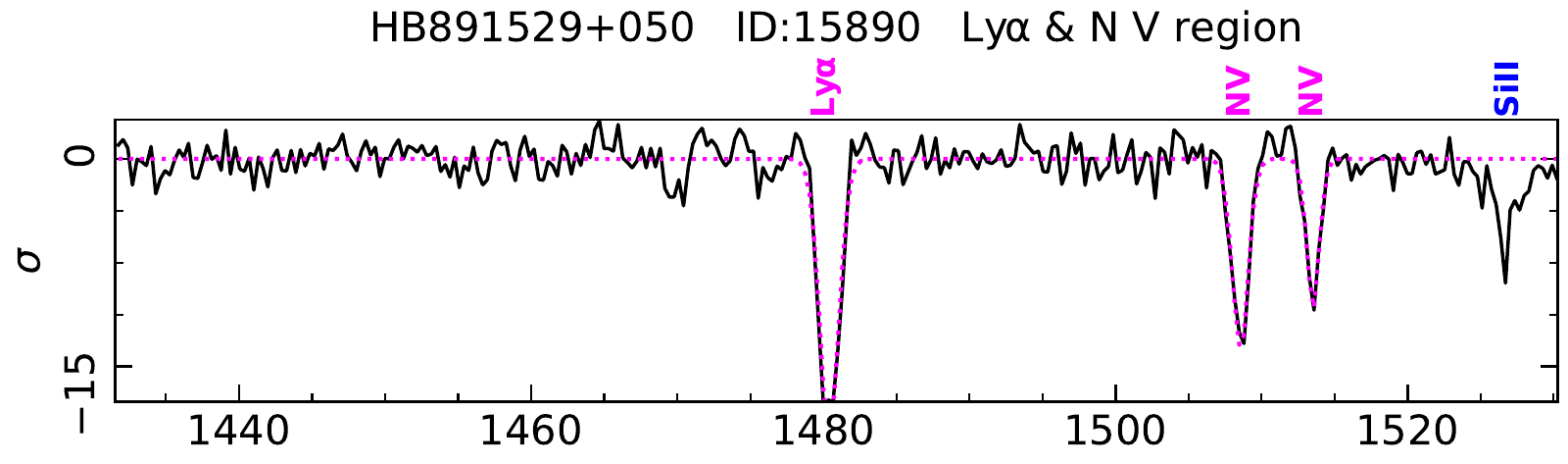}
}
\resizebox{0.94\hsize}{!}{
\includegraphics[angle=0]{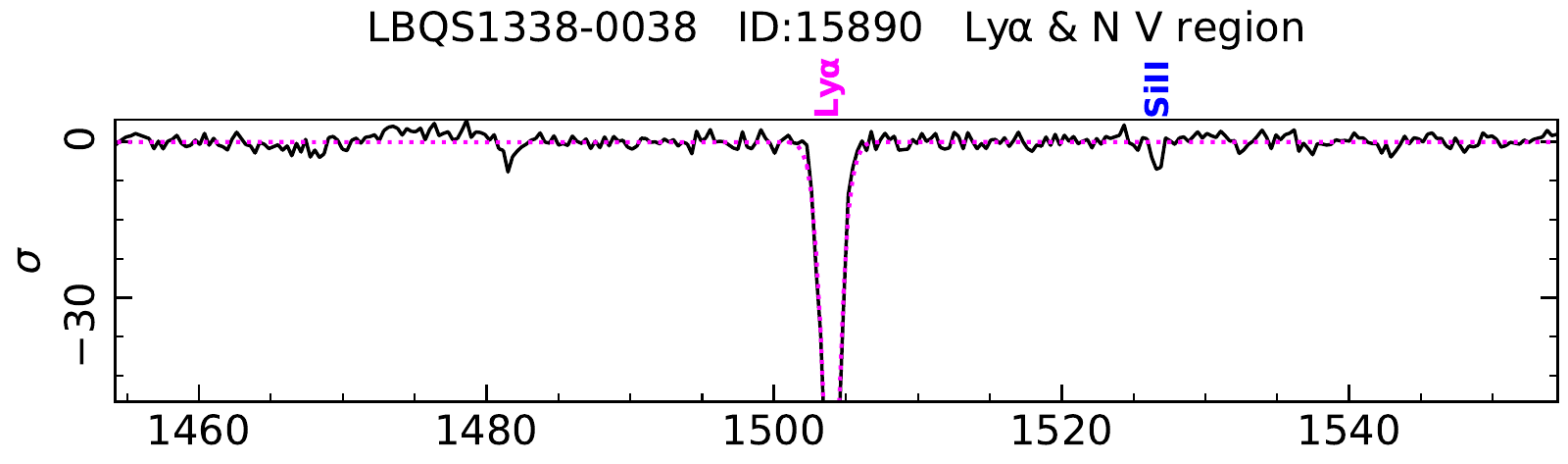}
\includegraphics[angle=0]{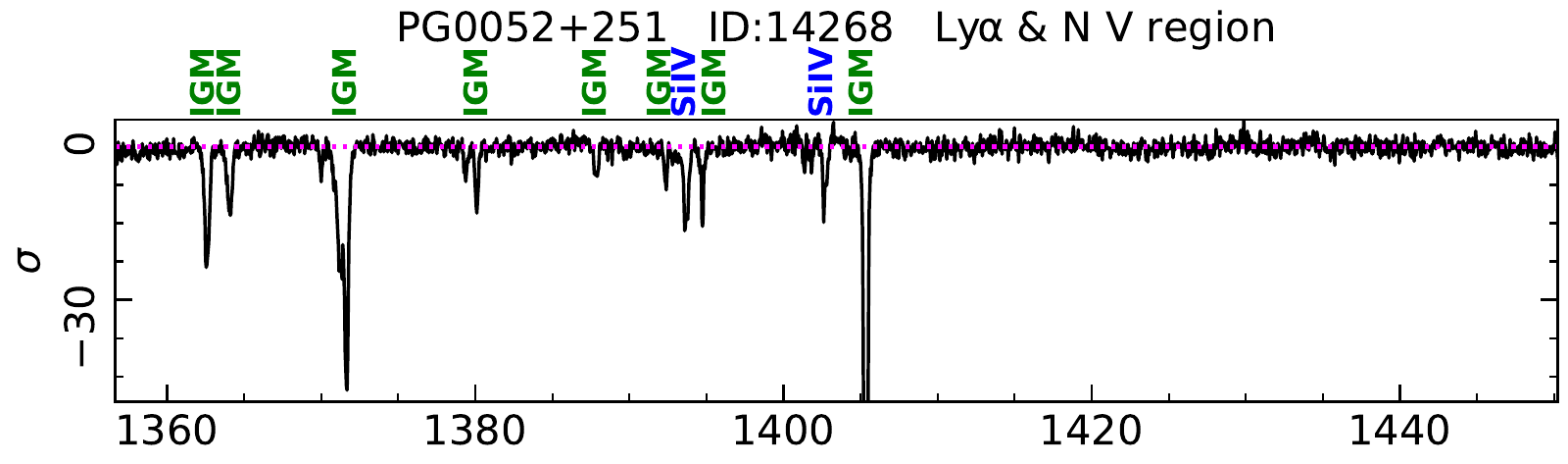}
}
\resizebox{0.94\hsize}{!}{
\includegraphics[angle=0]{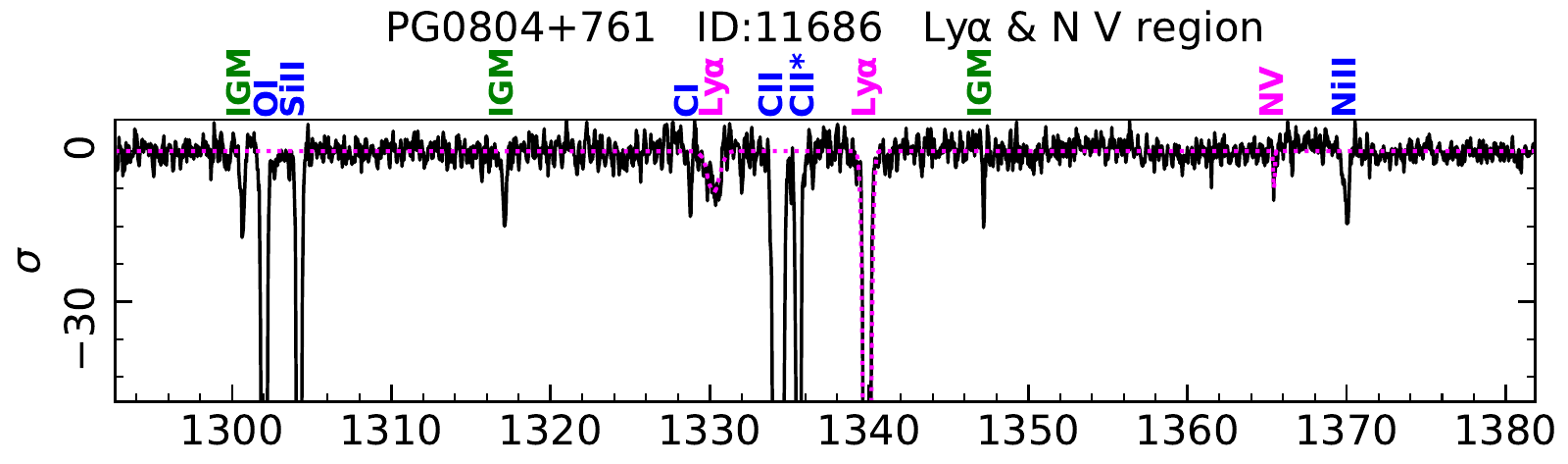}
\includegraphics[angle=0]{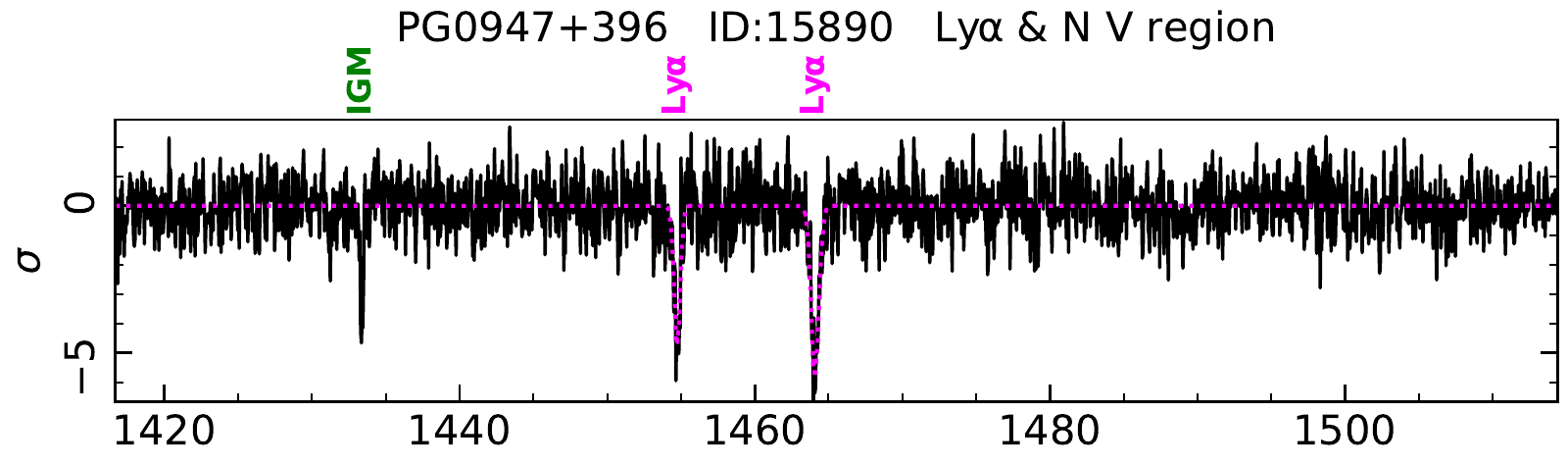}
}
\resizebox{0.94\hsize}{!}{
\includegraphics[angle=0]{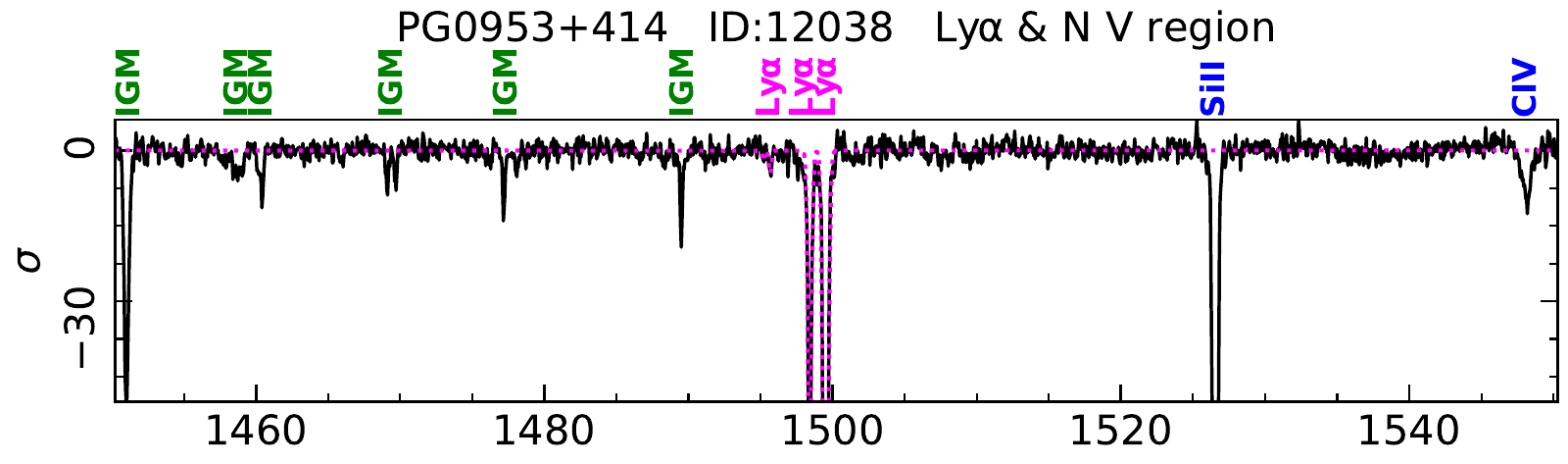}
\includegraphics[angle=0]{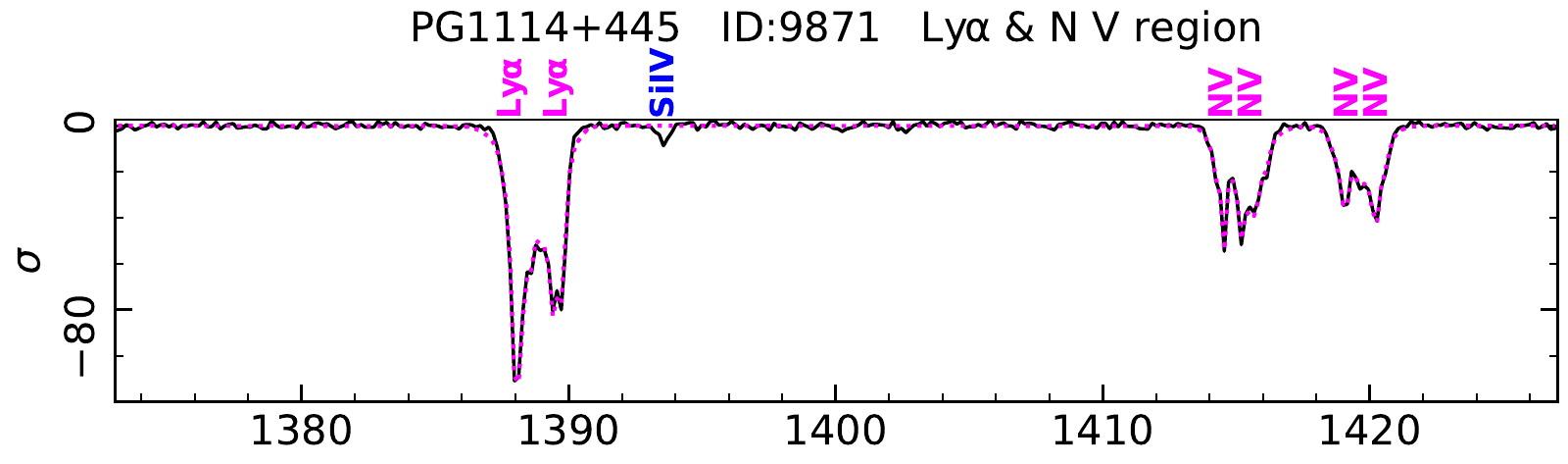}
}
\resizebox{0.94\hsize}{!}{
\includegraphics[angle=0]{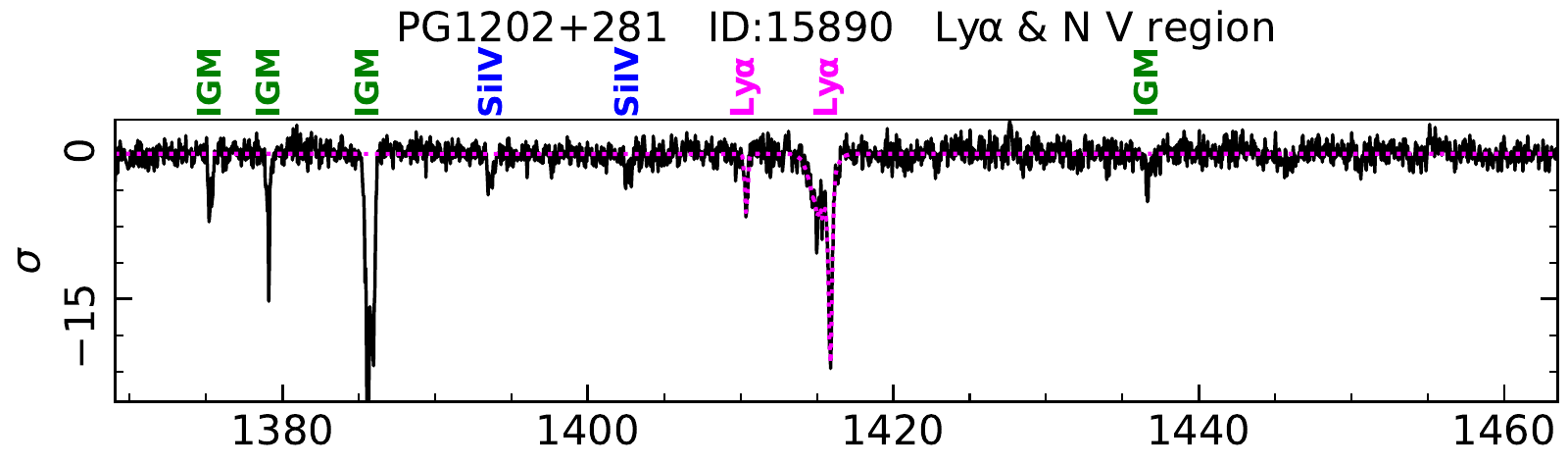}
\includegraphics[angle=0]{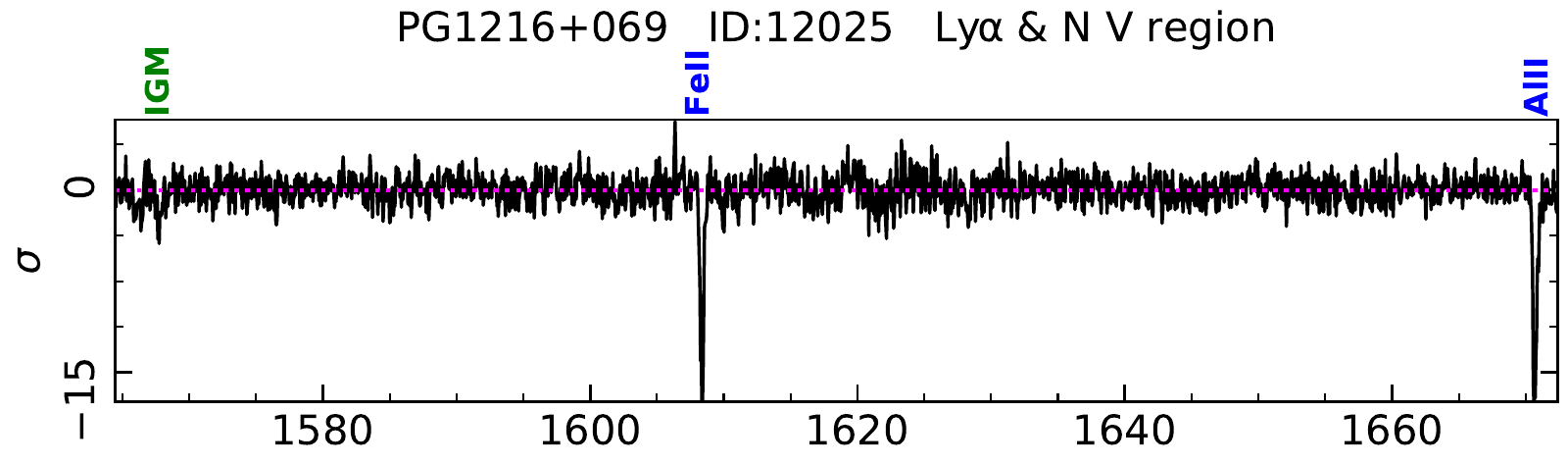}
}
\resizebox{0.94\hsize}{!}{
\includegraphics[angle=0]{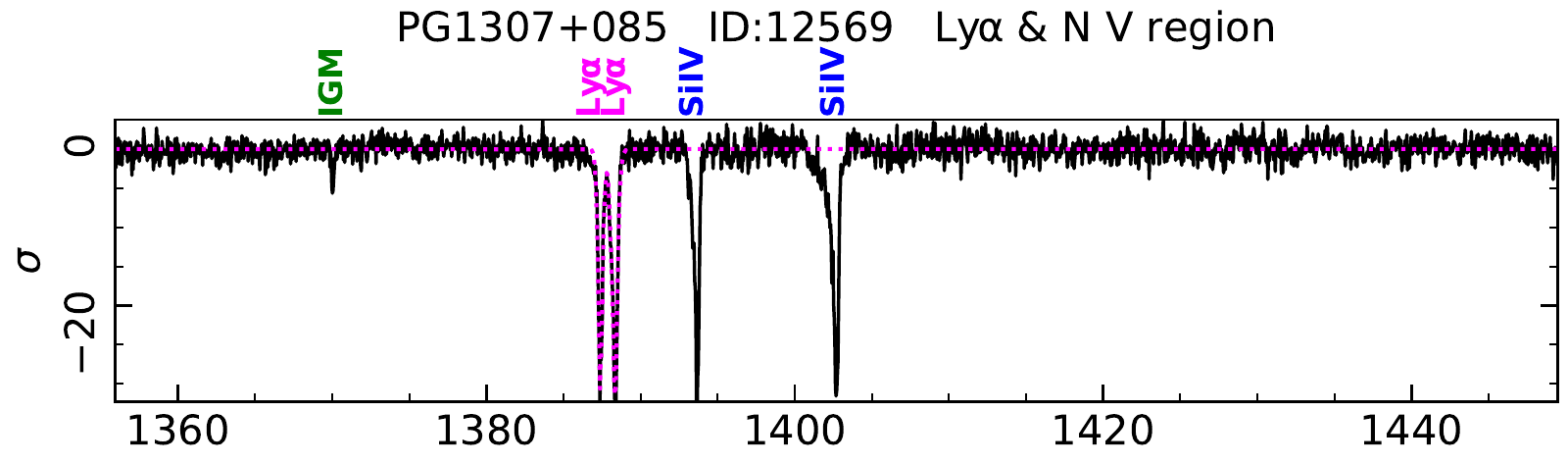}
\includegraphics[angle=0]{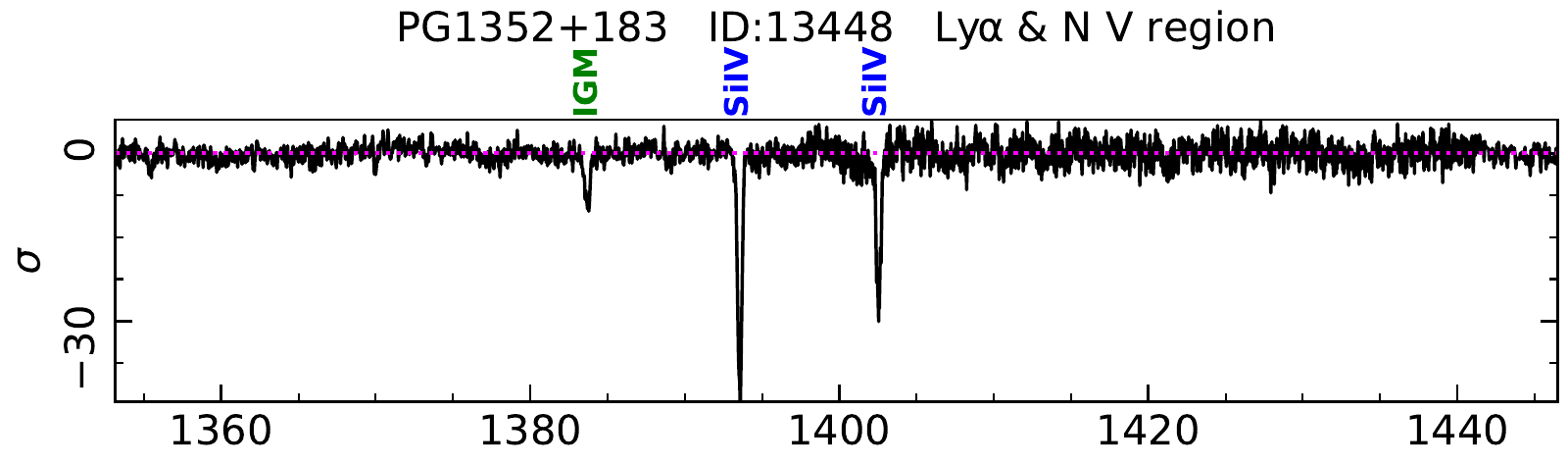}
}
\resizebox{0.94\hsize}{!}{
\includegraphics[angle=0]{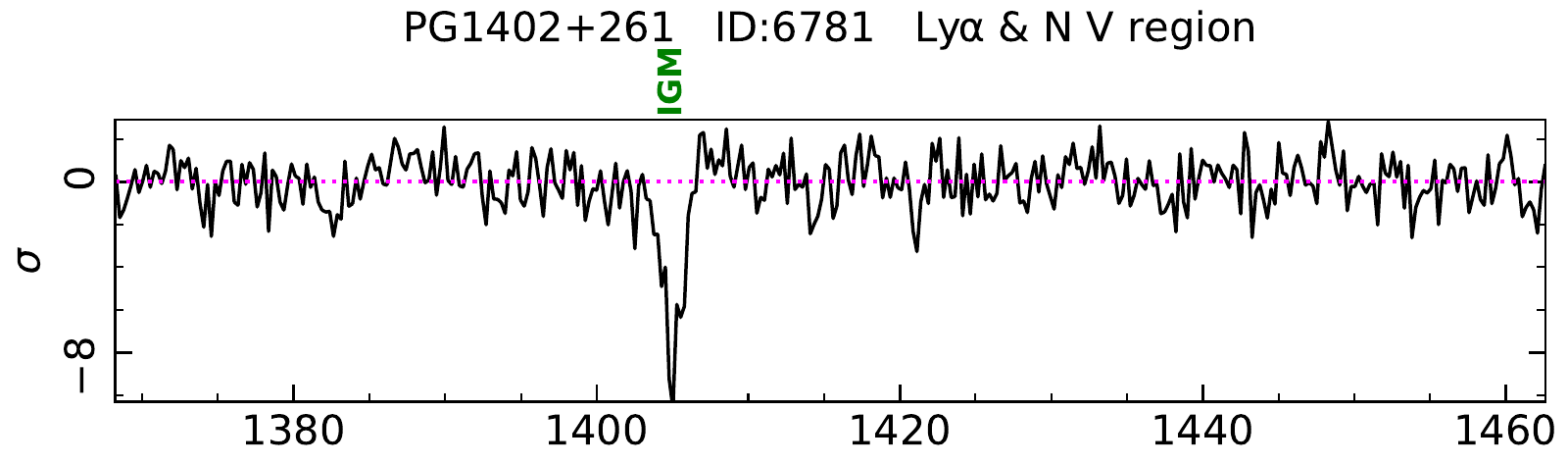}
\includegraphics[angle=0]{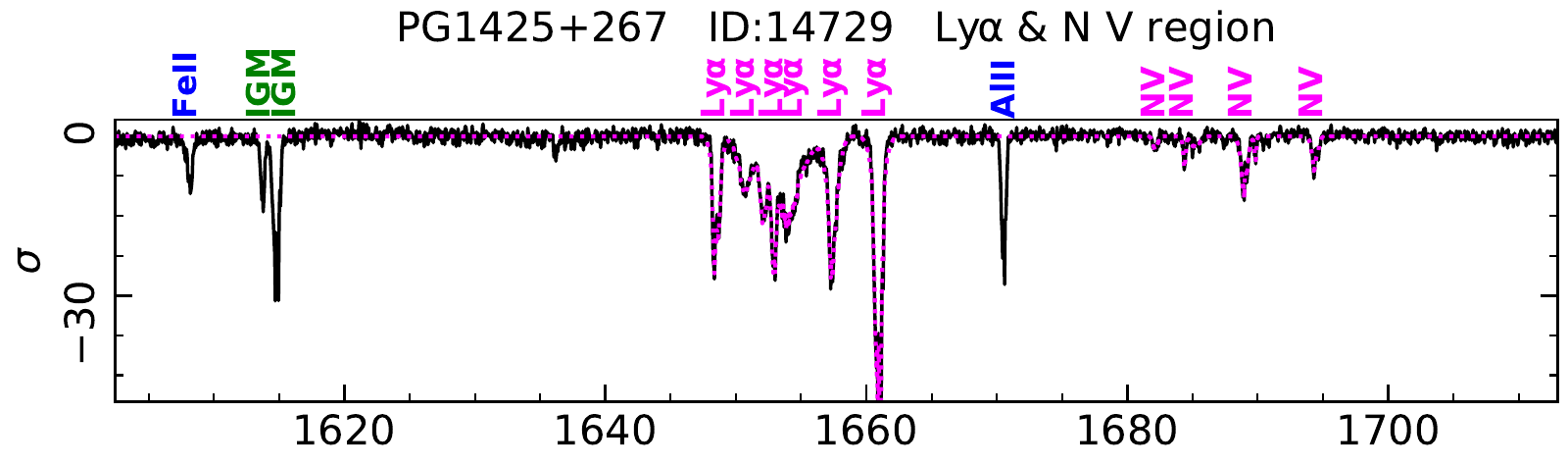}
}
\resizebox{0.94\hsize}{!}{
\includegraphics[angle=0]{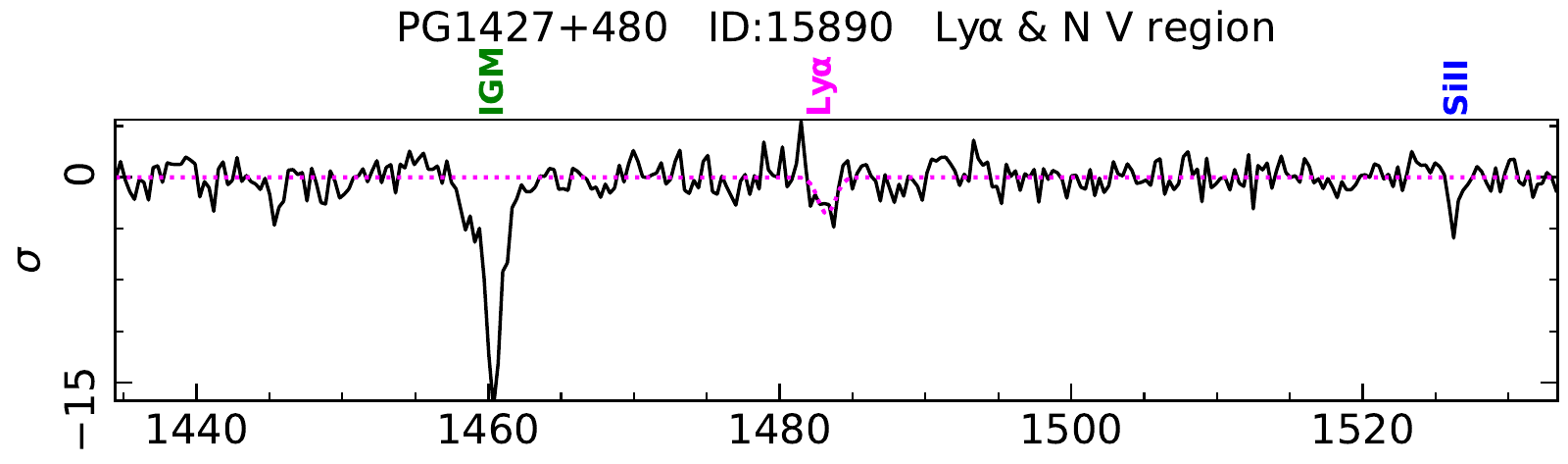}
\includegraphics[angle=0]{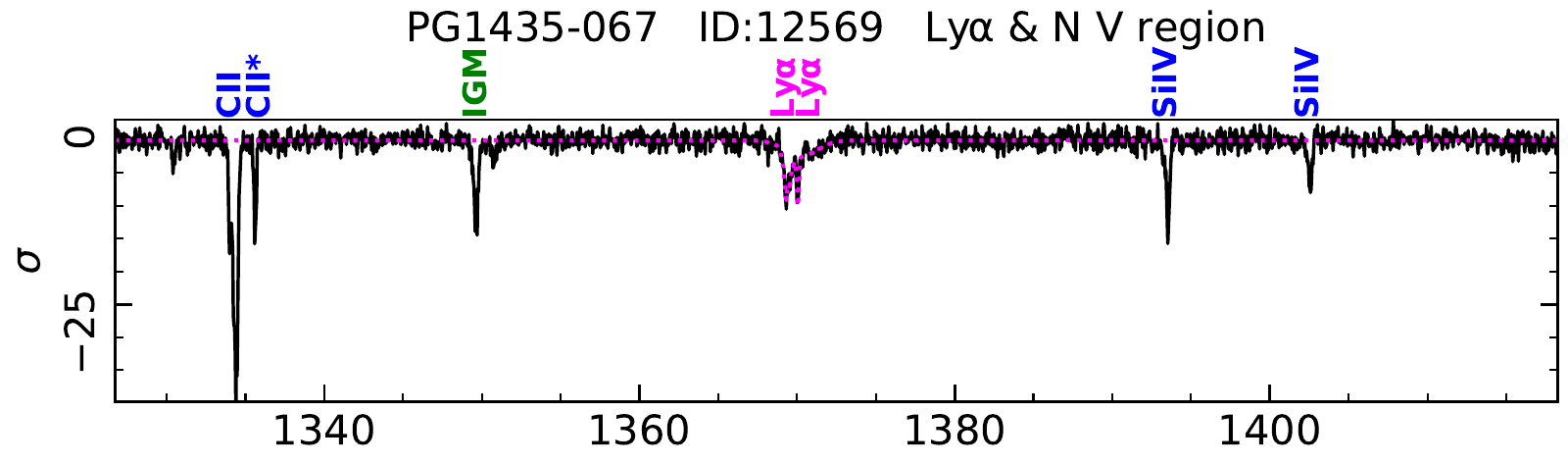}
}
\resizebox{0.94\hsize}{!}{
\includegraphics[angle=0]{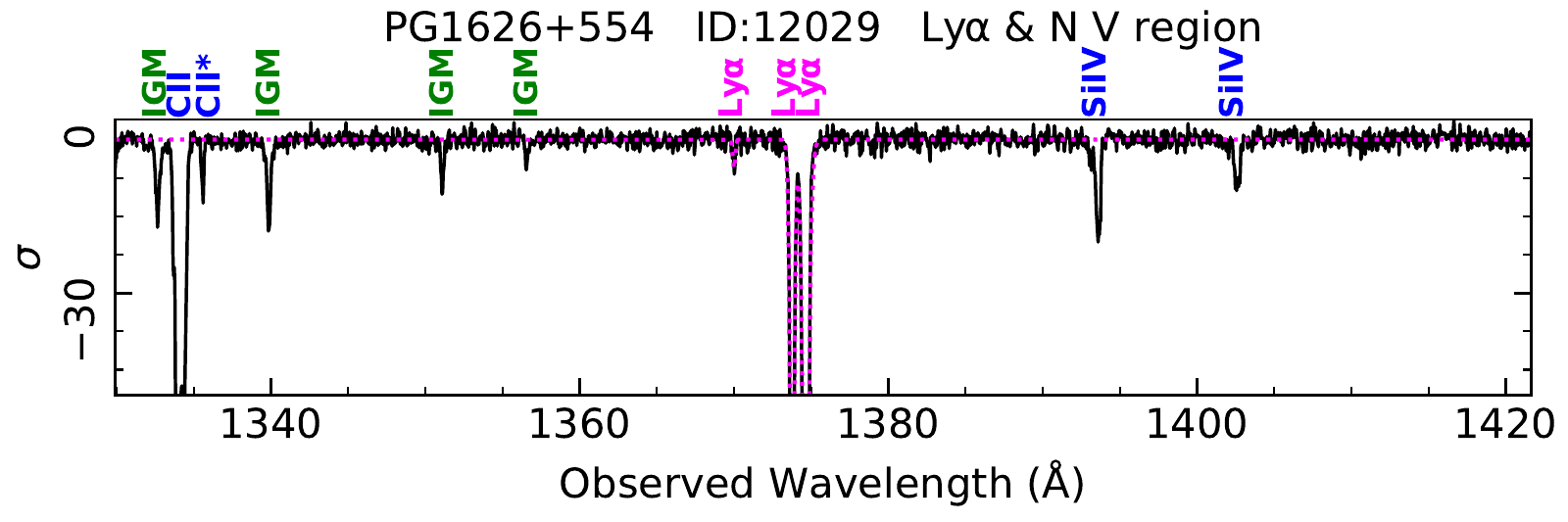}
\includegraphics[angle=0]{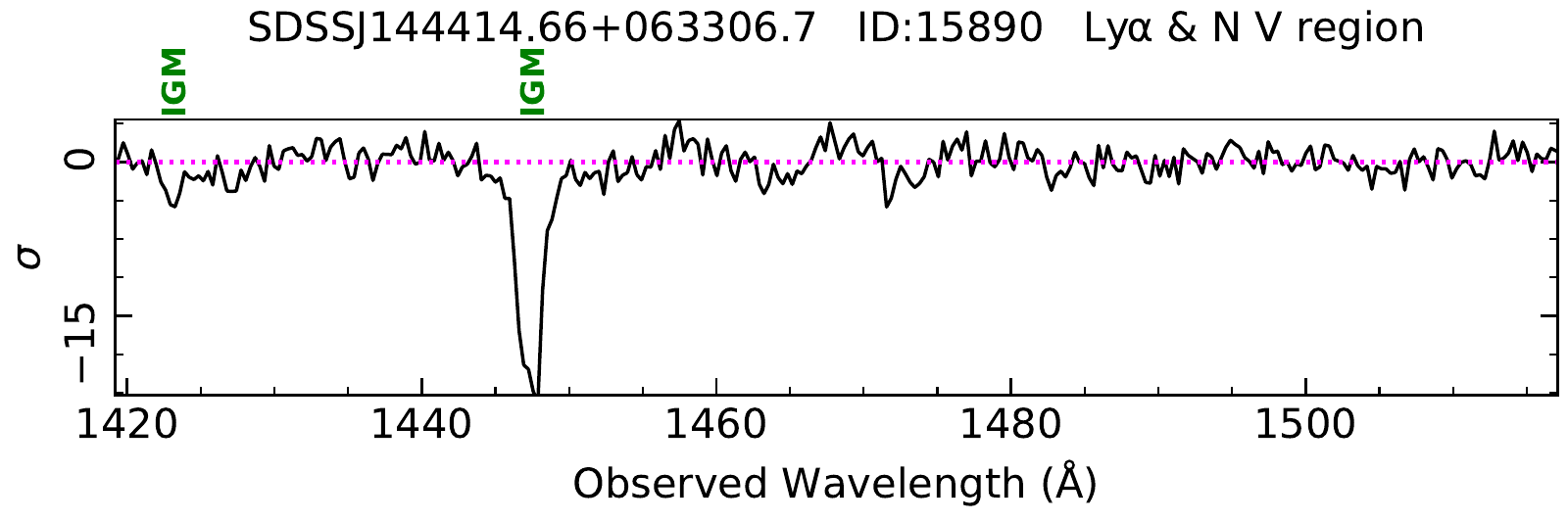}
}
\vspace{-0.2cm}
\caption{Absorption lines in the \lya and \nv region of the HST spectra of the \sub sample. The continuum and the emission lines are subtracted in the displayed data. The ISM lines are labeled in blue, IGM lines in green, and the intrinsic AGN lines in magenta. The best-fit model to the AGN absorption lines is shown as a dotted magenta line. The significance $\sigma$ is defined as (D$-$M)/E, where D is the data, M the continuum+emission line model without absorption, and E the error on the data.
\label{fig_abs_lya}}
%\vspace{0.3cm}
\end{figure*}

\begin{figure*}
\centering
\resizebox{0.94\hsize}{!}{
\includegraphics[angle=0]{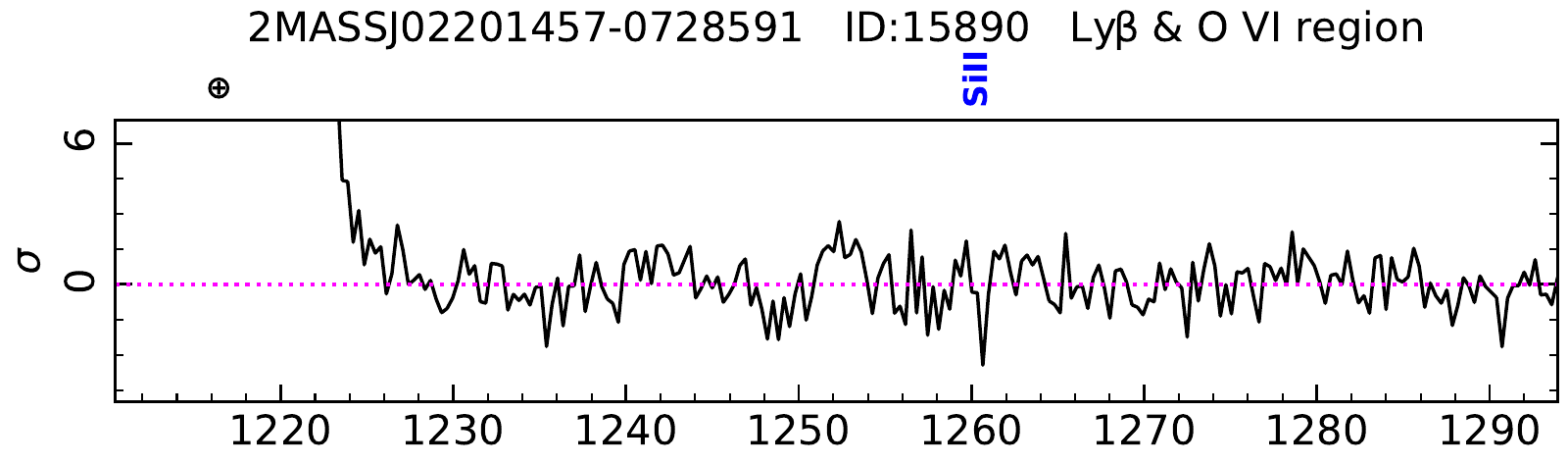}
\includegraphics[angle=0]{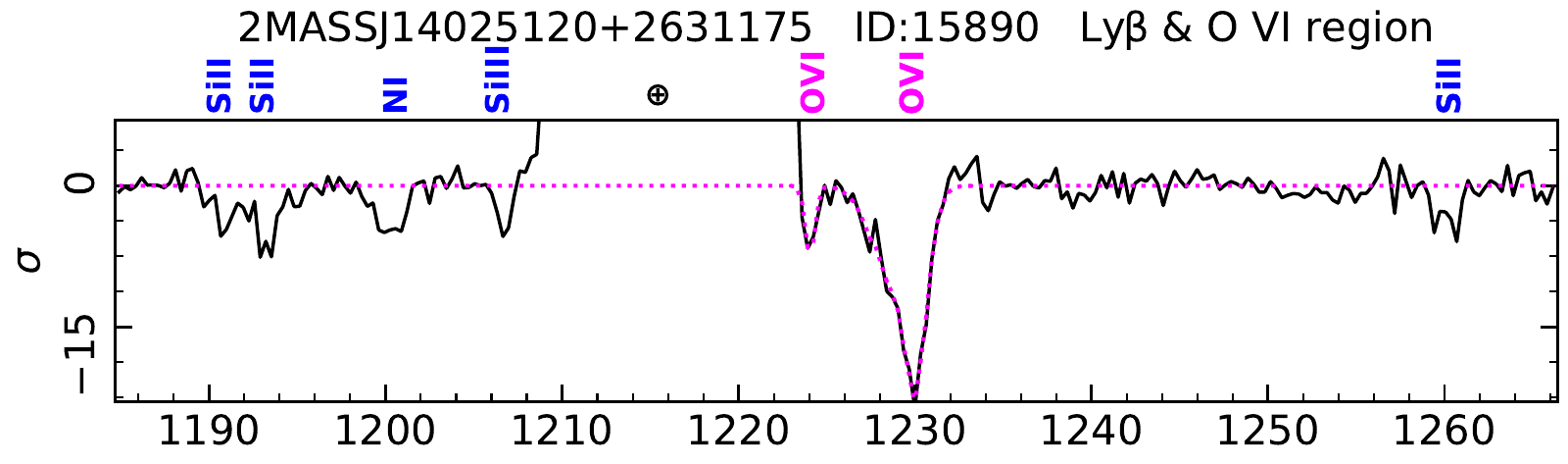}
}
\resizebox{0.94\hsize}{!}{
\includegraphics[angle=0]{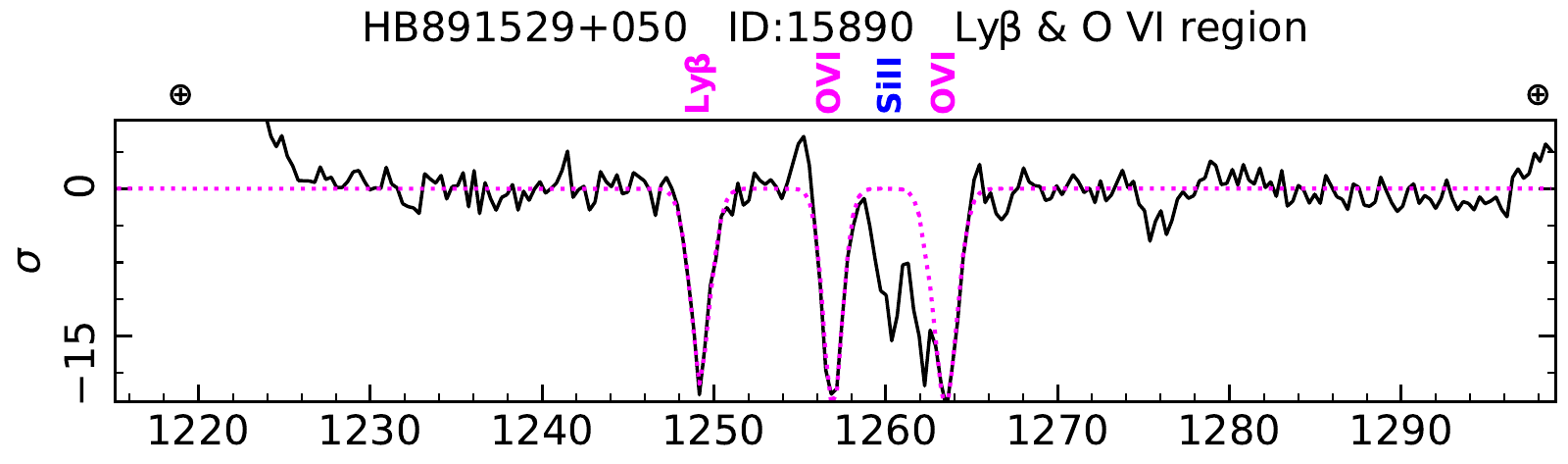}
\includegraphics[angle=0]{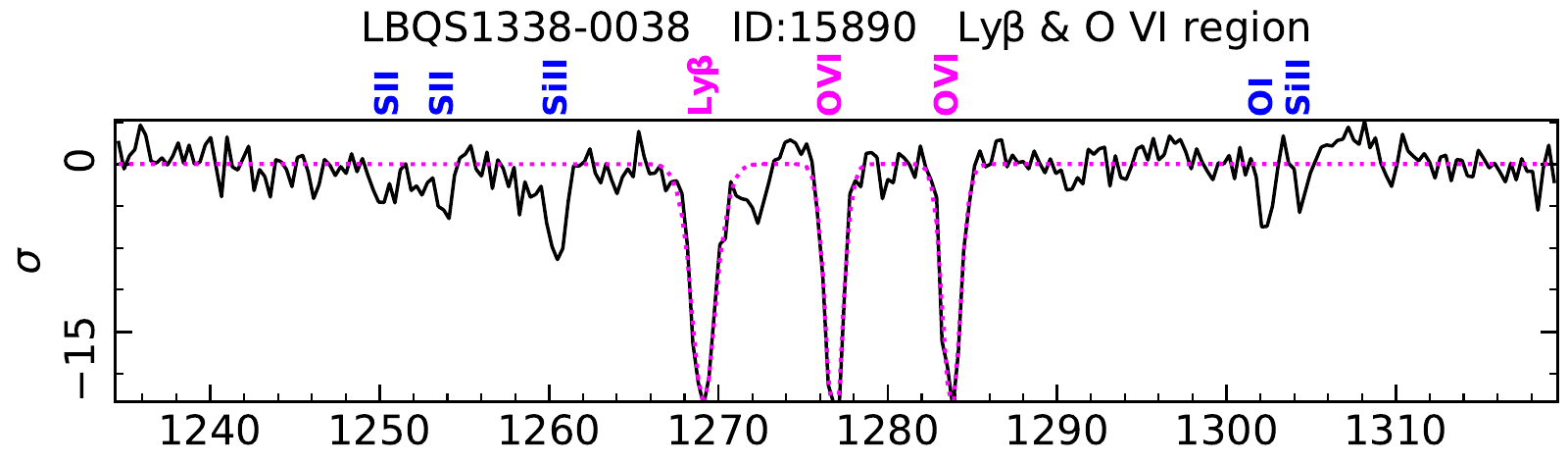}
}
\resizebox{0.94\hsize}{!}{
\includegraphics[angle=0]{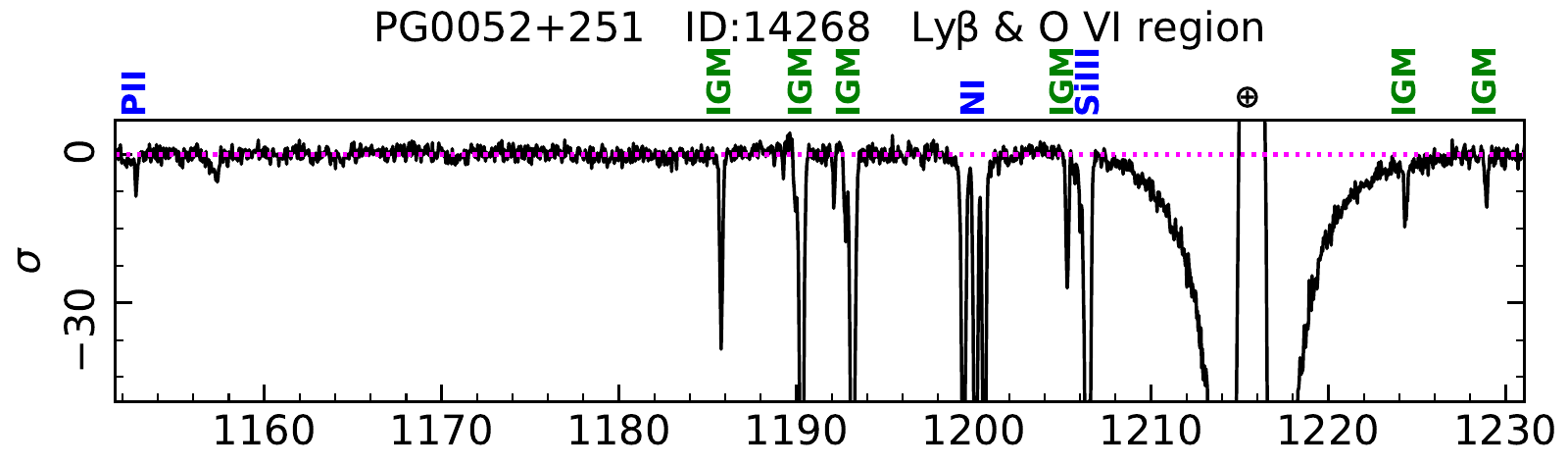}
\includegraphics[angle=0]{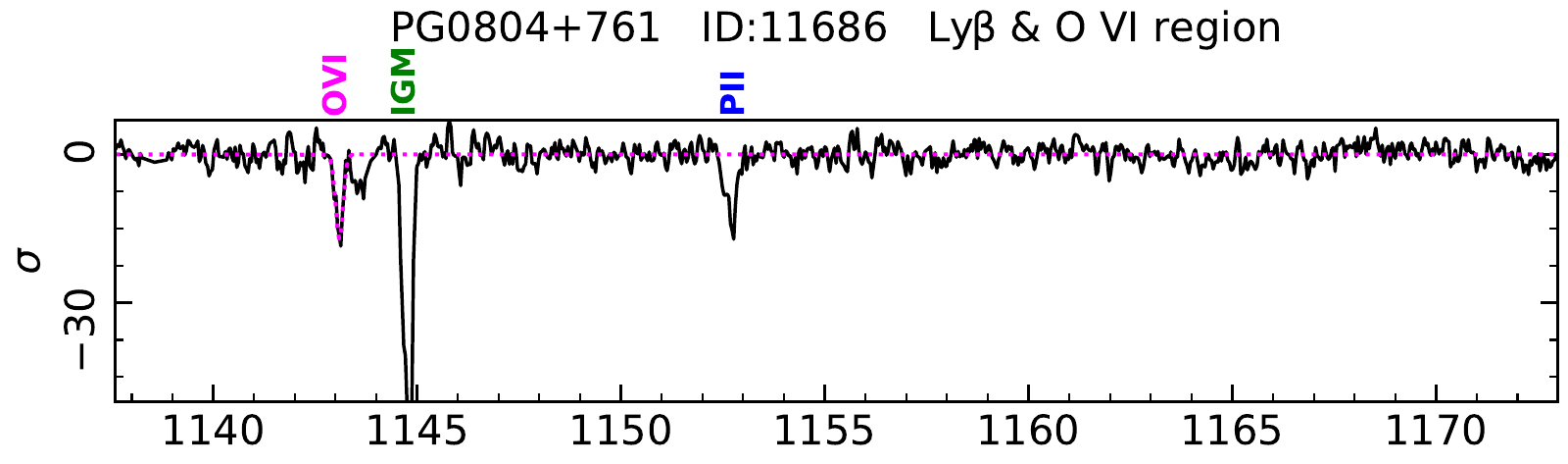}
}
\resizebox{0.94\hsize}{!}{
\includegraphics[angle=0]{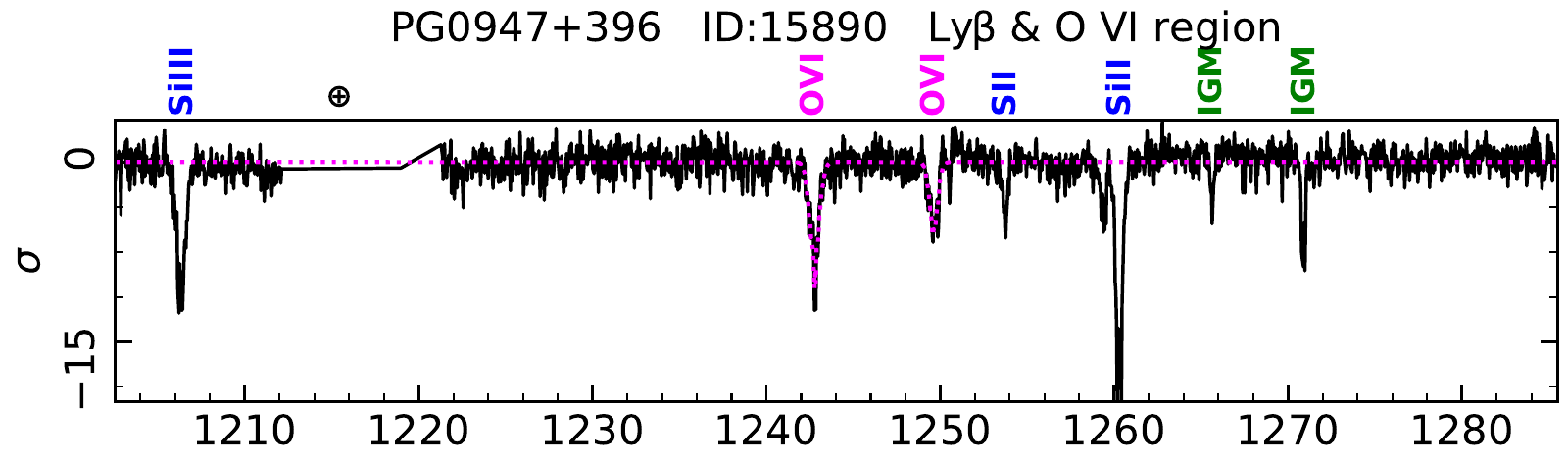}
\includegraphics[angle=0]{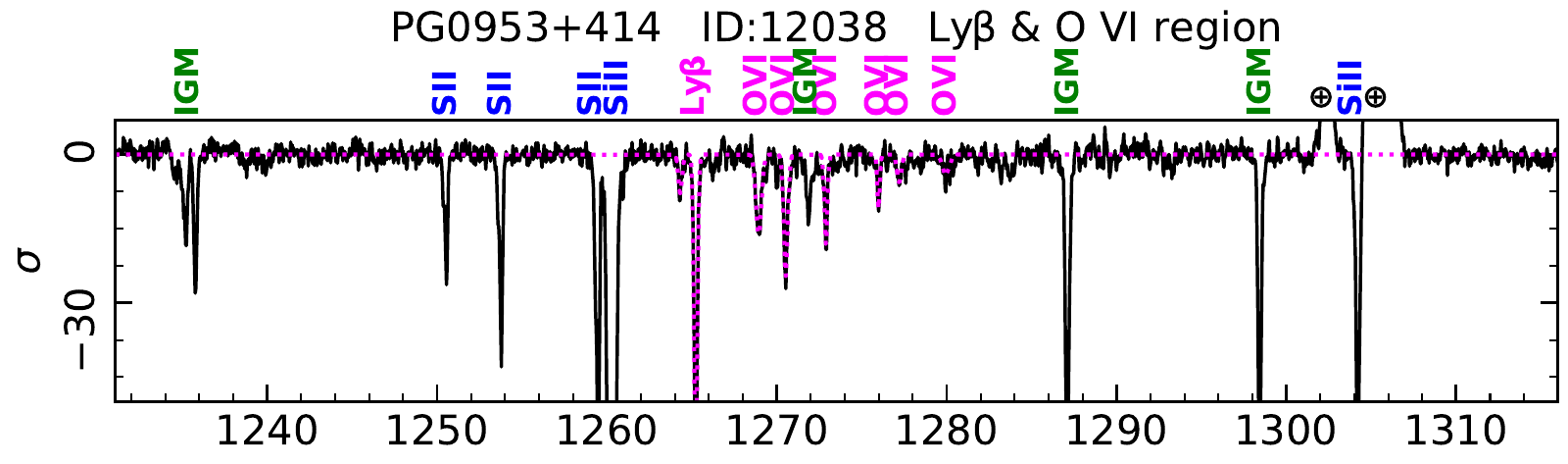}
}
\resizebox{0.94\hsize}{!}{
\includegraphics[angle=0]{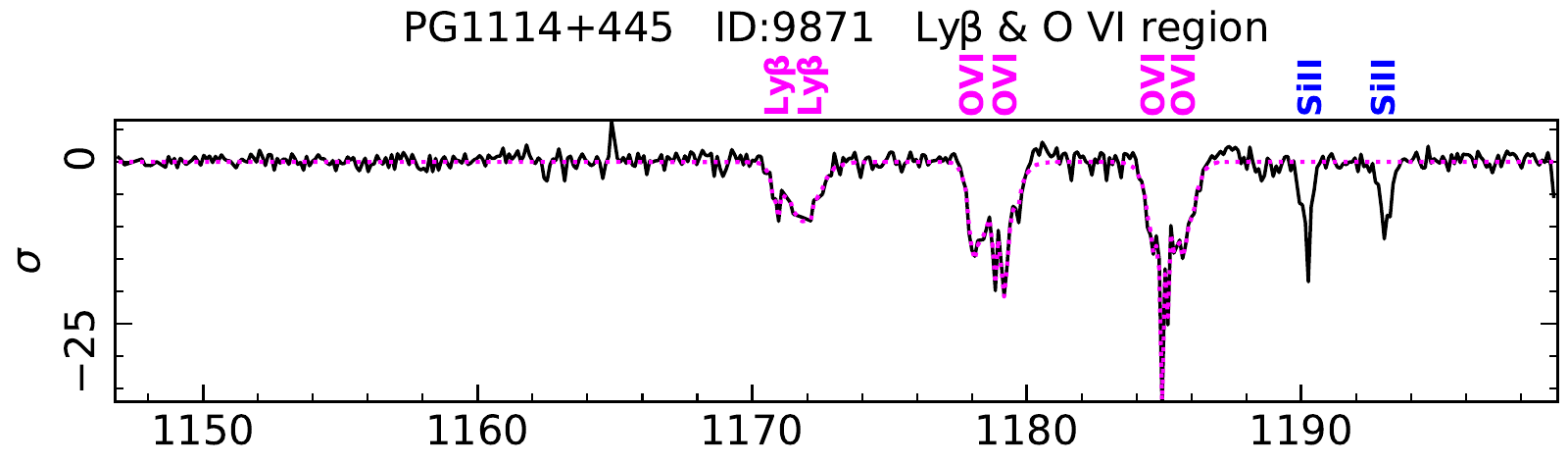}
\includegraphics[angle=0]{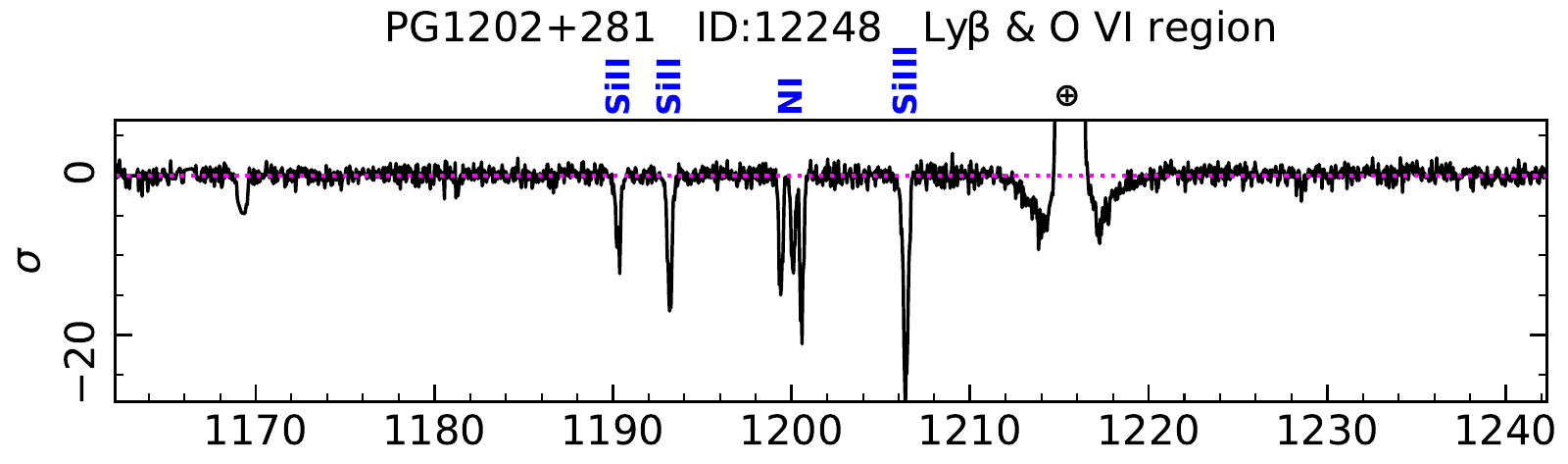}
}
\resizebox{0.94\hsize}{!}{
\includegraphics[angle=0]{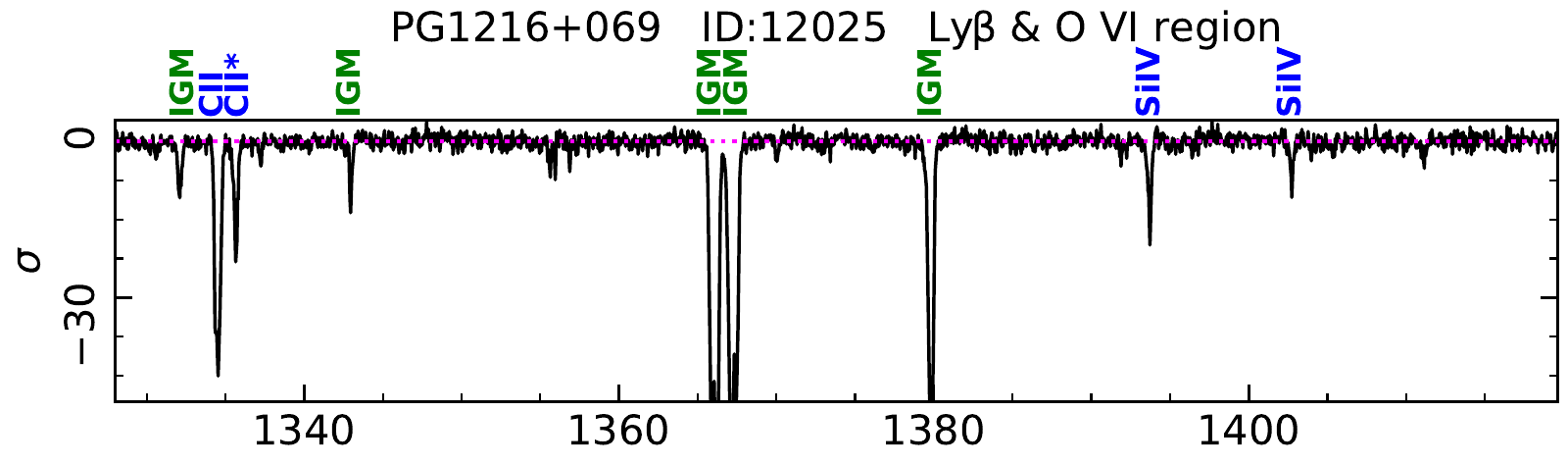}
\includegraphics[angle=0]{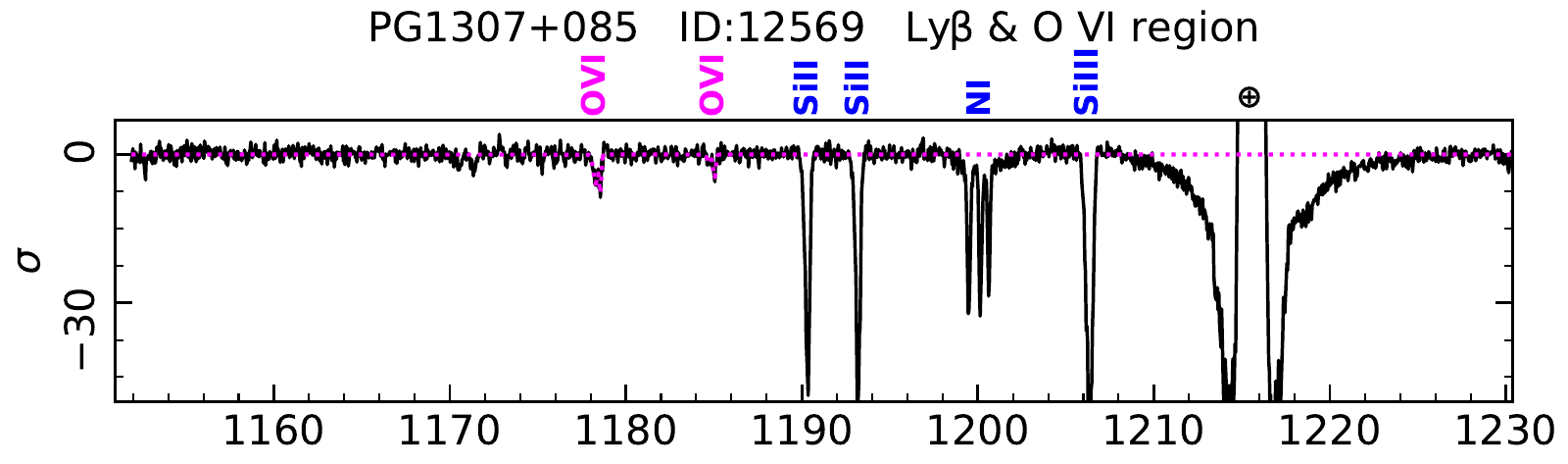}
}
\resizebox{0.94\hsize}{!}{
\includegraphics[angle=0]{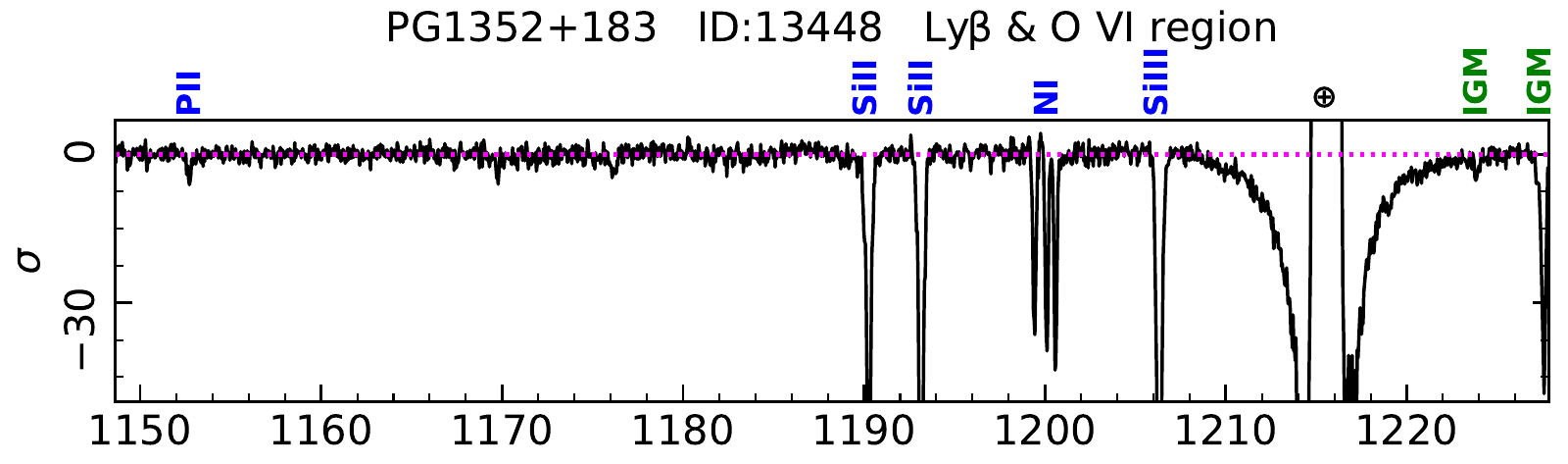}
\includegraphics[angle=0]{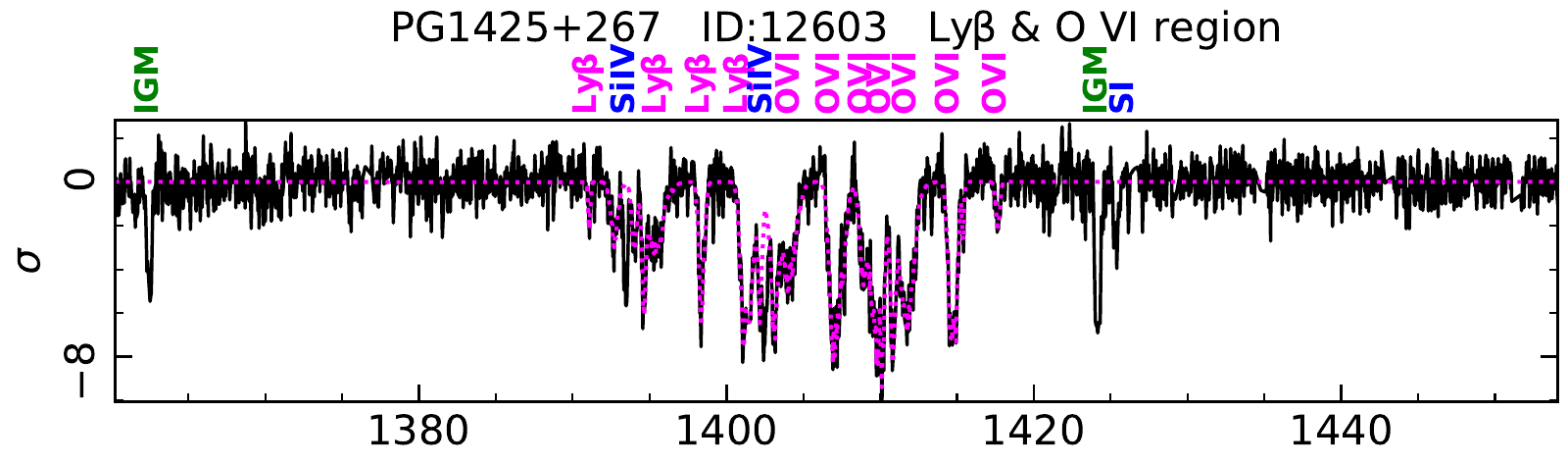}
}
\resizebox{0.94\hsize}{!}{
\includegraphics[angle=0]{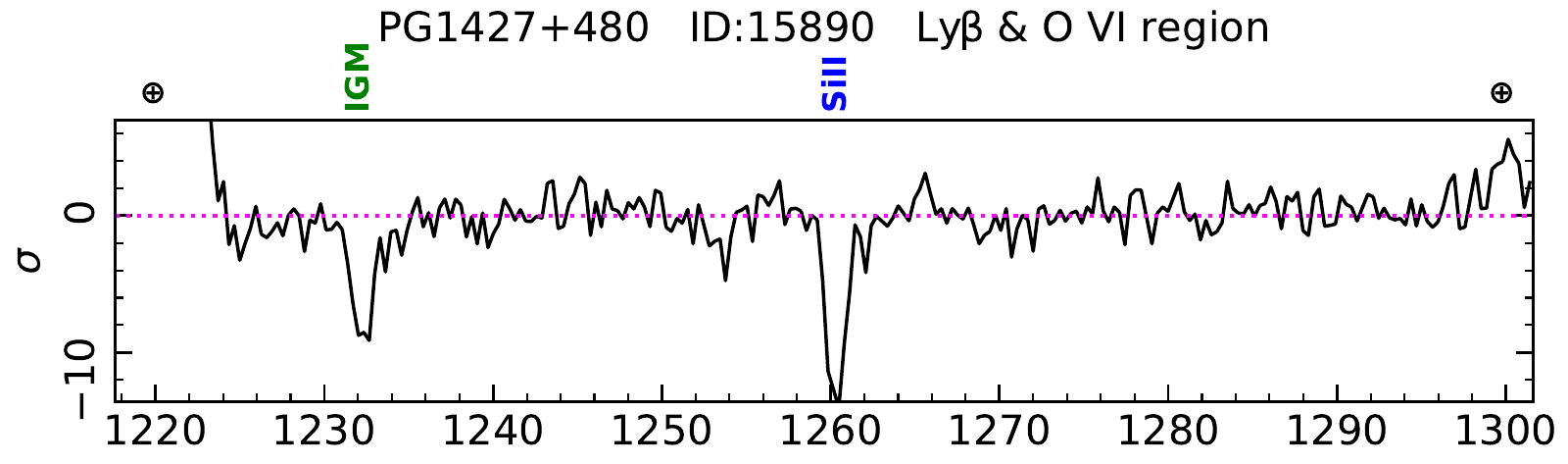}
\includegraphics[angle=0]{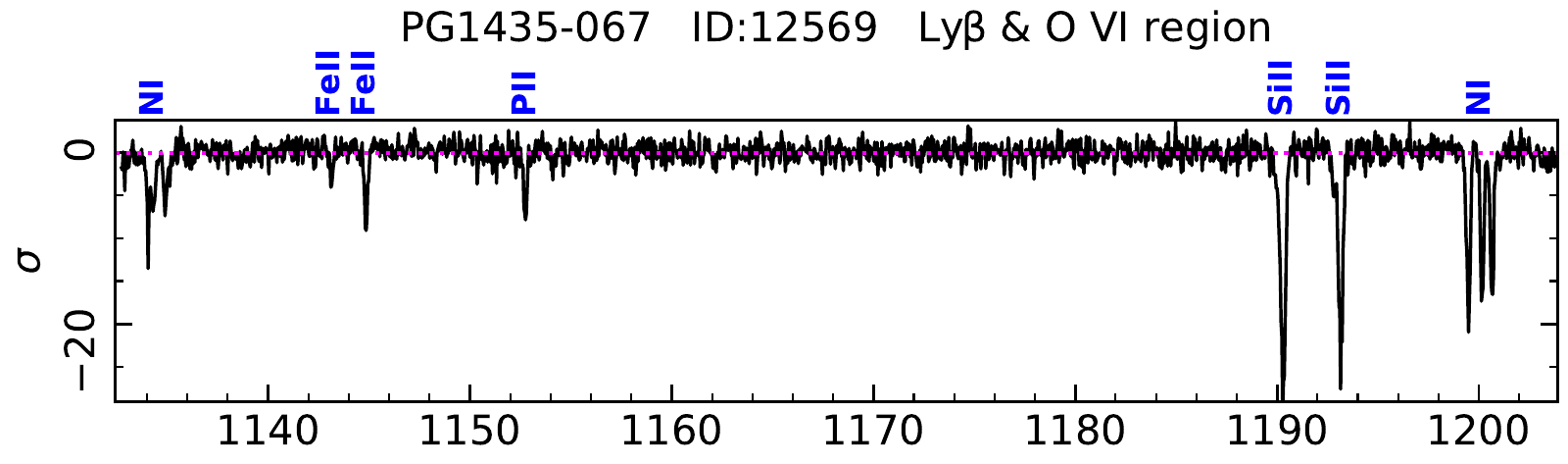}
}
\resizebox{0.94\hsize}{!}{
\includegraphics[angle=0]{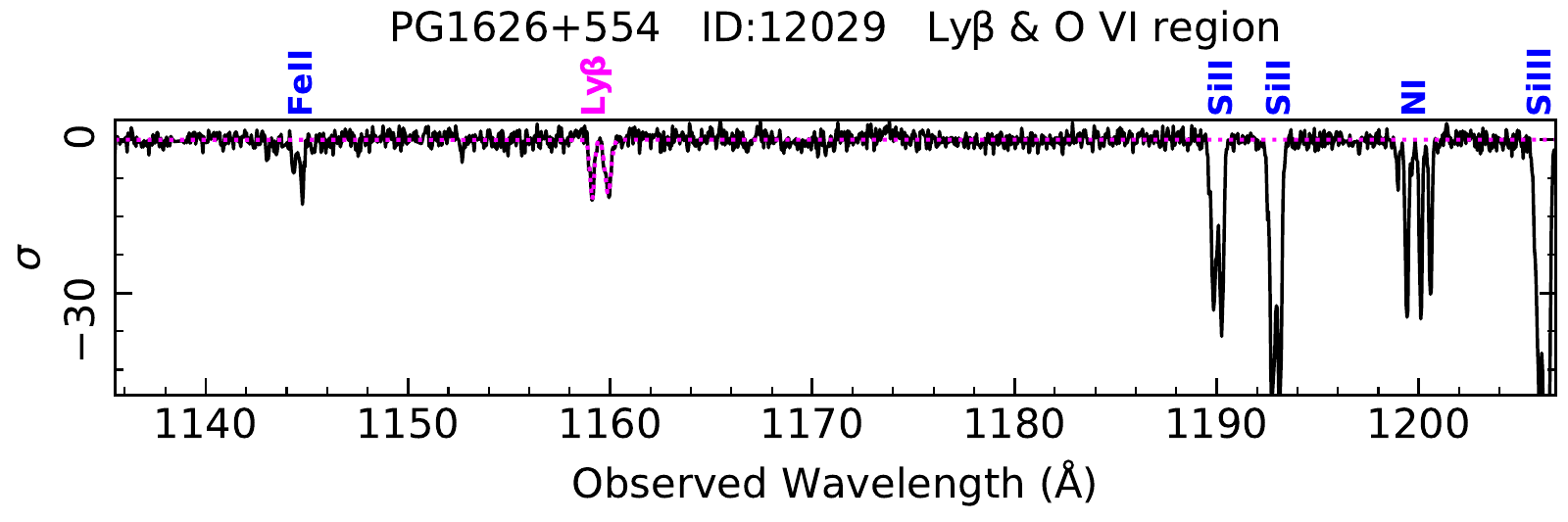}
\includegraphics[angle=0]{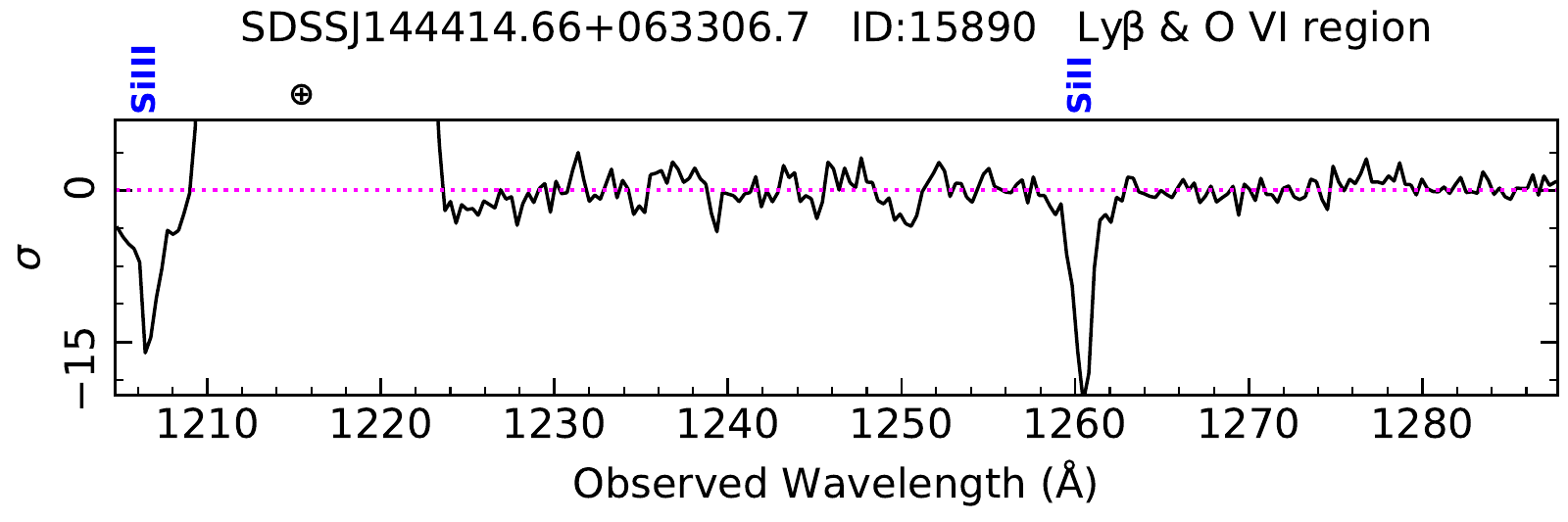}
}
\vspace{-0.2cm}
\caption{Absorption lines in the \lyb and \ovi region of the HST spectra of the \sub sample. The continuum and the emission lines are subtracted in the displayed data. The ISM lines are labeled in blue, IGM lines in green, and the intrinsic AGN lines in magenta. The geocoronal emission lines are indicated with the symbol $\oplus$. The best-fit model to the AGN absorption lines is shown as a dotted magenta line. The significance $\sigma$ is defined as (D$-$M)/E, where D is the data, M the continuum+emission line model without absorption, and E the error on the data.
\label{fig_abs_ovi}}
%\vspace{0.3cm}
\end{figure*}

\begin{figure*}
\centering
\resizebox{0.95\hsize}{!}{
\includegraphics[angle=0]{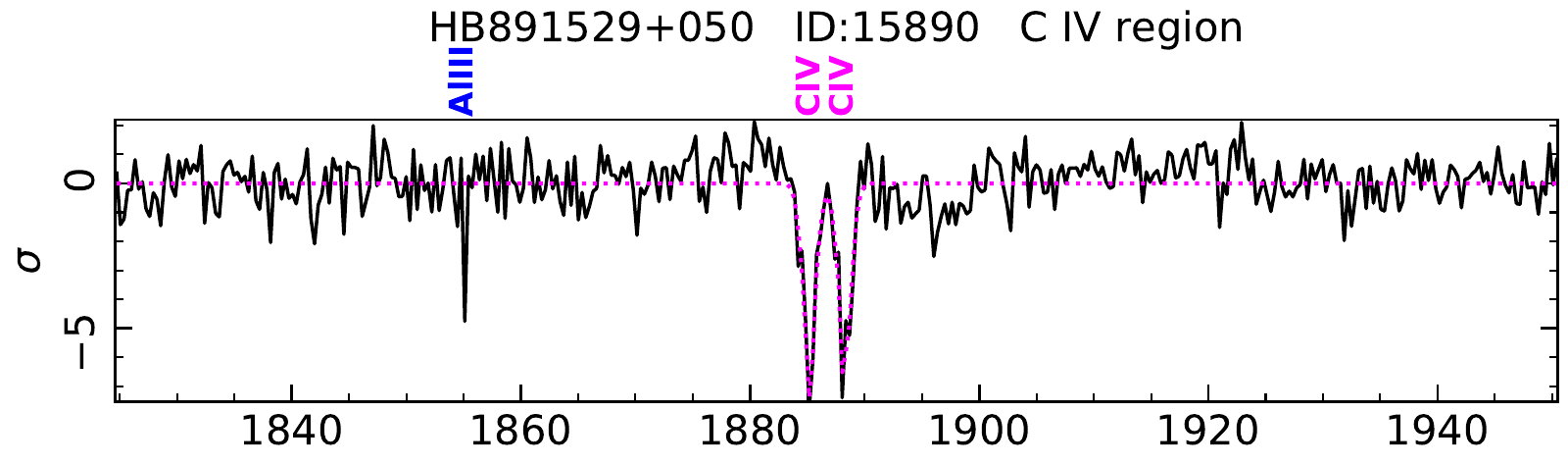}
\includegraphics[angle=0]{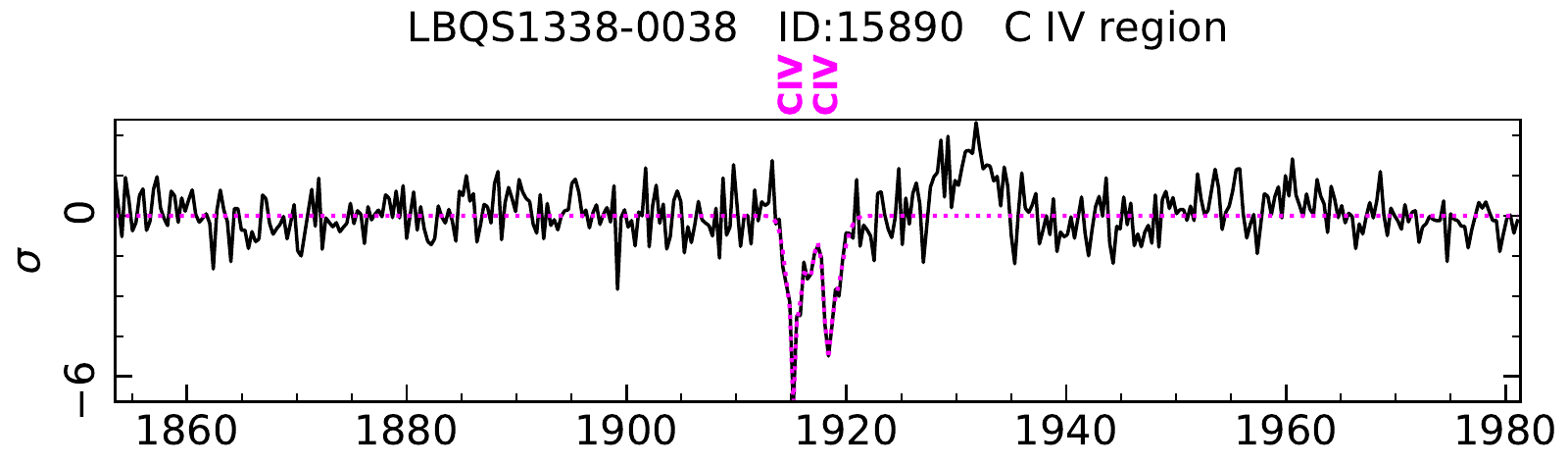}
}
\resizebox{0.95\hsize}{!}{
\includegraphics[angle=0]{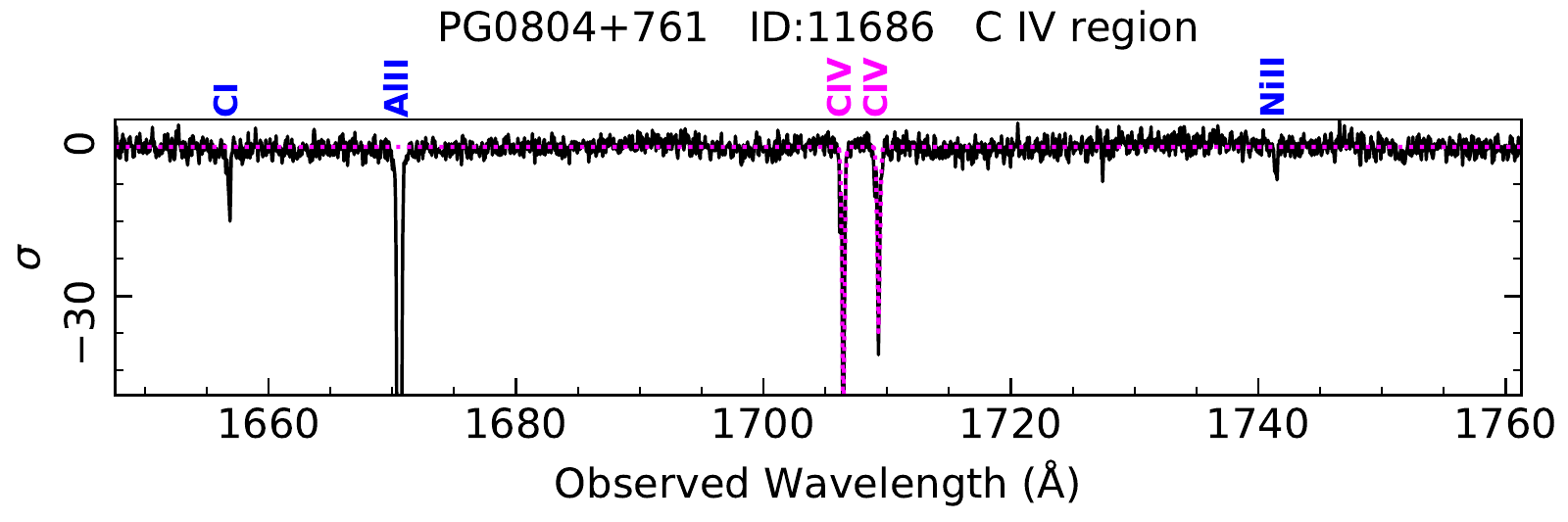}
\includegraphics[angle=0]{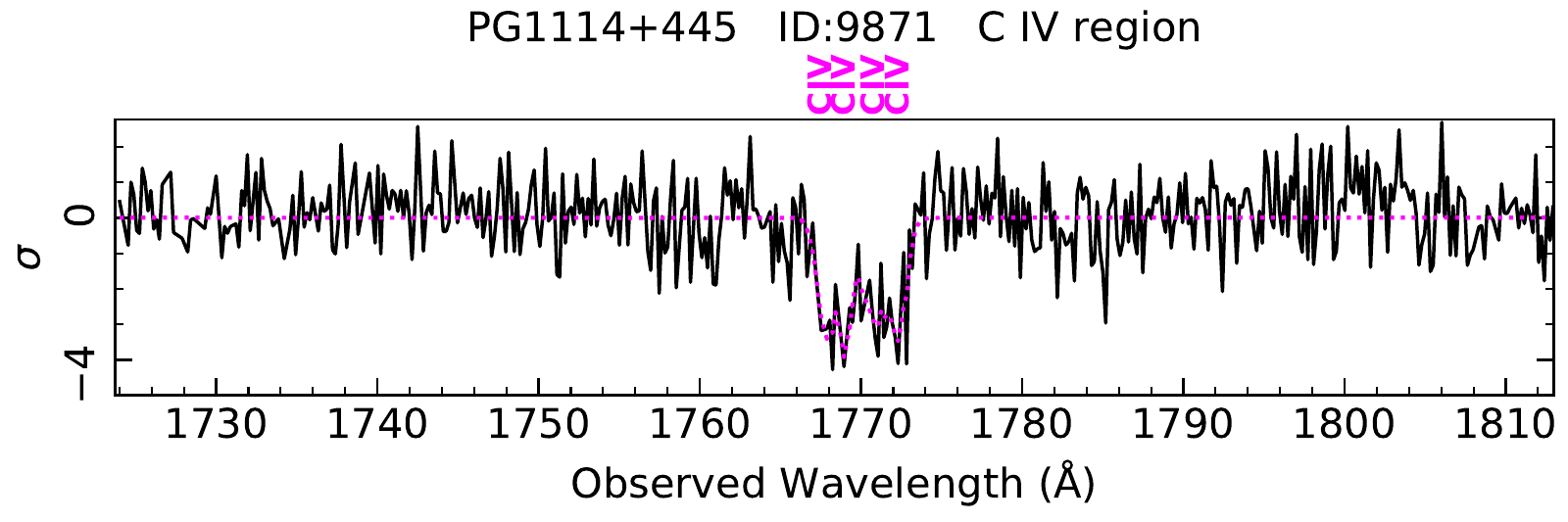}
}
%\vspace{-0.25cm}
\caption{Absorption lines in the \civ region of the HST spectra of the \sub sample. For the remaining objects that are not displayed in this figure, no significant absorption line of any kind is seen in the \civ region. The ISM lines are labeled in blue, IGM lines in green, and the intrinsic AGN lines in magenta. The best-fit model to the AGN absorption lines is shown as a dotted magenta line. The significance $\sigma$ is defined as (D$-$M)/E, where D is the data, M is the continuum+emission line model without absorption, and E is the error on the data.
\label{fig_abs_civ}}
%\vspace{0.3cm}
\end{figure*}

After fitting the continuum and the emission lines, we modeled all the intrinsic absorption lines. Each absorption line was modeled with a Gaussian function. We fit the parameters of each line, obtaining its equivalent width (EW), central wavelength (outflow velocity, $v$), and its FWHM. In the case of doublet lines (e.g., the \civ doublet), their line ratios are required to remain physically plausible: ranging from 2:1 to 1:1 ratio for the blue and red transitions, respectively. In rare cases where one of the intrinsic doublet lines fully overlaps with a strong non-intrinsic absorption line and cannot be de-blended, we gave the parameters only of  the other uncontaminated line of the doublet. For those doublet lines that their line ratios are <~2:1 (indicating line saturation), the higher $N_{\rm ion}$ that is inferred from the red transition is used in our statistical analysis of the sample, to minimize the effect of line saturation. The absorption lines and their best-fit models are shown in Fig. \ref{fig_abs_lya} (\lya and \nv region), Fig. \ref{fig_abs_ovi} (\lyb and \ovi region), and Fig. \ref{fig_abs_civ} (\civ region). To save space in these figures, panels that show no significant absorption line of any kind are excluded. The corresponding best-fit parameters of all the intrinsic absorption lines are provided in Table \ref{table_abs} (spanning two pages) in Appendix \ref{sect_append}.

%============================
% TABLE: Asymmetry parameter table
%
\begin{table}[!tbp]
\begin{minipage}[t]{\hsize}
\setlength{\extrarowheight}{3pt}
\setlength{\tabcolsep}{3.5pt}
\caption{Parameters of the \civ emission line in the HST spectra of the \sub AGN sample.}
\centering
\footnotesize
\renewcommand{\footnoterule}{}
\begin{tabular}{c c c c c}
\hline \hline
Object & $a_f$ & $a_p$ & $v_{\rm mean}$ & FWHM$_{\rm mean}$ \\
\hline
2MASS J02201457-0728591   & $1.19$ & $-0.09$ & $+220$  & $6330$ \\
2MASS J14025120+2631175   & $0.83$ & $+0.12$ & $+100$  & $6190$ \\
HB89 1257+286             & N/A    & N/A     & N/A     & N/A \\
HB89 1529+050               & $1.01$ & $-0.00$ & $-530$  & $8320$ \\
LBQS 1338-0038            & $1.37$ & $-0.18$ & $+680$  & $5180$ \\
PG 0052+251                   & $0.93$ & $+0.03$ & $-260$  & $6710$ \\
PG 0804+761               & $0.96$ & $+0.02$ & $-310$  & $5790$ \\
PG 0947+396                   & $0.99$ & $+0.00$ & $-150$  & $7170$ \\
PG 0953+414                   & $0.98$ & $+0.02$ & $-340$  & $6500$ \\
PG 1114+445               & $0.98$ & $+0.01$ & $-40$   & $6660$ \\
PG 1202+281                   & $1.06$ & $-0.05$ & $-300$  & $7340$ \\
PG 1216+069                   & $0.91$ & $-0.05$ & $+230$  & $4970$ \\ 
PG 1307+085                   & $0.86$ & $+0.12$ & $-670$  & $5940$ \\   
PG 1352+183                   & $0.77$ & $+0.14$ & $-1040$ & $5540$ \\
PG 1402+261               & $0.83$ & $+0.11$ & $-1050$ & $6450$ \\
PG 1416-129               & $0.88$ & $+0.04$ & $-460$  & $7000$ \\
PG 1425+267                   & N/A    & N/A     & N/A     & N/A \\
PG 1427+480                   & $1.15$ & $-0.10$ & $+60$   & $5900$ \\
PG 1435-067                   & $0.95$ & $+0.04$ & $-320$  & $6580$ \\
PG 1626+554                   & $1.08$ & $-0.08$ & $+160$  & $7060$ \\
SDSS J144414.66+063306.7        & $1.15$ & $-0.10$ & $+160$  & $6890$ \\
\hline
\end{tabular}
\end{minipage}
\tablefoot{
The $a_f$ as the "flux asymmetry" parameter and $a_p$ as the "profile asymmetry" parameter, as defined in Sect. \ref{sect_fit}. The $v_{\rm mean}$ and FWHM$_{\rm mean}$ are the flux-weighted mean velocity shift and width of the line, respectively, in \kms. The parameters in this table are calculated using the best-fit model parameters given in Table \ref{table_emission}. For the two objects labeled with "N/A" the HST spectra do not cover the \civ emission
line.}
\label{table_asym}
\end{table}
%============================

%============================
% FIG: Lya and N V absorption residuals
%
%============================
%
%============================
% FIG: Ly-beta and O VI absorption residuals
%
%============================
%
%============================
% FIG: C IV absorption residuals
%
%============================

%============================
% FIG: UFO search
%
\begin{figure*}
\centering
\resizebox{0.95\hsize}{!}{
\includegraphics[angle=0]{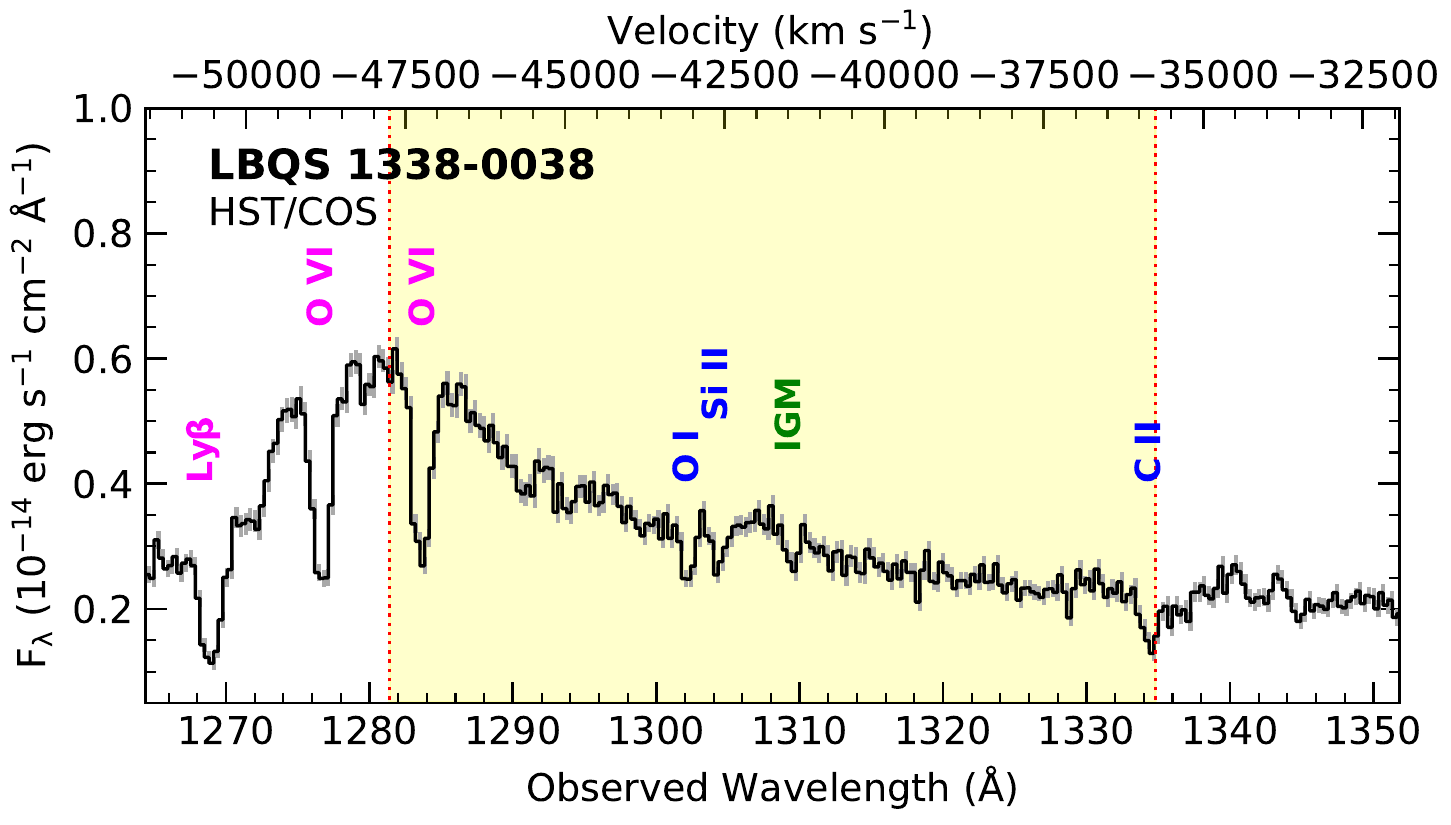}\hspace{0.7cm}
\includegraphics[angle=0]{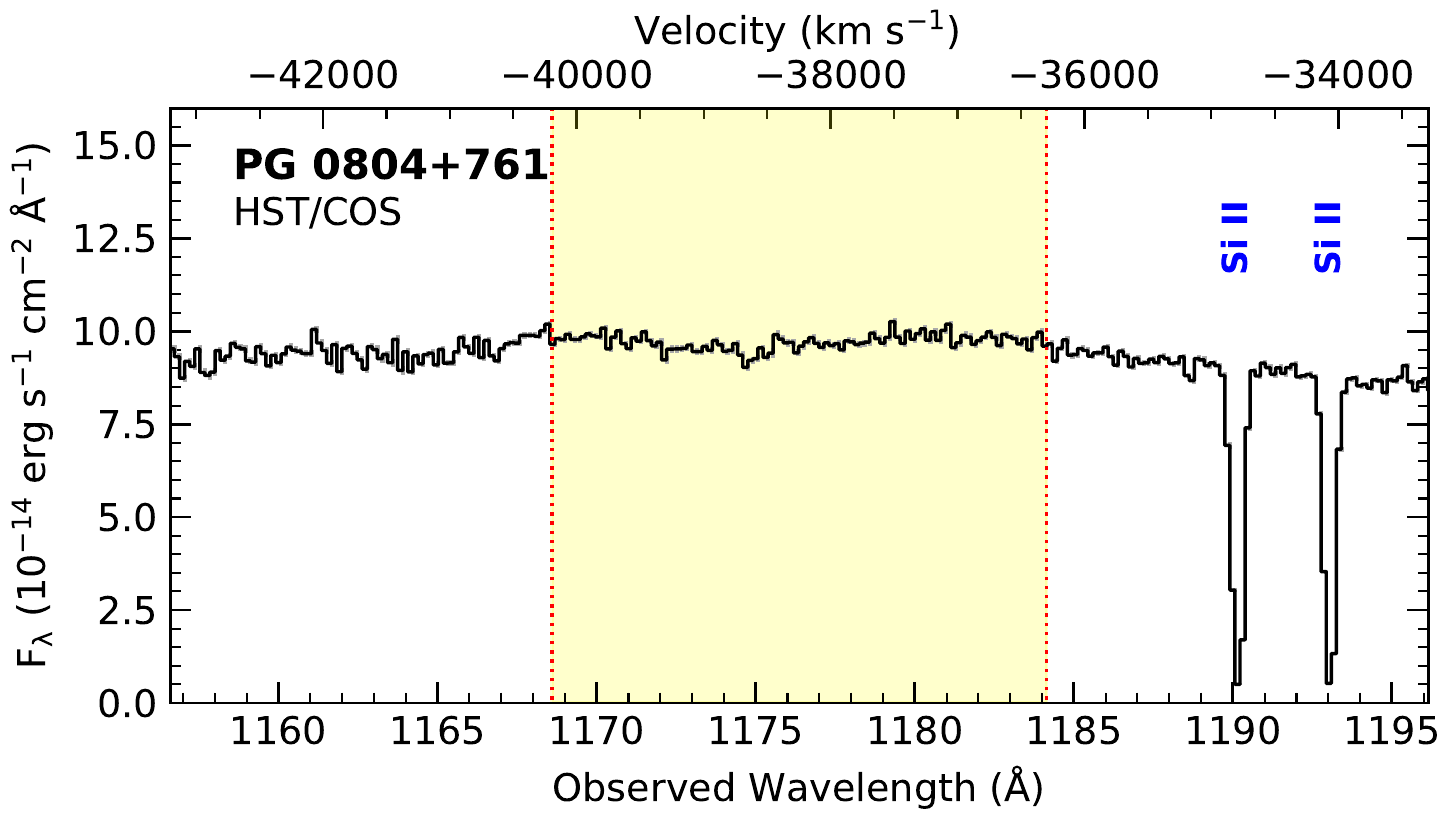}
}
\resizebox{0.95\hsize}{!}{
\includegraphics[angle=0]{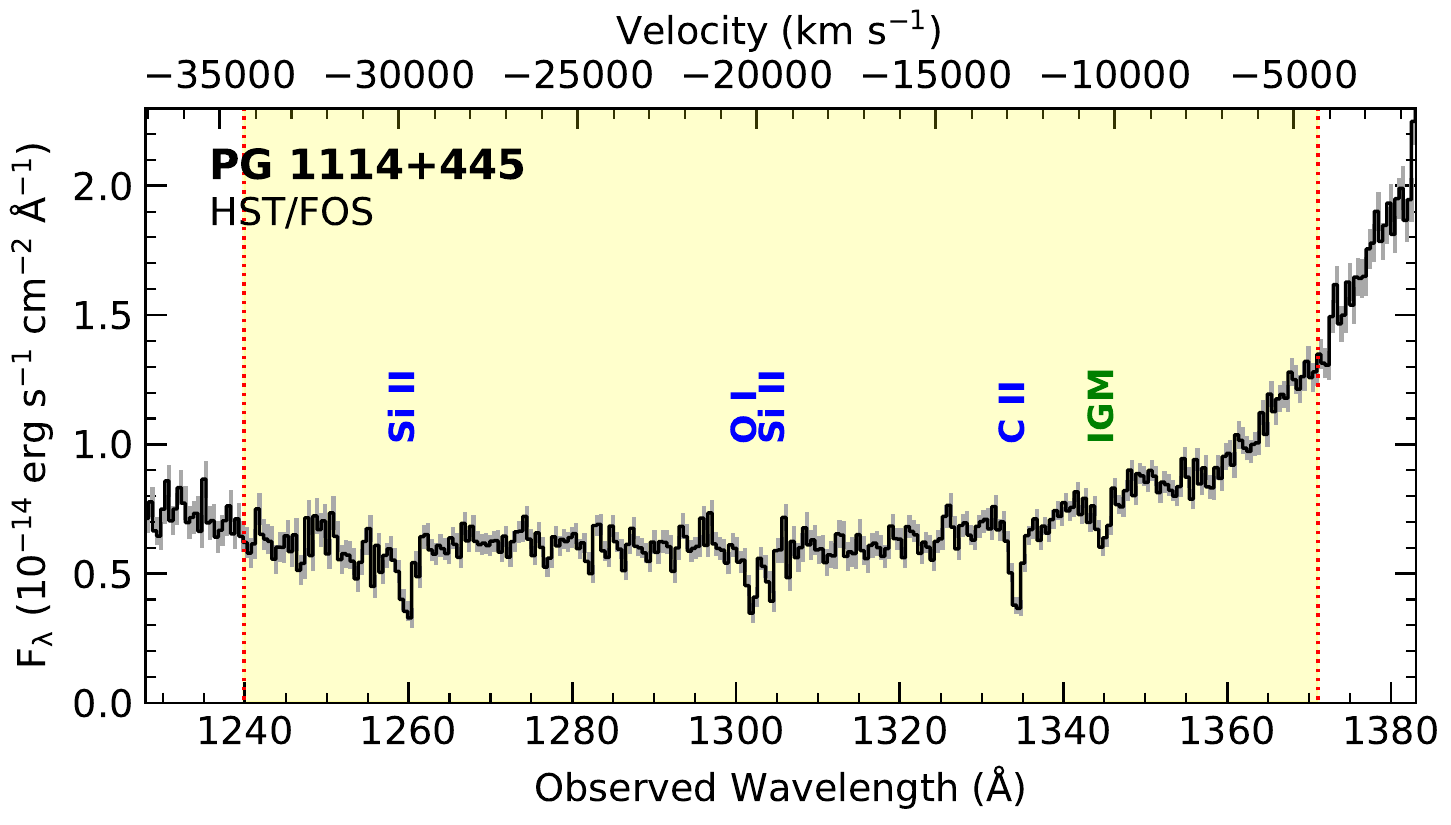}\hspace{0.7cm}
\includegraphics[angle=0]{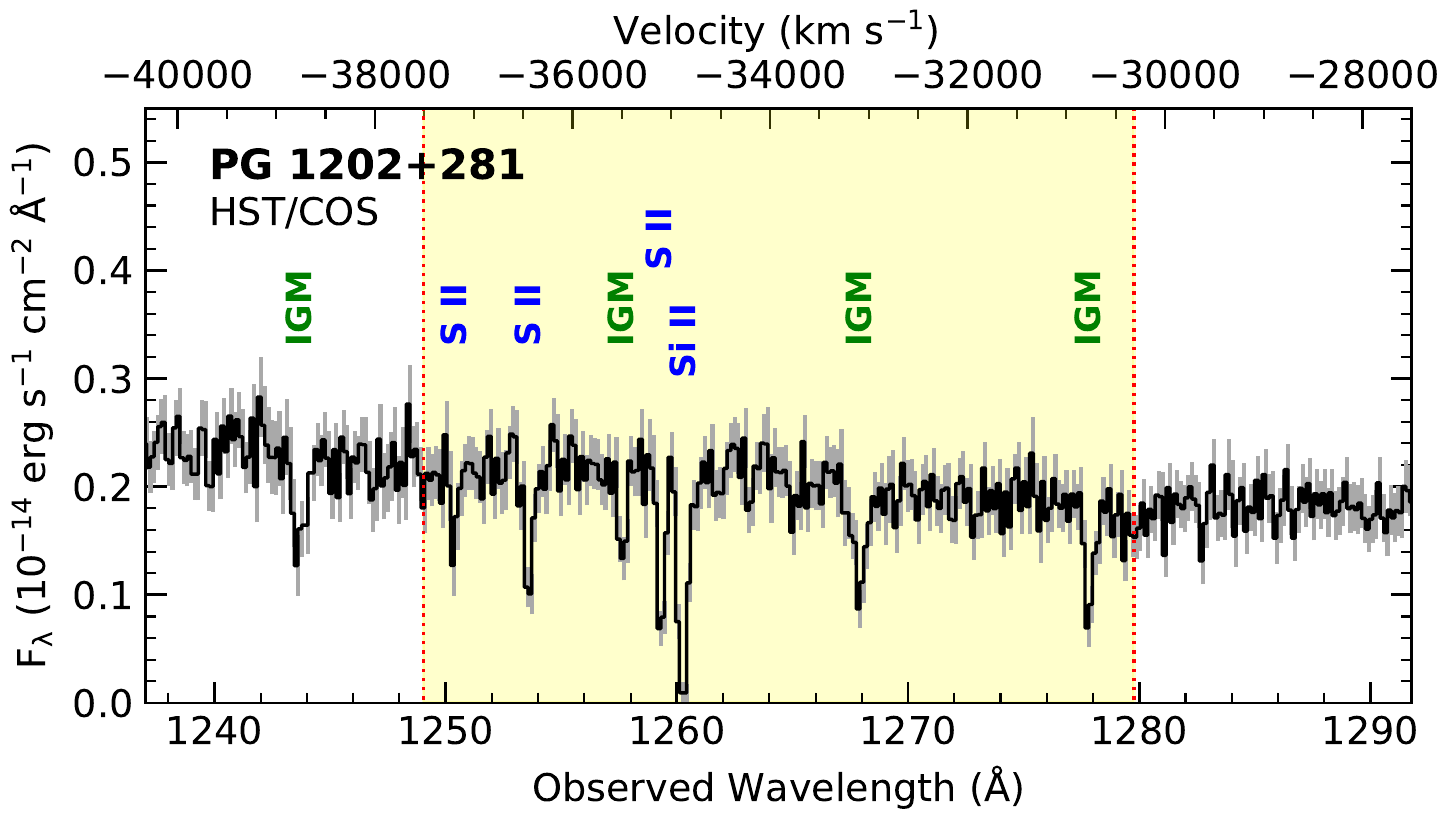}
}
\caption{Regions of the HST spectra where the blueshifted \lya counterparts of the X-ray UFOs (\paperI) would be expected. In each panel, the region highlighted in yellow between the vertical dotted lines corresponds to the velocity uncertainty of the X-ray UFO. The significant UV absorption features are labeled. The ISM lines are labeled in blue, IGM lines in green, and the intrinsic AGN absorption lines in magenta. No apparent UV counterparts to the X-ray UFOs are detected in the HST spectra. The intrinsic absorption lines in LBQS 1338-0038 ({\it top-left panel}) are from the warm absorber and have moderate velocities. The parameters of the AGN absorption lines are provided in Table \ref{table_abs}. We note that for three of the objects with X-ray UFO in \paperI, the UV counterpart cannot be checked and thus are not included in this figure: the heavily reddened 2MASS~J10514425+3539306 and 2MASS J16531505+2349427 do not have any HST UV spectra, and in the case of PG~0947+396, the potential \lya counterpart of the X-ray UFO would fall outside the detection range of the HST data.
\label{fig_ufo}}
%\vspace{0.6cm}
\end{figure*}
%============================

In the \xmm study of the \sub sample (\paperI), statistically significant Fe~K absorption by UFOs, at ${\gtrsim {95\%}}$ confidence level according to Monte Carlo simulations, was found in seven targets: 2MASS~J10514425+3539306, 2MASS~J16531505+2349427, LBQS~1338-0038, PG~0804+761, PG~0947+396, PG~1114+445, and PG~1202+281. We used the measured outflow velocity of the X-ray UFOs (\paperI) and the redshift of the targets in Table \ref{table_log} (which are also used in \paperI) to calculate where any corresponding UV absorption counterpart would appear in the HST spectra. As described in Sect. \ref{sect_sample}, the heavily-reddened 2MASS~J10514425+3539306 and 2MASS J16531505+2349427 do not have any HST UV spectra. Also, in the case of PG~0947+396, the potential \lya counterpart of the X-ray UFO would fall outside the detection range of the available HST data and thus cannot be investigated. For the four remaining targets, we show the HST spectral regions corresponding to potential \lya counterparts in Fig. \ref{fig_ufo}. In previous UV UFO studies, the strongest UV signature of X-ray UFOs has been seen as \lya \citep{Kris18a,Mehd22b} at the same velocity as the X-ray UFO. In Fig. \ref{fig_ufo}, the highlighted regions between the vertical dotted lines correspond to the velocity uncertainty of the X-ray UFO. All absorption lines are identified in Fig. \ref{fig_ufo} and we find no relativistically blueshifted \lya counterpart to the X-ray UFOs.

The velocity shift of the intrinsic absorption lines in the HST spectra of the \sub sample ranges from $+600$~\kms to $-3300$~\kms (Table \ref{table_abs}). All the absorption lines are relatively narrow ({40 < FWHM < 680}~\kms) and no apparent broad-absorption lines are found in the spectra. We discuss the UV outflow properties of the \sub HST sample in Sect. \ref{sect_wind}.

For each intrinsic UV absorption line, we calculated an estimate for its column density, $N,$ (see the techniques described in \citealt{Sava91}). Assuming the line is optically thin and lies on the linear part of the curve-of-growth, its equivalent width (EW) can be used to calculate a minimum column density for the absorbing ion ($N_{\rm ion}$) according to:
\begin{equation}
N_{\rm ion} ({\rm cm}^{-2}) = \frac{m_{\rm e}\, c^2\, \tau}{\pi\, e^2\, f\, \lambda^2}  \approx \frac{1.13 \times 10^{20}}{f\, \lambda^2}\, {\rm EW}(\mbox{\AA}),
\end{equation}
where $m_{\rm e}$ is electron mass, $c$ the speed of light, $\tau$ the optical depth, $e$ the electric charge, $f$ the oscillator strength, and $\lambda$ the laboratory wavelength of the line. For each line, we used the $\lambda$ and $f$ values provided by the Atomic Spectra Database v5.9 \citep{Kram21} of National Institute of Standards and Technology (NIST). Our calculation of $N_{\rm ion}$ assumes the UV absorber has a full covering fraction ({\cf = 1}) and, thus, $N_{\rm ion}$ is considered a minimum column density. In this work where we carry out an empirical analysis of the absorption lines for a sample of objects, this approach is practical and adequate for the purpose of estimating the column densities of the UV ions.

%============================
% FIG: Bar plot
%

\begin{figure}
\centering
\resizebox{0.93\hsize}{!}{\includegraphics[angle=0]{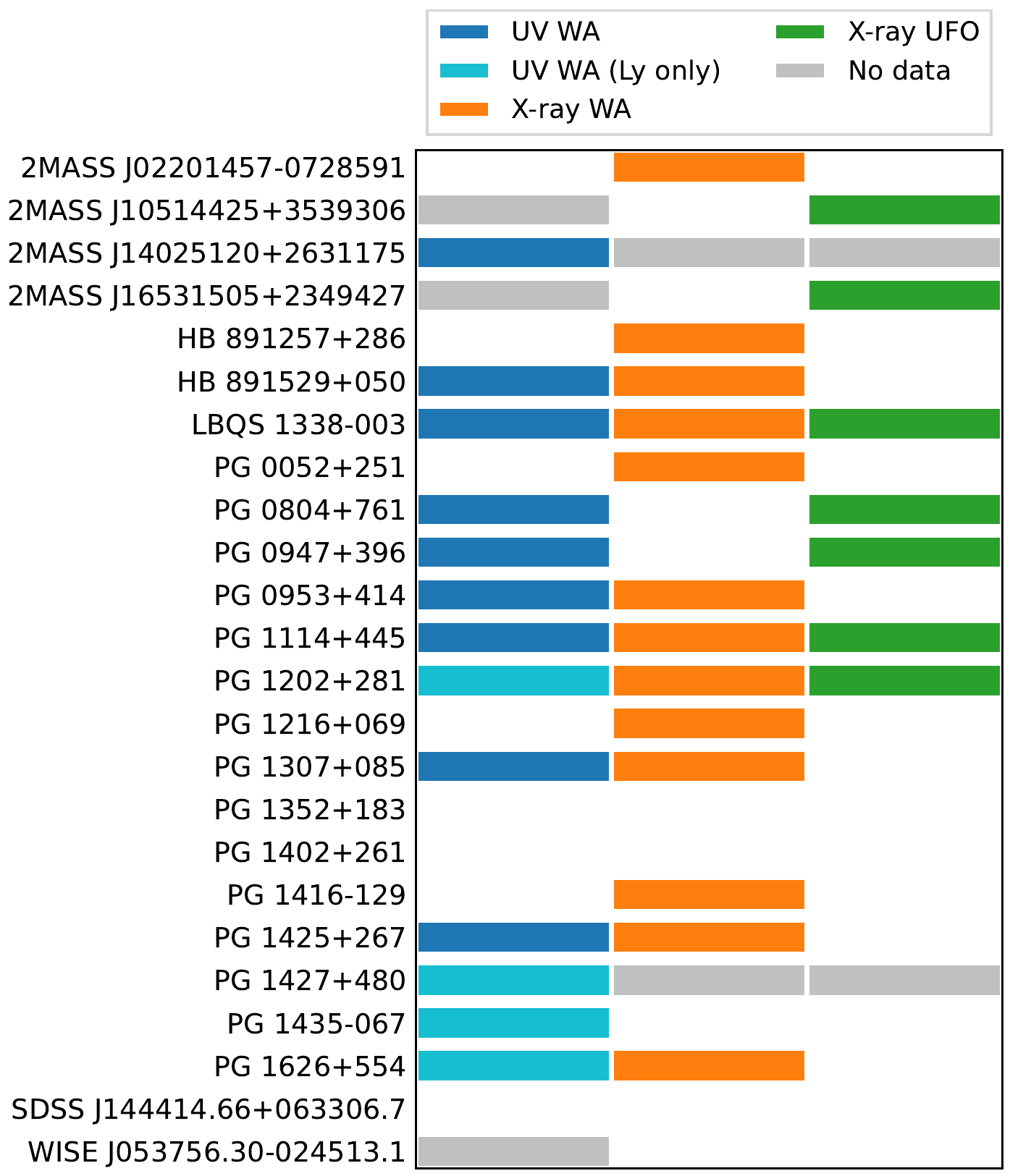}}
\caption{Chart illustrating what outflows are found in each AGN. Gray space means the data is unavailable. White space means WA or UFO is not found in the data.
\label{fig_bar}}
%\vspace{0.5cm}
\end{figure}
%============================

After completing the spectral analysis and parametrization of all the intrinsic emission and absorption lines, we assessed the statistical properties of our results and examined the relations between different parameters. The chart in Fig. \ref{fig_bar} illustrates the types of outflows that are found for each AGN. The histograms of Fig. \ref{fig_hist} illustrate the distribution of the \ion{C}{iv} $a_p$ and $a_f$ asymmetry parameters, as well as the velocities, $v_{\rm mean}$ and FWHM$_{\rm mean}$, for the \sub HST sample. In Fig. \ref{fig_uv_rel}, using the best-fit parameters of the intrinsic absorption lines for the \sub sample (Table \ref{table_abs}), the column density, $N_{\rm ion}$, of each ion (top panel), and its turbulent velocity, $\sigma_v$, (bottom panel), are plotted versus the velocity shift $v$ (i.e., flow velocity). The turbulent velocity, $\sigma_v$, is calculated from the line's measured FWHM (Table \ref{table_abs}) according to ${\sigma_v=\mathrm{FWHM}/\sqrt{\ln 256}}$. The displayed uncertainty on $N_{\rm ion}$ in Fig. \ref{fig_uv_rel} corresponds to the fractional uncertainty of the line's EW (Table \ref{table_abs}). In Fig. \ref{fig_kplot}, we show the kinematic plot (k-plot) for the \sub UV measurements. The k-plot, introduced in \cite{Gasp18}, is a useful diagnostic tool for exploring the kinematics of inflows and outflows in the CCA feeding and feedback mechanism. Finally, in Fig. \ref{fig_rel}, we show how our obtained parameters of the intrinsic UV absorption are related (or unrelated) to other AGN properties: the bolometric luminosity, $L_{\rm bol}$, the Eddington ratio, ${L_{\rm bol} / L_{\rm Edd}}$, and its redshift. The Spearman’s rank correlation coefficient ($r_s$) and the corresponding null hypothesis p-value probability ($p_{\rm null}$) from the two-sided t-test are calculated for the data points shown in Figs. \ref{fig_uv_rel} and \ref{fig_rel}. We discuss and interpret the statistical results and the relations between the AGN outflow parameters in Sect. \ref{sect_rel}.

%%%%%%%%%%%%%%%%%%%%%%%%%%%%%%%%%%%%%%%%%%%%%%%%%%%%%%%%%%%%%%%%%%%%%%%%%%%%%%%%%%%%%%%%%%%%%%%%%%%%%%%
%%%%%%%%%%%%%%%%%%%%%%%%%%%%%%%%%%%%%%%%%%%%%%%%%%%%%%%%%%%%%%%%%%%%%%%%%%%%%%%%%%%%%%%%%%%%%%%%%%%%%%%
%%%%%%%%%%%%%%%%%%%%%%%%%%%%%%%%%%%%%%%%%%%%%%%%%%%%%%%%%%%%%%%%%%%%%%%%%%%%%%%%%%%%%%%%%%%%%%%%%%%%%%%
\section{Discussion} 
\label{sect_discuss}

%%%%%%%%%%%%%%%%%%%%%%%%%%%%%%%%%%%%%%%%%%%%%%%%%%%%%%%%%%%%%%%%%%%%%%%%%%%%%%%%%%%%%%%%%%%%%%%%%%%%%%%
\subsection{UV properties of ionized outflows in AGN at intermediate redshifts}
\label{sect_wind}

Ionized outflows have been studied via X-ray and UV spectroscopy in mainly the local Seyfert-1 galaxies at redshift $z < 0.1$ (e.g., \citealt{Blu05,Dunn07,Crens12,Laha14}). The \sub campaign extends the X-ray (\paperI) and UV (this paper) spectroscopy of the outflows to higher redshifts (0.1--0.4) and luminosities ($10^{45}$--$10^{46}$~\ergs). We derived parameters of the AGN outflows that are detected in the UV band. Below we compare our findings and statistics on the properties of the \sub sample with previous UV sample studies of ionized outflows in Seyfert galaxies \citep{Cren99,Dunn07,Crens12} and quasars/QSOs \citep{Misa07,Cull19,Veil22}. Similar to our study, using HST/COS observations of the QUEST (Quasar/ULIRG Evolutionary Study) sample, \cite{Veil22} investigated the properties of ionized \nv and \ovi outflows in quasars at ${z \lesssim 0.3}$.

Using our HST observations of the \sub sample we looked for UV spectral signatures of different types of AGN outflows: narrow and low-velocity absorption lines (i.e., the warm-absorber outflows), broad and intermediate-velocity absorption lines (i.e., the obscuring disk winds), and broad and narrow relativistic absorption lines (i.e., UFOs). We find that the characteristics and parameters of all the intrinsic absorption lines in the \sub AGN sample are consistent with warm-absorber outflows. The absorption lines in the \sub sample are relatively narrow ({40 < FWHM < 680}~\kms) with velocity shifts ranging from $+600$ to $-3300$~\kms. We note that only one component of one object (Comp. 1 of PG~0804+761) shows positive $v$ (inflow); hence, apart from this outlier, the components of all other objects only show  outflows. We find that the column densities and velocities of the UV absorbers in the \sub sample (Table \ref{table_abs}) are generally comparable to those found in the Seyfert galaxies \citep{Cren99} and quasars \citep{Veil22}.

We do not detect significant \siiv absorption in any of the QSOs in the \sub HST sample, whereas narrow \siiv absorption lines from warm absorbers are sometimes seen in Seyfert-1 galaxies (e.g., \citealt{Math97,Gab05}). The reason for this may be due to an ionization effect \citep{Math97}, where \siiv originates from a lower ionization phase than the other UV ions we see in the \sub sample (\civ, \nv, and \ovi). The photoionization computations of \citet{Meh16b} show that \siiv concentration peaks at significantly lower ionization parameters than those of the other aforementioned higher ionization ions. Since the \sub targets are more luminous than typical Seyfert-1 galaxies, for comparable WA parameters there will be less \siiv in spectra of the \sub QSOs. Hence, due to the lower ionic column density, \siiv may not produce detectable absorption features.

Previous UV surveys of ionized outflows in Seyfert galaxies suggest that they have a global covering fraction of about 0.5 \citep{Dunn07,Crens12}. In our investigation we find that 13 out of 21 AGN (about 60\%) exhibit intrinsic UV absorption lines. This includes targets that only show intrinsic \ion{H}{i} absorption (\lya and \lyb lines). If one sets a criterion for ionized absorption (i.e., showing either \civ, \nv, or \ovi absorption lines), then 9 out of 21 (about 40\%) targets show intrinsic ionized outflows (see Fig. \ref{fig_bar}). This is comparable to those of Seyfert galaxies \citep{Dunn07,Crens12} and the recent results of \citet{Veil22}, where they find 60\% of the quasars show \nv and \ovi outflows. Similarly, \citet{Cull19} found the fraction of quasars with at least one intrinsic absorption system is about 40\%, and \citet{Misa07} reported that half of their observed quasars contain intrinsic narrow absorption lines. Thus, the ionized outflows in the \sub HST sample and similar quasars are likely to have a geometry and global covering fraction similar to those of local Seyfert galaxies. Since the outflow parameters and physical characteristics of ionized outflows in the AGN spanning low to intermediate redshifts (${z < 0.4}$) and luminosities ($L_{\rm bol} < 10^{46}$~\ergs) are consistent with each other, this points to a common mechanism for the launch and driving of the AGN outflows.

%============================
% FIG: histograms figure
%
\begin{figure}
\centering
\resizebox{0.83\hsize}{!}{\includegraphics[angle=0]{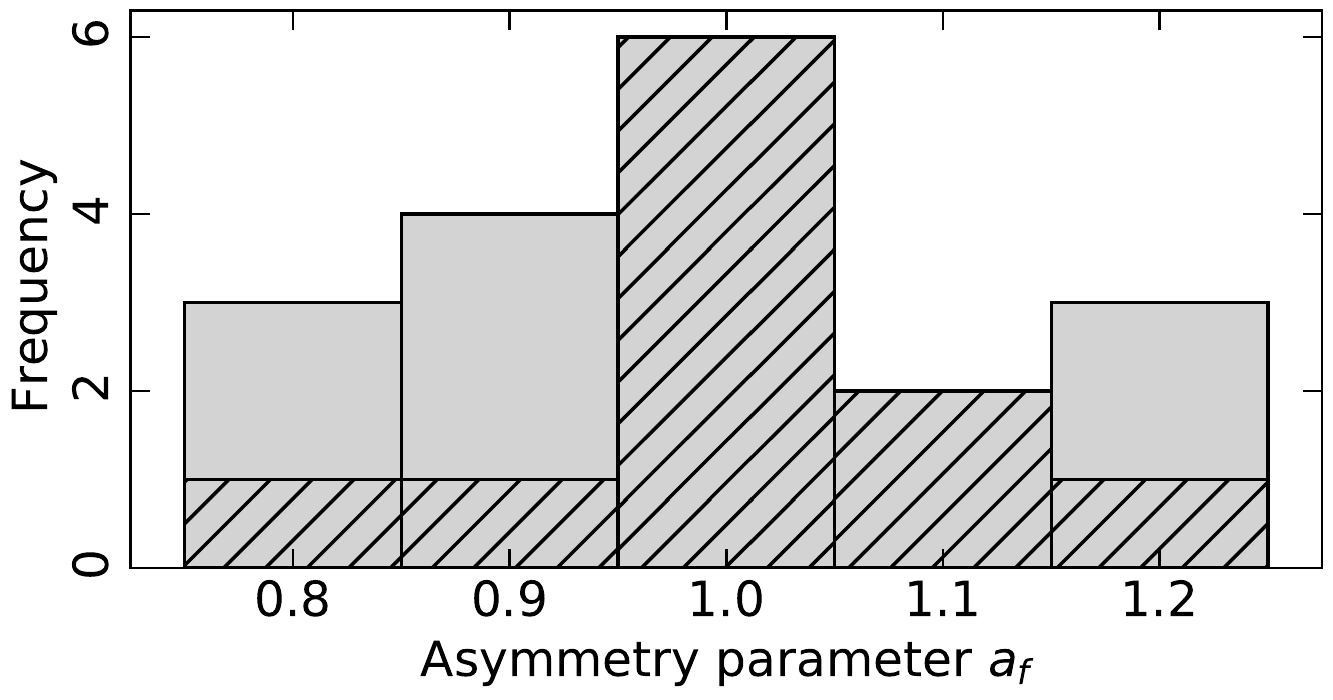}}\vspace{0.1cm}
\resizebox{0.83\hsize}{!}{\includegraphics[angle=0]{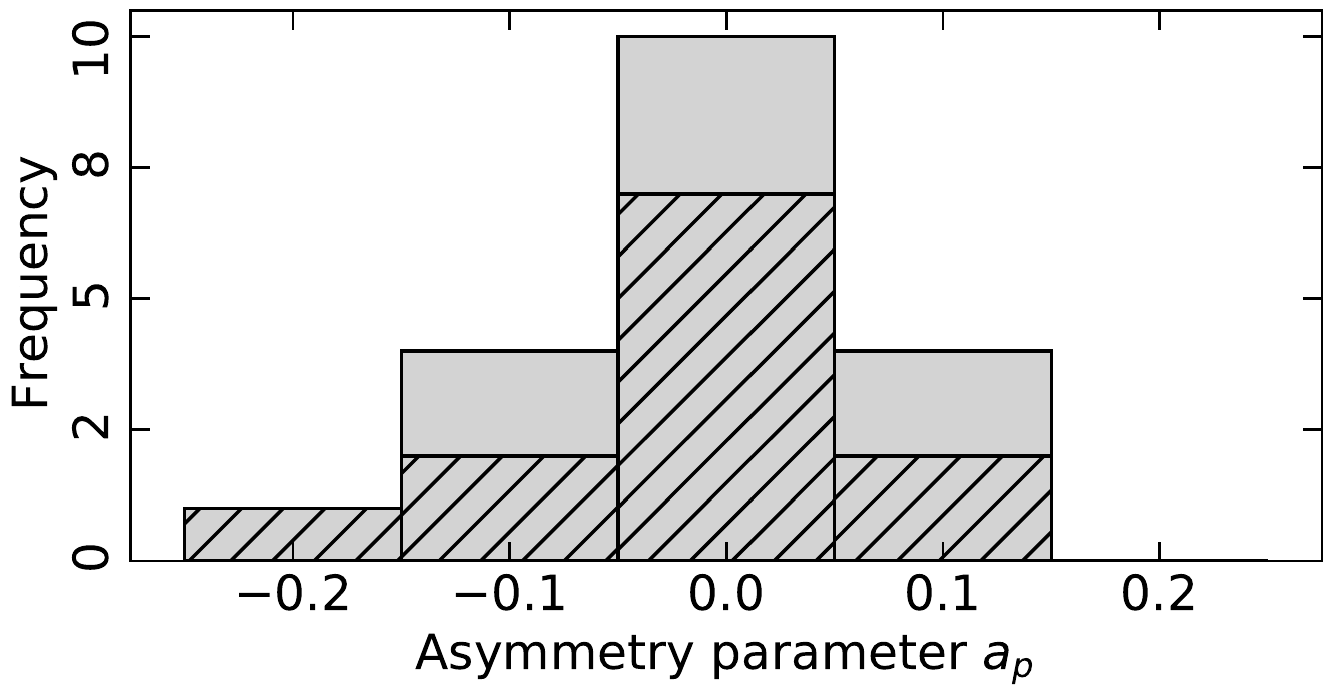}}\vspace{0.1cm}
\resizebox{0.83\hsize}{!}{\includegraphics[angle=0]{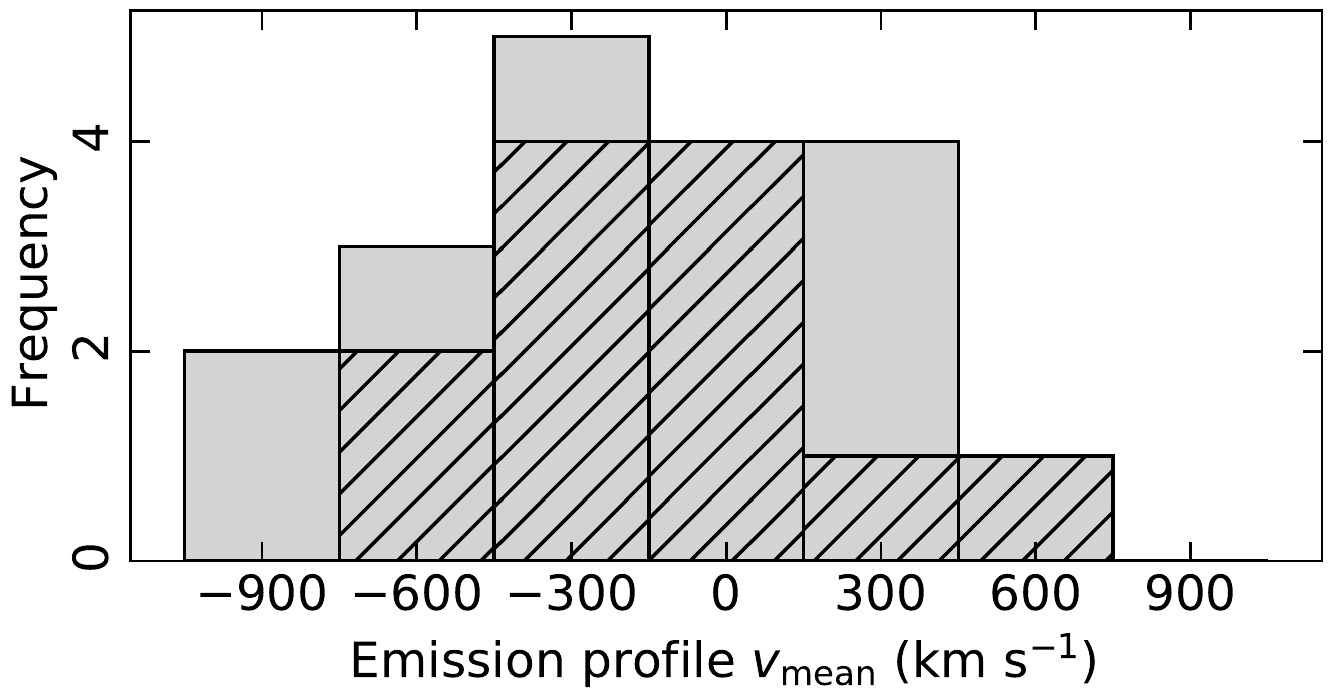}}\vspace{0.1cm}
\resizebox{0.83\hsize}{!}{\includegraphics[angle=0]{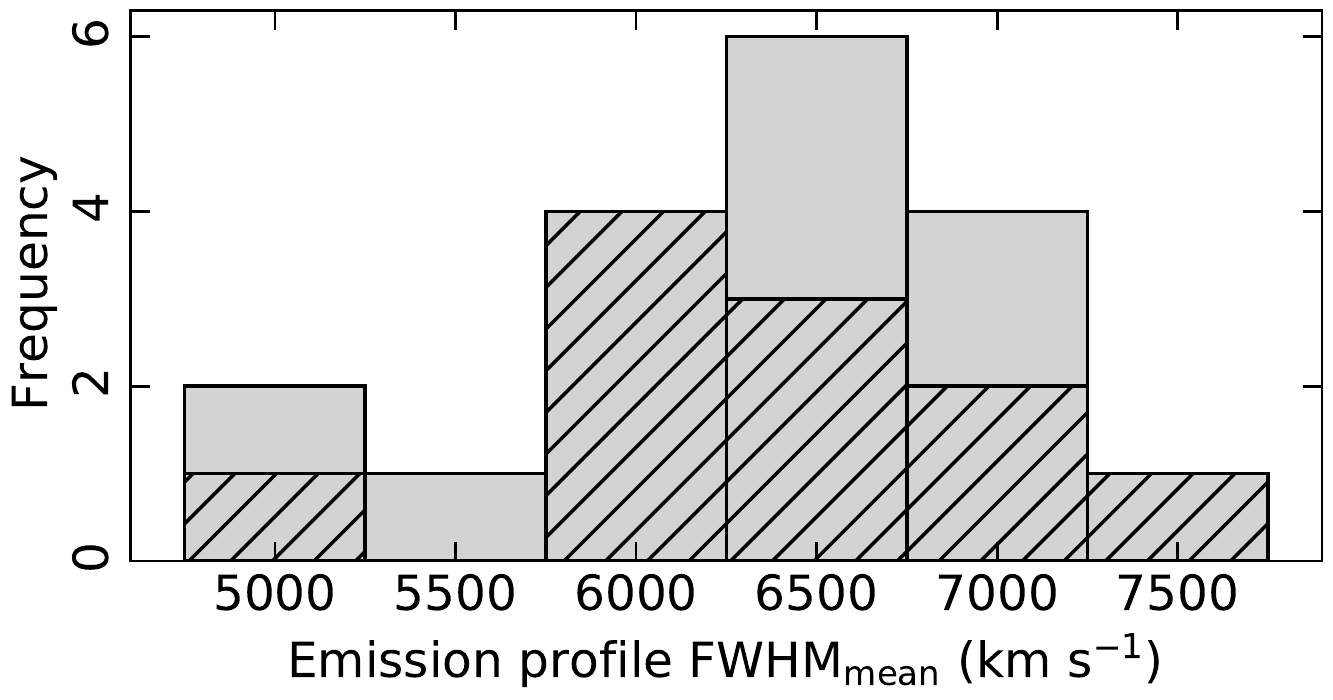}}
\caption{Distributions of the parameters of the \civ emission line in the \sub HST sample. The histograms are produced using the derived parameters given in Table \ref{table_asym}. The targets that have intrinsic absorption by ionized outflows are indicated with hatched lines.
\label{fig_hist}}
\end{figure}
%============================

%============================
% FIG: UV WA scatter plot
%
\begin{figure}
\centering
\resizebox{0.98\hsize}{!}{\includegraphics[angle=0]{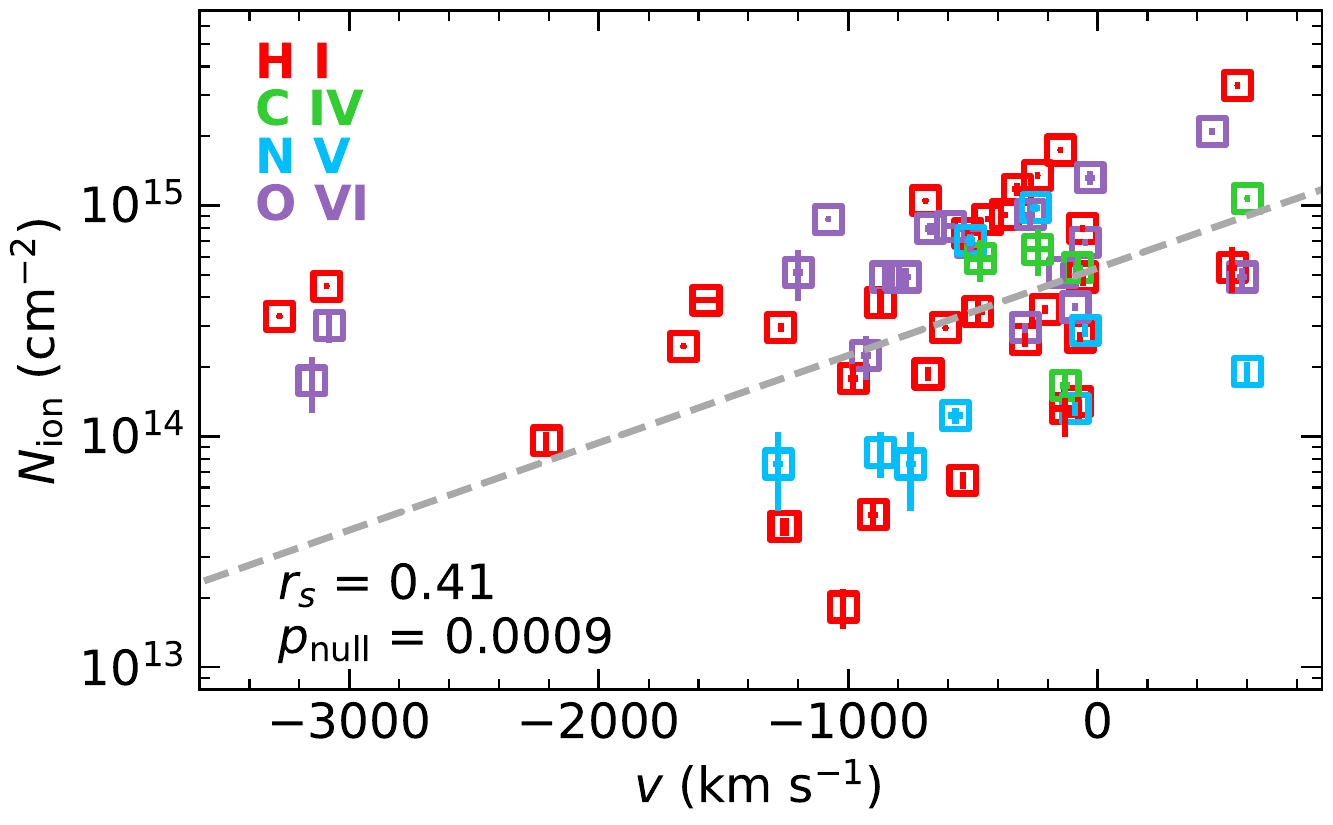}}
\resizebox{0.98\hsize}{!}{\includegraphics[angle=0]{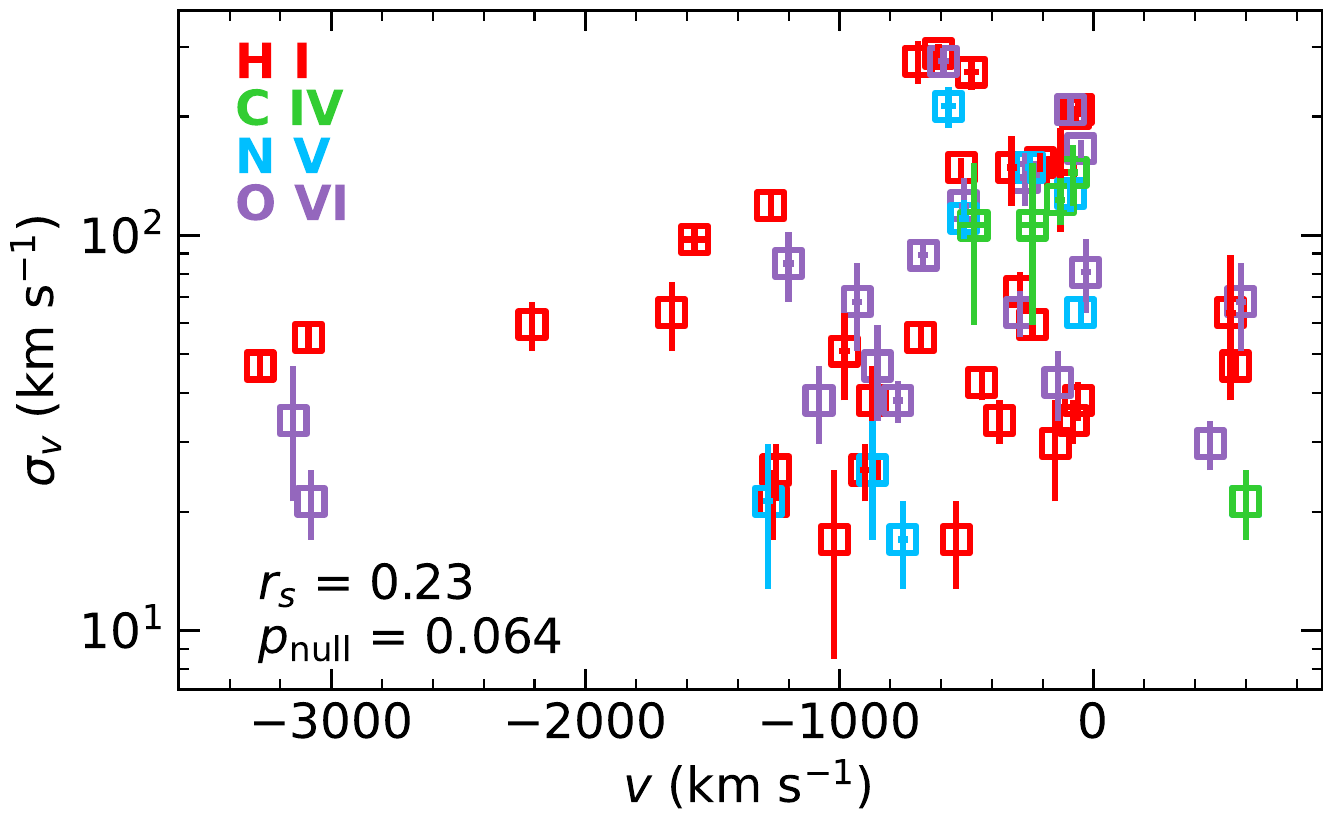}}
\caption{Column density $N_{\rm ion}$ ({\it top panel}) and turbulent velocity, $\sigma_v$, ({\it bottom panel}) plotted versus the flow velocity, $v,$ of the intrinsic UV absorption lines detected in the \sub HST sample. The data points correspond to the best-fit parameters of Table \ref{table_abs}. The $\sigma_v$ is calculated from the FWHM of the absorption line according to ${\sigma_v=\mathrm{FWHM}/\sqrt{\ln 256}}$. The Spearman’s rank correlation coefficient, $r_s$, and the null hypothesis probability, ${p_{\rm null}}$, are given in the inset. There is a statistically significant correlation between $N_{\rm ion}$ and $v$, even including the four outlier data points on the far left, which belong to only one object (PG~1307+085). All data points with a positive value for $v$ (i.e., inflow) belong to only one component of one object (Comp. 1 of PG~0804+761). All other targets show only outflows.
\label{fig_uv_rel}}
\end{figure}
%============================

%============================
% FIG: k-plot
%
\begin{figure}
\centering
\resizebox{0.98\hsize}{!}{\includegraphics[angle=0]{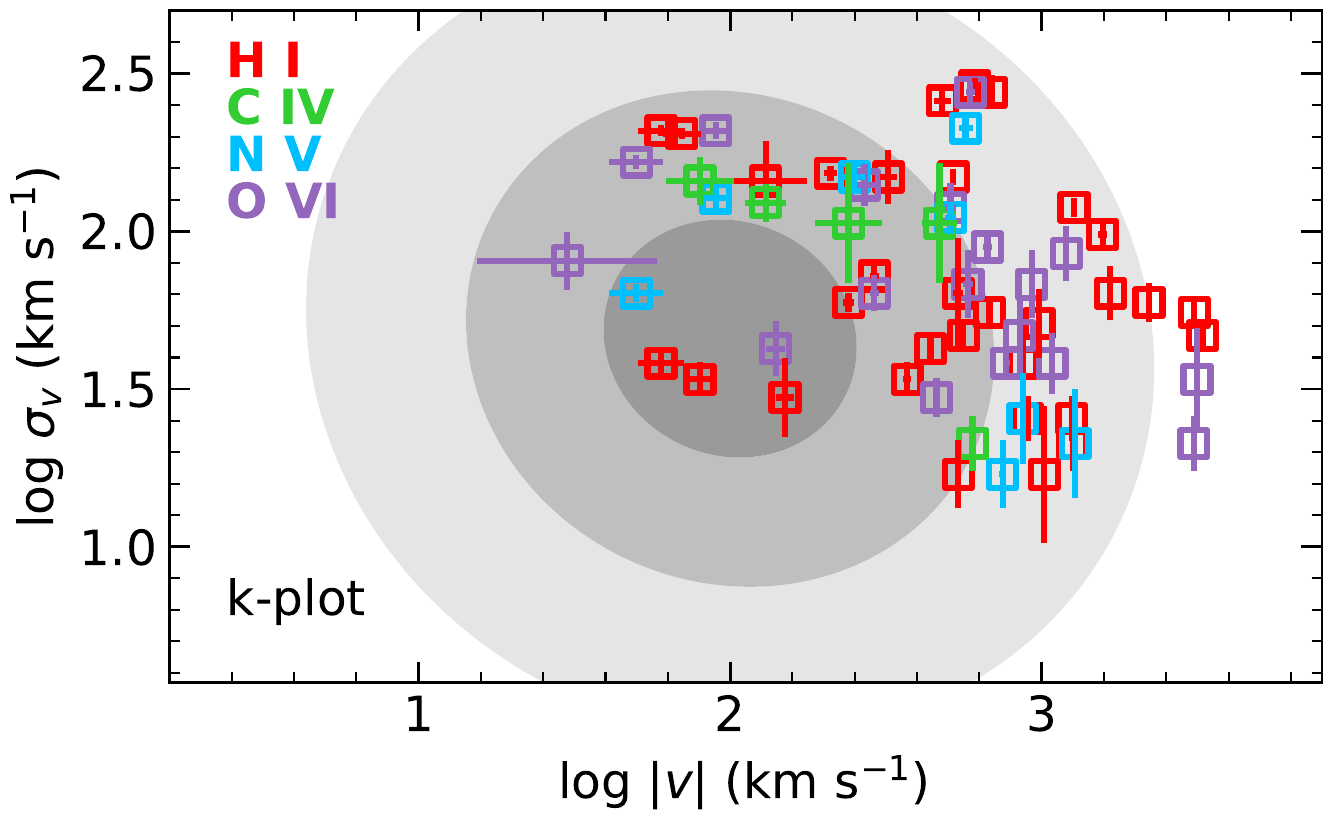}}
\caption{Diagnostic kinematic plot (k-plot) for the intrinsic UV absorbers in the \sub sample, where the log of turbulent velocity $\sigma_v$ is plotted versus the log of the flow velocity $v$. The \sub measurements are over-plotted on the 1-3$\sigma$ confidence contours of the CCA simulations from \citet{Gasp18}.
\label{fig_kplot}}
\end{figure}
%============================

The asymmetry of emission lines can provide useful diagnostic information, serving as an indicator of disk wind activity in AGN \citep{Coat16} and even in X-ray binaries \citep{Mata22}. Broad and blueshifted absorption lines, such as the UV counterparts of obscuring disk winds, can modify the shape of the emission lines; see, for instance, the case of NGC~5548 \citep{Kaas14,Kris19b,Mehd22c}. Such absorption would produce discernible asymmetry in the observed line profiles. As shown in Table \ref{table_asym} and the histograms of Fig. \ref{fig_hist}, the \civ line profiles in the \sub sample do not deviate substantially from being symmetrical. Also, the frequency distributions of the asymmetry parameters are symmetrical (Fig. \ref{fig_hist}) and peak at ${a_f = 1}$ and ${a_p = 0.0}$ (i.e., peaking at fully symmetrical values). Figure \ref{fig_hist} also shows that there are no links between the asymmetry parameters and the presence of narrow UV outflows in the \sub sample. This lack of significant asymmetry in the \sub sample suggests that broad and blueshifted absorption lines do not modify the \civ profile. This is consistent with our spectroscopic search for intrinsic narrow and broad lines in the COS spectra. We do not find BALs in the 21 targets of the \sub HST sample, which is consistent with the ${\sim 5}$\% finding of \citet{Cull19}.

%============================
% FIG: all scatter plots
%
\begin{figure*}
\centering
\resizebox{0.98\hsize}{!}{
\includegraphics[angle=0]{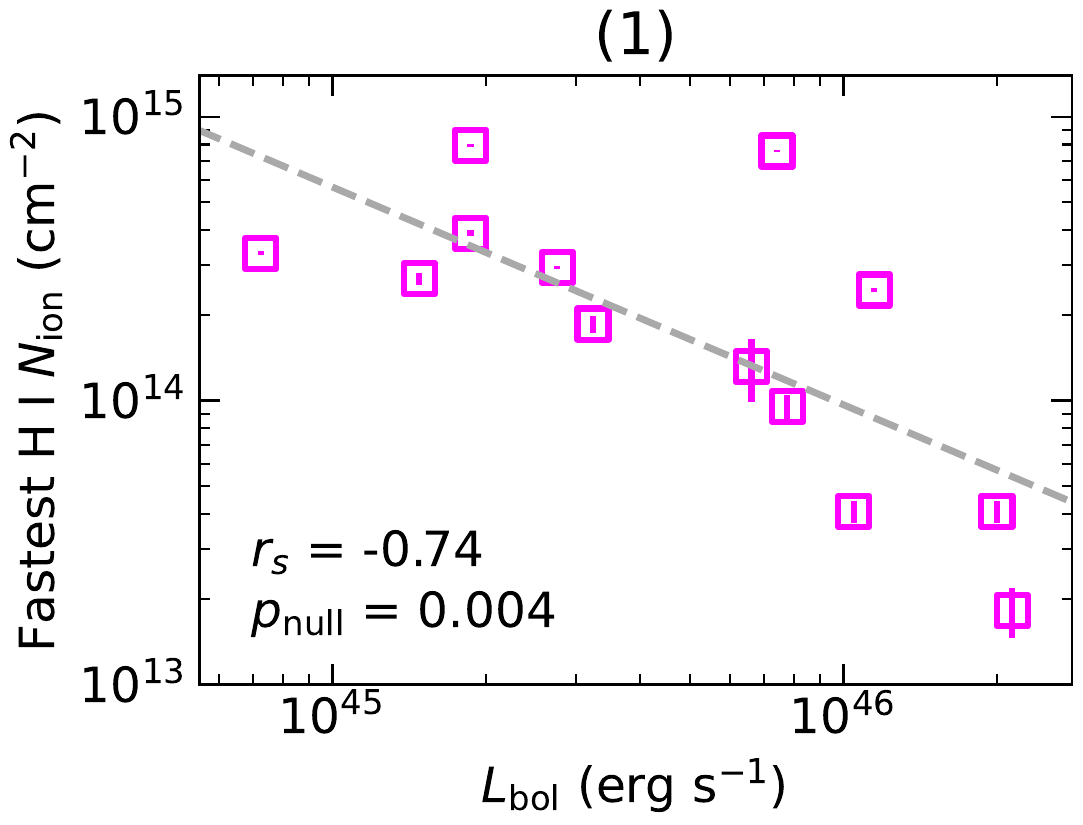}
\includegraphics[angle=0]{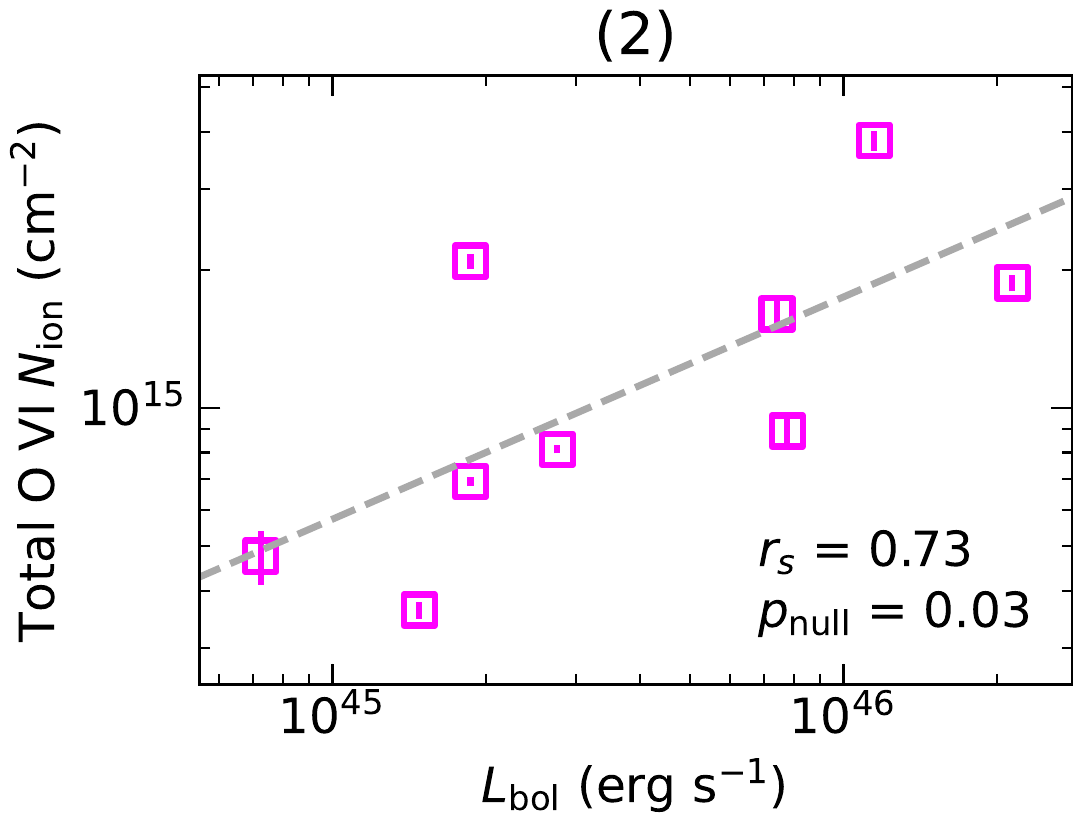}
\includegraphics[angle=0]{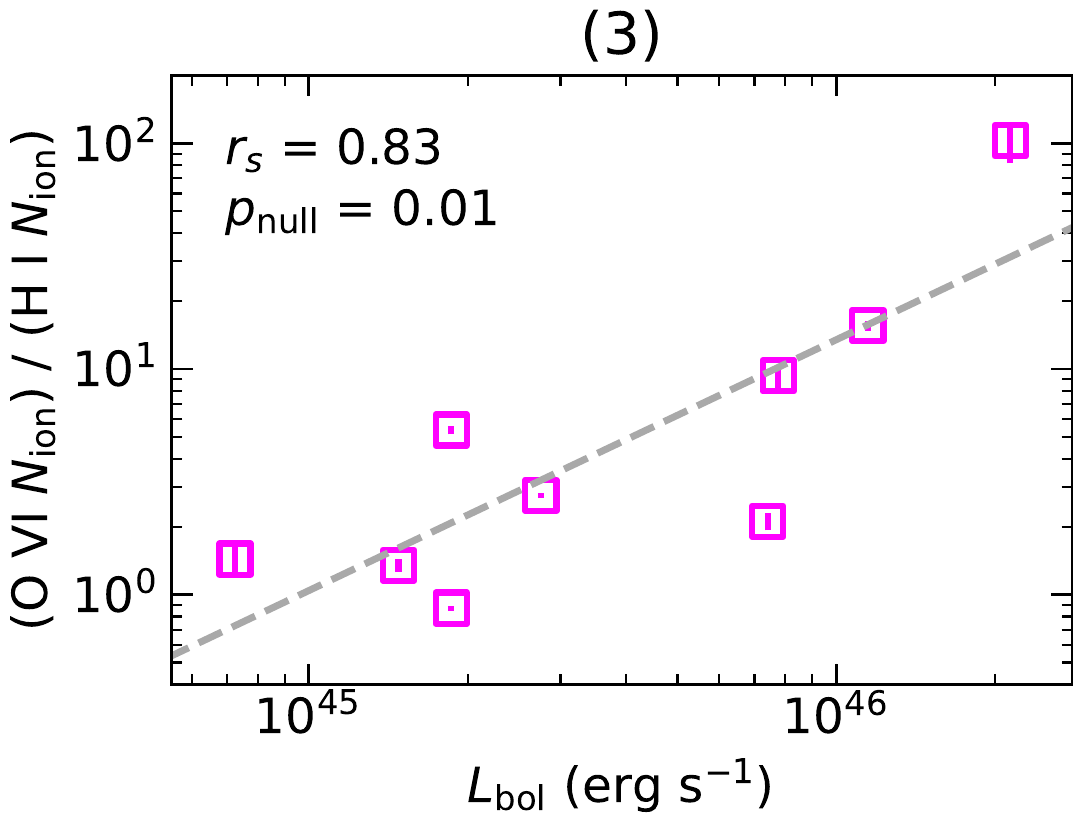}
}
\resizebox{0.98\hsize}{!}{
\includegraphics[angle=0]{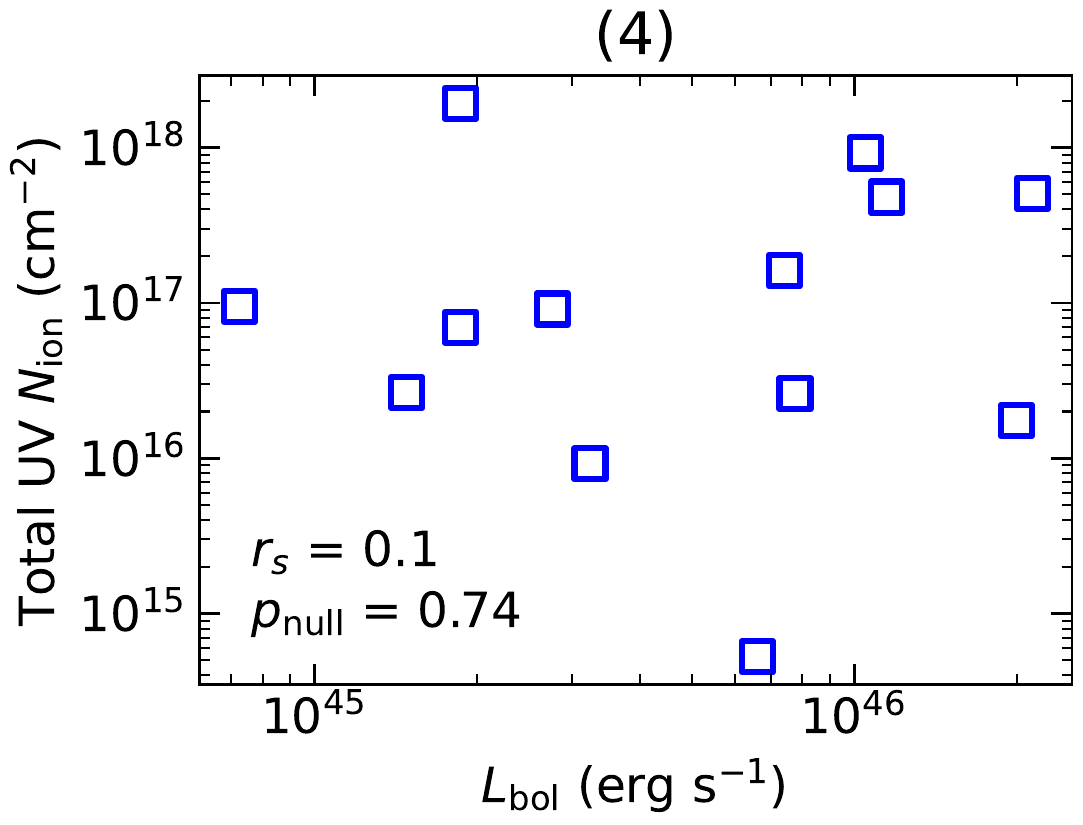}
\includegraphics[angle=0]{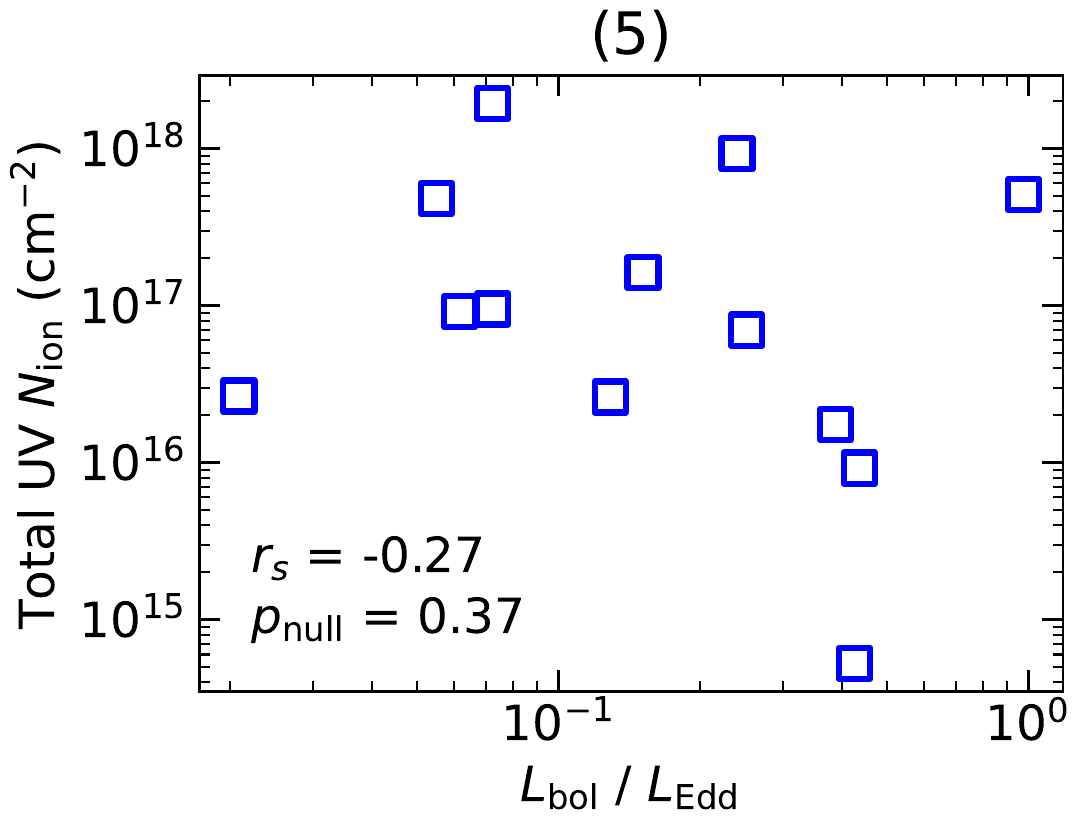}
\includegraphics[angle=0]{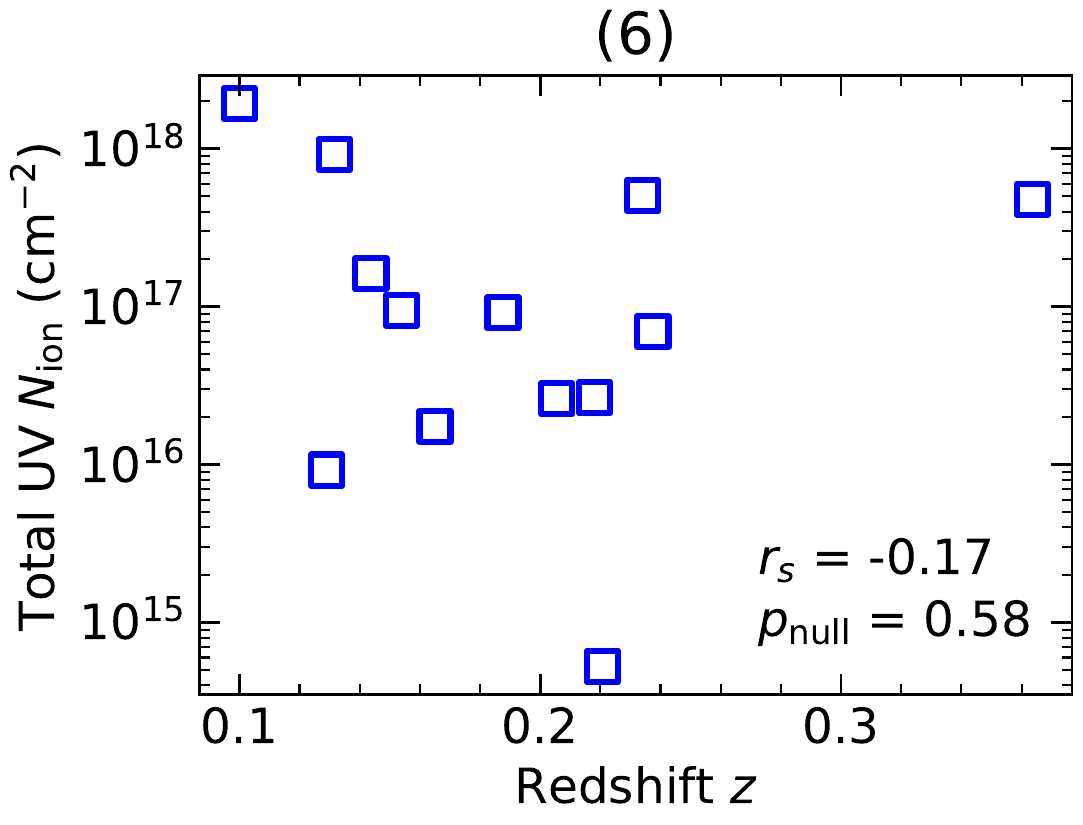}
}
\resizebox{0.98\hsize}{!}{
\includegraphics[angle=0]{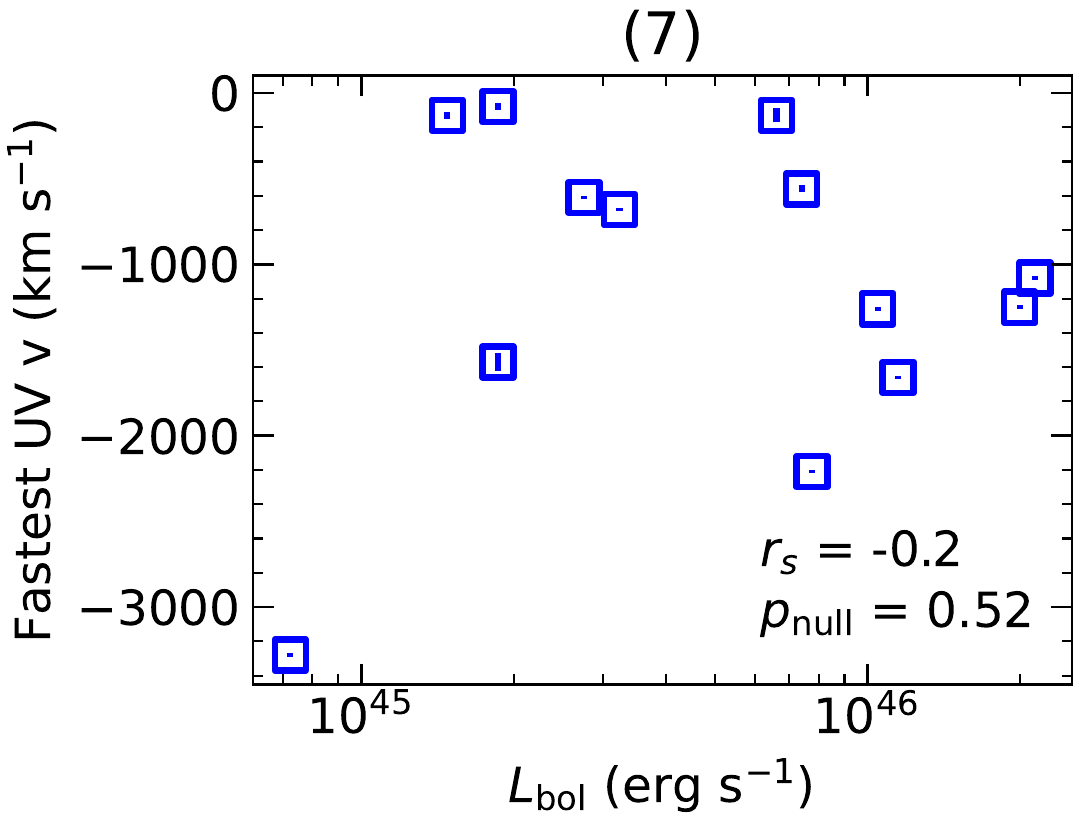}
\includegraphics[angle=0]{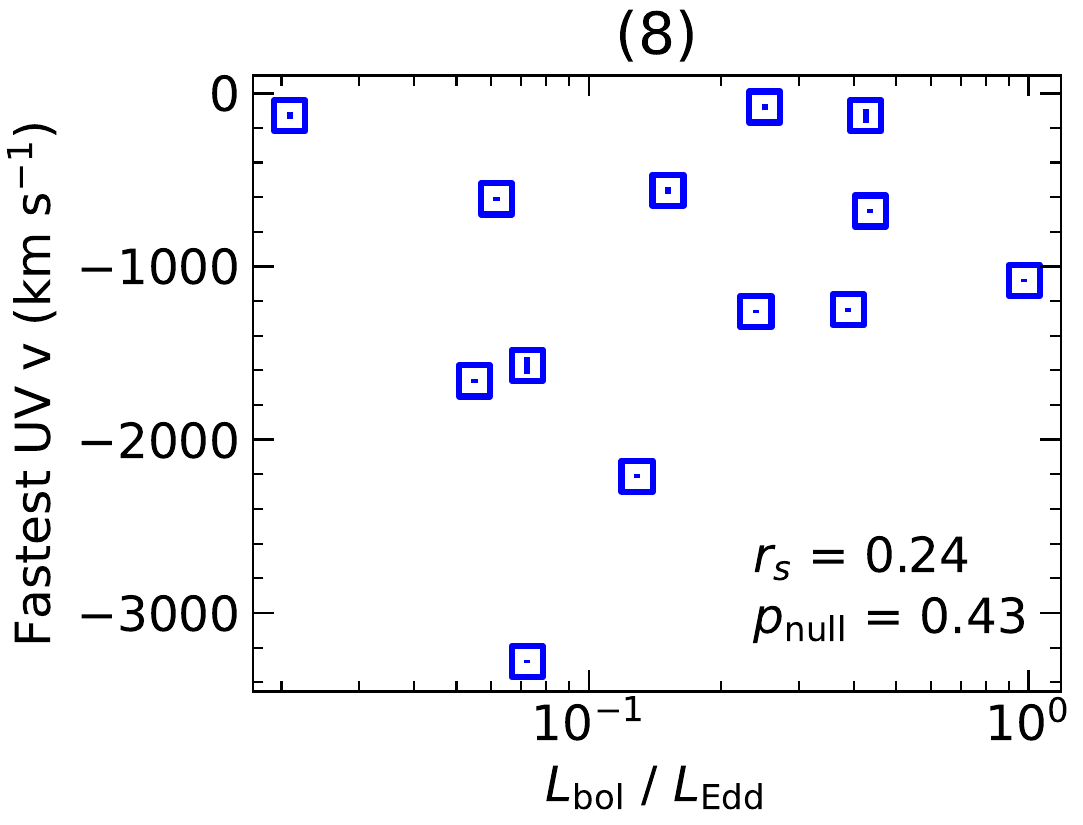}
\includegraphics[angle=0]{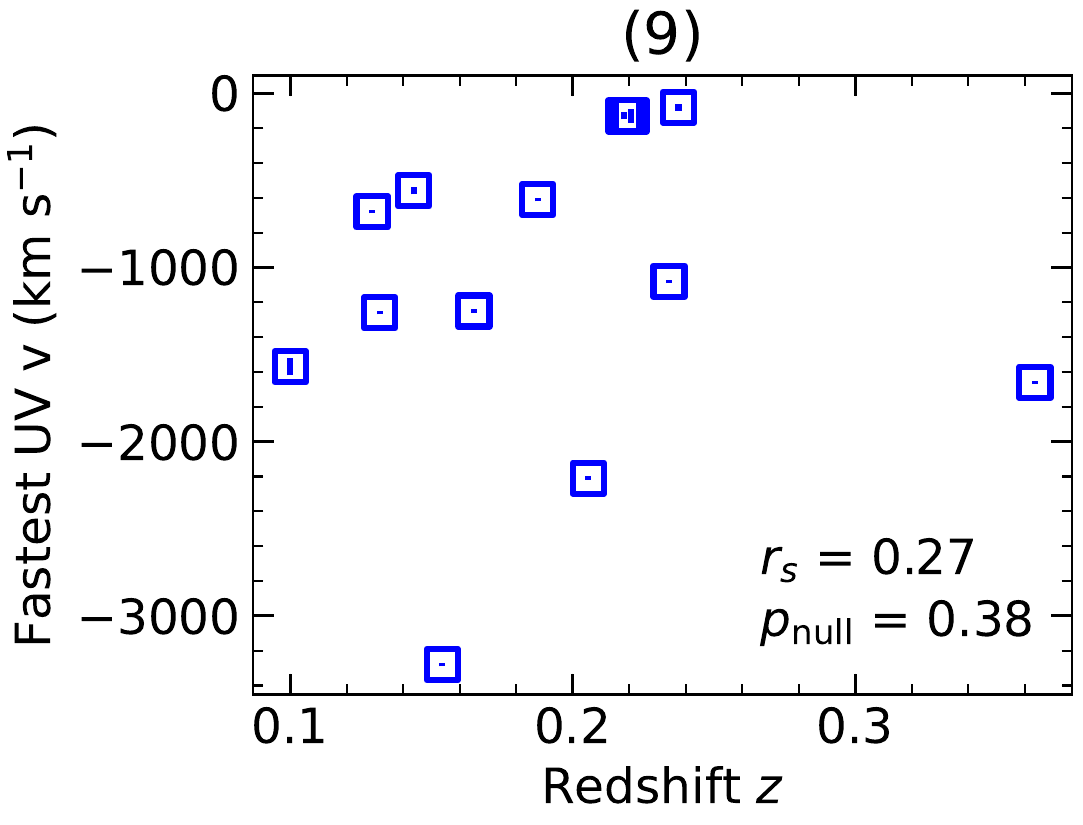}
}
\caption{Relations between parameters of the UV outflows and other AGN parameters in the \sub HST sample. In the corner of each panel, the Spearman's rank correlation coefficient, $r_s$, and the null hypothesis probability, $p_{\rm null}$, are given. In panels with statistically significant correlations (${p_{\rm null} < 0.05}$) the data points are displayed in magenta and in those with no significant correlation in blue. The UV measurement uncertainties are displayed in this figure, however, they are significantly smaller than the explored scales.
\label{fig_rel}}
%\vspace{0.6cm}
\end{figure*}
%============================

In our analysis of the HST spectra of the \sub sample, we searched for UV absorption counterparts to the X-ray UFOs (Fig. \ref{fig_ufo}). Importantly, any unidentified line candidate must be verified that it is not an IGM line. Any intervening IGM \lya line can only appear at redshifted wavelengths $> 1215.7$~\AA. Therefore, in the case of \iras \citep{Mehd22b}, due to its relatively low redshift (${z = 0.06040}$, \citealt{deGr92}), its UFO \lya feature at 1192 \AA\ could not possibly be an IGM line. However, in the case of the \sub sample, the targets are at higher redshifts (Table \ref{table_log}), and so IGM lines can potentially be observed over a wider region of the HST spectrum. So both a UFO and the IGM can produce lines at wavelengths $> 1216$~\AA. In the case of some \sub targets with an X-ray UFO, we detected lines in the regions where \lya counterparts of the X-ray UFOs are expected (see Fig. \ref{fig_ufo}). However, these lines are identified as either the warm absorber or the ISM lines and the remaining lines cannot be ruled out as IGM lines. Because there are no accompanying ionized UFO lines (either \civ, \nv, or \ovi), and the unidentified lines are consistent with characteristics of IGM lines (including verification in some cases in the literature, see Sect. \ref{sect_iden}), we conclude that these unidentified lines are most likely IGM lines. 

The lack of detections for UV counterparts to the X-ray UFOs is most likely because X-ray UFOs that produce \ion{Fe}{xxv} and \ion{Fe}{xxvi} absorption (like in the \sub sample) are too highly ionized to produce significant absorption lines in the UV band (ionization parameter ${\log \xi \gtrsim 3.7}$, \citealt{Kris18b}) or due to variability between the non-contemporaneous X-ray and UV observations (see \citealt{Kris18b}). Although in the case of LBQS~1338-0038, the HST and \xmm observations are relatively close to each other (four days apart). Furthermore, if the X-ray UFO is near the accretion disk, it may not cover enough of the UV source, which is larger than the compact X-ray source and this means that it may not produce detectable UV absorption. For more details, we refer to the recent behavior of the disk wind in NGC~5548 \citep{Mehd22c}. Interestingly, all targets with an X-ray UFO (\paperI) that have HST data (five targets) demonstrate the presence of UV ionized outflows (Fig. \ref{fig_bar}). While five targets is a limited number, this is nonetheless in agreement with the suggestion that the warm absorbers and UFOs are ultimately physically connected to each other, as shown by recent theoretical and numerical AGN feedback entrainment models (e.g., \citealt{Gasp17}). However, this does not imply that warm absorbers are exclusively formed by UFOs, as there can be additional origins and mechanisms that give rise to the warm absorbers, as discussed in the following section. 

%%%%%%%%%%%%%%%%%%%%%%%%%%%%%%%%%%%%%%%%%%%%%%%%%%%%%%%%%%%%%%%%%%%%%%%%%%%%%%%%%%%%%%%%%%%%%%%%%%%%%%%
\subsection{Relations between UV ionized outflows and other properties of the AGN}
\label{sect_rel}

Understanding the scaling relations of AGN winds and finding correlations between winds and AGN properties, can provide useful insights into the driving mechanism and impact of AGN winds (see e.g., \citealt{Fior17}). Interestingly, both galaxy-scale winds and UFOs have been found to show relations between their outflow velocity and the bolometric luminosity of the AGN, with $L_{\rm bol} \propto v^{5}$ \citep{Fior17}. The far-UV spectroscopy of type-1 AGN with the Far Ultraviolet Spectroscopic Explorer (FUSE, \citealt{Moos00}) also found a tendency in the maximum velocity of the intrinsic UV absorbers increasing with the source luminosity, while spanning four orders of magnitude in luminosity \citep{Kris06}. For the \sub sample, we investigated such relations using the results of our spectral modeling with HST. In Fig. \ref{fig_uv_rel}, we examine the relations between the parameters of the UV outflows, whereas in Fig. \ref{fig_rel}, we check for relations between the UV outflows and other AGN properties. The Spearman’s rank correlation coefficient ($r_s$) and the corresponding null hypothesis p-value ($p_{\rm null}$) are given in the insets of panels in Figs. \ref{fig_uv_rel} and \ref{fig_rel}. We note that apart from the panels and relations shown in Fig. \ref{fig_rel}, we also checked other combinations of parameters and their relations were found not to be statistically significant.

For the \sub HST sample, we do not find a significant correlation between the maximum UV outflow velocity and the bolometric luminosity of the AGN (Fig. \ref{fig_rel}, panel 7). One possibility is that our sample surveys an overly limited a span of luminosity ($10^{45}$--$10^{46}$ \ergs) for the relation between $v$ and $L_{\rm bol}$ to become apparent. However, another explanation is that the UV warm absorbers, with their moderate outflow velocities, do not directly follow the velocity-luminosity relations that are primarily seen for the more powerful UFOs and galactic-scale winds \citep{Fior17}. Also, in \citet{Fior17}, we can see that the warm absorber data are those exhibiting the least significant correlation and, on their own, they do not display the $L_{\rm bol} \propto v^{5}$ relation, as in our findings. The lack of significant $v$-\lbol correlation for the warm absorbers in the \sub sample may be because of additional origins (e.g., from the torus, \citealt{Krol01}) as well as additional formation processes (e.g., cloud condensation in CCA, \citealt{Gasp17b}) that may characterize warm absorbers, in contrast to the more powerful disk winds by comparison.

In Fig. \ref{fig_kplot}, the k-plot for the intrinsic UV absorbers in the \sub sample is compared with the CCA simulations of \citet{Gasp18}. The data points that fall within the 1-2$\sigma$ confidence CCA contours (i.e., the two inner gray zones) are likely associated to the CCA condensation phase. Those data points that fall outside the 1-2$\sigma$ confidence counters (right-hand region of the plot) are likely to be associated with the AGN ejection phase. In the CCA scenario, the relatively slow outflows (within the 1-2$\sigma$) contours can still be condensing \citep{Gasp18}. The results of Fig. \ref{fig_kplot} show that the UV absorbers in the \sub sample are potentially consistent with the CCA scenario, suggesting that both CCA condensation/feeding and ejection/feedback are taking place in these AGN. The lack of any significant correlation between $v$ and $\sigma_v$ (Fig. \ref{fig_uv_rel}, bottom panel) is a signature of such superposing processes (i.e., condensation and feedback). 

Interestingly, we find significant statistical correlations ($p_{\rm null} < 0.05$) between the column density ($N$) of the UV ions and $L_{\rm bol}$  values of the AGN in the \sub sample. Our results show that the $N_{\rm H\,I}$ value of the fastest outflow component decreases with $L_{\rm bol}$ (Fig. \ref{fig_rel}, panel 1), while the total column density of \ovi increases with $L_{\rm bol}$ (Fig. \ref{fig_rel}, panel 2). The ratio of these \ion{O}{vi} to \ion{H}{i} column densities is significantly correlated with $L_{\rm bol}$ (Fig. \ref{fig_rel}, panel 3). In a recent \chandra study of the ionization distribution in nine Seyfert-1 galaxies, \citet{Kesh22} found that log of the total column density of the X-ray outflow is anti-correlated with the log of the X-ray luminosity. This is similar behavior to the $N_{\rm H\,I}$--\lbol relation we found for the \sub QSO sample. Our $N_{\rm ion}$--\lbol relations are likely to be a manifestation of the photoionization process in AGN, where, toward higher source luminosities, the wind becomes more ionized, resulting in weaker absorption by the neutral and low-ionization ions (such as \ion{H}{i}) and stronger absorption by high-ionization ions (such as \ovi). This trend of increasing column density with AGN luminosity that we see for \ovi may also be attributed to the CCA scenario (e.g., \citealt{Gasp13b,Gasp17b}), in which stronger cloud condensation and raining onto the SMBH (also boosting AGN luminosity or power) results in higher column densities along the line of sight \citep{Gasp17b}.

Finally, we find a significant correlation between the outflow velocity and the column density of the UV absorbing ions for the \sub HST sample (Fig. \ref{fig_uv_rel}). As the outflow velocity of the UV ions increases, their column density decreases. This correlation also holds statistical significance for both neutral and ionized ions. The data points with positive $v$ (i.e., inflow) in Fig. \ref{fig_uv_rel} belong to only one component of one object (Comp. 1 of PG 0804+761) and their inclusion does not significantly alter the measured correlations. The observed relation between the outflow velocity and the column density can be explained by a mechanical power that evacuates the UV-absorbing medium \citep{Gasp11,Sado17}. Overall, this $N$--$v$ relation, and the $N$--$L_{\rm bol}$ relation that we found, are consistent with AGN feedback simulations that show a composition of radiative and mechanical processes, such as  general relativistic, radiative magnetohydrodynamic simulations \citep{Gasp17}. In a follow-up \sub paper, our team plans to combine the X-ray (\paperI) and UV (this paper) outflow results of the \sub sample with those of AGN at other luminosities and redshifts, with an aim to further study the observed relations and investigate the AGN feedback models.  

%%%%%%%%%%%%%%%%%%%%%%%%%%%%%%%%%%%%%%%%%%%%%%%%%%%%%%%%%%%%%%%%%%%%%%%%%%%%%%%%%%%%%%%%%%%%%%%%%%%%%%%
%%%%%%%%%%%%%%%%%%%%%%%%%%%%%%%%%%%%%%%%%%%%%%%%%%%%%%%%%%%%%%%%%%%%%%%%%%%%%%%%%%%%%%%%%%%%%%%%%%%%%%%
%%%%%%%%%%%%%%%%%%%%%%%%%%%%%%%%%%%%%%%%%%%%%%%%%%%%%%%%%%%%%%%%%%%%%%%%%%%%%%%%%%%%%%%%%%%%%%%%%%%%%%%

\section{Conclusions} 
\label{sect_concl}

In this UV spectroscopic study with HST, we determined the parameters of ionized outflows in a sample of 21 QSOs at intermediate redshifts (${0.1 < z <0.4}$) and bolometric luminosities (${10^{45} < L_{\rm bol} < 10^{46}}$~\ergs). The spectroscopic characteristics and parameters of the ionized outflows in the \sub sample are found to be comparable to those seen in the less luminous Seyfert-1 galaxies in the local universe. We find 60\% of our targets show the presence of intrinsic outflowing \ion{H}{i} absorption, while 40\% exhibit ionized outflows seen as absorption by either \ion{C}{iv}, \ion{N}{v}, or \ion{O}{vi}. In two-thirds of these, the UV outflow exhibits multiple velocity components. All the absorption lines in the \sub sample are relatively narrow ({40 \kms < FWHM < 680}~\kms), with outflow velocities ranging up to $-3300$~\kms. In targets with an X-ray UFO that have HST data, we find no significant UV absorption counterpart; this is most likely due to the UFO gas being too highly ionized to produce UV absorption and/or due to its low covering fraction of the UV source. However, all \sub targets with an X-ray UFO show presence of UV outflows at lower velocities, which is in agreement with a physical connection between these two types of outflows, as suggested by current theoretical and hydrodynamical simulations of self-regulated AGN feeding and feedback (see the review by \citealt{Gasp20}). 

Broad and blueshifted UV absorption lines, such as those associated with BAL quasars or the UV counterparts of X-ray obscuring disk winds, are not detected in the \sub HST sample. Moreover, our analysis of the asymmetry parameters of the \ion{C}{iv} emission lines in the \sub sample shows that they are relatively symmetrical. Our investigation of the UV spectral signatures of different types of AGN outflows concludes that all the UV ionized outflows in the \sub sample are consistent with warm-absorber outflows and potentially CCA condensation, as in the case of those typically seen in Seyfert-1 galaxies. 

The assessment of the results of our HST spectral modeling reveals interesting relations between the UV ionized outflows and other properties of the AGN. We find that the column density \ion{H}{i} of the fastest UV absorber component decreases with the bolometric luminosity of the AGN, while the total \ion{O}{vi} column density increases with the bolometric luminosity. This is likely a manifestation of the photoionization process in AGN, where, toward higher AGN luminosities, the wind becomes more ionized, resulting in weaker UV absorption by neutral or low-ionization ions (\ion{H}{i}) and stronger absorption by high-ionization ions (\ion{O}{vi}). In addition, we find that as the outflow velocity of the UV ions increases, their column density decreases. This may be interpreted as a consequence of a mechanically-powered wind that evacuates the UV-absorbing medium. Overall, the observed relations we find for the \sub HST sample are consistent with both radiative and mechanical outflow mechanisms, which theoretical simulations have shown act together in AGN feedback.

%%%%%%%%%%%%%%%%%%%%%%%%%%%%%%%%%%%%%%%%%%%%%%%%%%%%%%%%%%%%%%%%%%%%%%%%%%%%%%%%%%%%%%%%%%%%%%%%%%%%%%%
%%%%%%%%%%%%%%%%%%%%%%%%%%%%%%%%%%%%%%%%%%%%%%%%%%%%%%%%%%%%%%%%%%%%%%%%%%%%%%%%%%%%%%%%%%%%%%%%%%%%%%%
%%%%%%%%%%%%%%%%%%%%%%%%%%%%%%%%%%%%%%%%%%%%%%%%%%%%%%%%%%%%%%%%%%%%%%%%%%%%%%%%%%%%%%%%%%%%%%%%%%%%%%%
\begin{acknowledgements}
This work was supported by NASA through grants for HST program numbers 15890 and 15673 from the Space Telescope Science Institute, which is operated by the Association of Universities for Research in Astronomy, Incorporated, under NASA contract NAS5-26555. This research has made use of the NASA/IPAC Extragalactic Database (NED), which is funded by the National Aeronautics and Space Administration and operated by the California Institute of Technology. M. B. is supported by the European Union’s Horizon 2020 research and innovation programme Marie Skłodowska-Curie grant No 860744 (BID4BEST). G. M. and all the Italian co-authors acknowledge support and fundings from Accordo Attuativo ASI-INAF n. 2017-14-H.0. E. B. is supported by a Center of Excellence of the Israeli Science Foundation (grant No. 2752/19). B. D. M. acknowledges support from a Ramón y Cajal Fellowship (RYC2018-025950-I) and the Spanish MINECO grant PID2020-117252GB-I00. M. Gaspari acknowledges partial support by HST GO-15890.020/023-A, the {\it BlackHoleWeather} program, and NASA HEC Pleiades (SMD-1726). M. Giustini is supported by the ``Programa de Atracci\'on de Talento'' of the Comunidad de Madrid, grant number 2018-T1/TIC-11733. A. L. L. acknowledges support from CONACyT grant CB-2016-01-286316. R. M. acknowledges financial support from the ASI-INAF agreement n. 2022-14-HH.0. M. P. is supported by the Programa Atracci\'on de Talento de la Comunidad de Madrid via grant 2018-T2/TIC-11715, and acknowledges support from the Spanish Ministerio de Econom\'ia y Competitividad through the grant ESP2017-83197-P, and PID2019-106280GB-I00. G. P. acknowledges funding from the European Research Council (ERC) under the European Union's Horizon 2020 research and innovation programme (grant agreement No 865637). P. O. P. acknowledges financial support from the CNES french spatial agency and from the National High Energy Programme (PNHE) of the French CNRS. S. B. acknowledges financial support from the PRIN MIUR project `Black Hole winds and the Baryon Life Cycle of Galaxies: the stone-guest at the galaxy evolution supper', contract \#2017PH3WAT. We thank the anonymous referee for the constructive comments.
\end{acknowledgements}

%%%%%%%%%%%%%%%%%%%%%%%%%%%%%%%%%%%%%%%%%%%%%%%%%%%%%%%%%%%%%%%%%%%%%%%%%%%%%%%%%%%%%%%%%%%%%%%%%%%%%%%
%%%%%%%%%%%%%%%%%%%%%%%%%%%%%%%%%%%%%%%%%%%%%%%%%%%%%%%%%%%%%%%%%%%%%%%%%%%%%%%%%%%%%%%%%%%%%%%%%%%%%%%
%%%%%%%%%%%%%%%%%%%%%%%%%%%%%%%%%%%%%%%%%%%%%%%%%%%%%%%%%%%%%%%%%%%%%%%%%%%%%%%%%%%%%%%%%%%%%%%%%%%%%%%
\newpage
\bibliographystyle{aa}
\bibliography{references}{}
%%%%%%%%%%%%%%%%%%%%%%%%%%%%%%%%%%%%%%%%%%%%%%%%%%%%%%%%%%%%%%%%%%%%%%%%%%%%%%%%%%%%%%%%%%%%%%%%%%%%%%%
%%%%%%%%%%%%%%%%%%%%%%%%%%%%%%%%%%%%%%%%%%%%%%%%%%%%%%%%%%%%%%%%%%%%%%%%%%%%%%%%%%%%%%%%%%%%%%%%%%%%%%%
%%%%%%%%%%%%%%%%%%%%%%%%%%%%%%%%%%%%%%%%%%%%%%%%%%%%%%%%%%%%%%%%%%%%%%%%%%%%%%%%%%%%%%%%%%%%%%%%%%%%%%%
\begin{appendix}

\section{Parameters of the UV emission and absorption lines in the SUBWAYS HST sample}
\label{sect_append}

\begin{figure*}
\centering
\resizebox{0.95\hsize}{!}{
\includegraphics[angle=0]{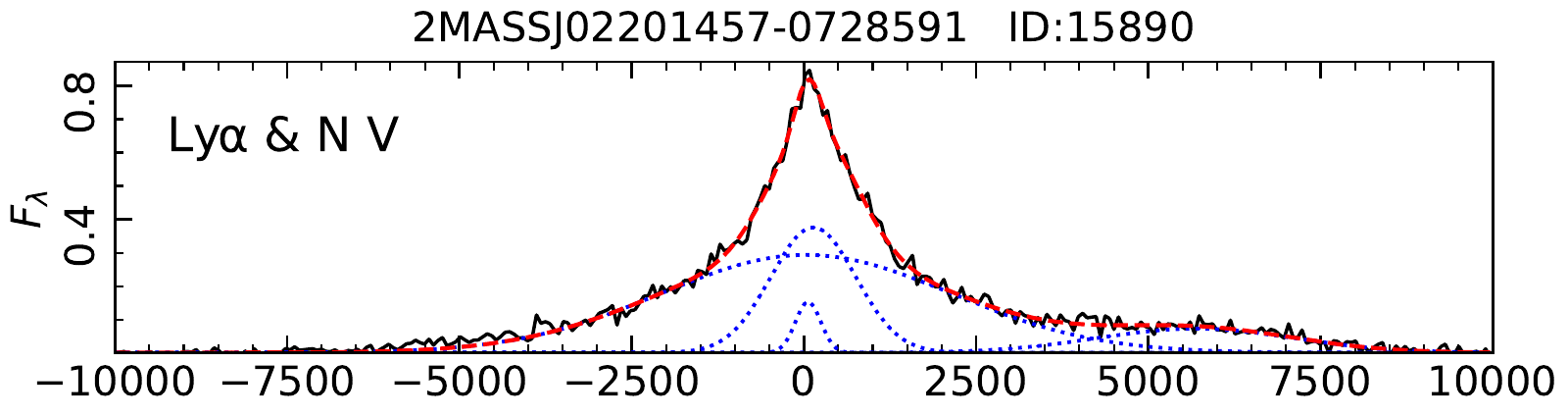}
\includegraphics[angle=0]{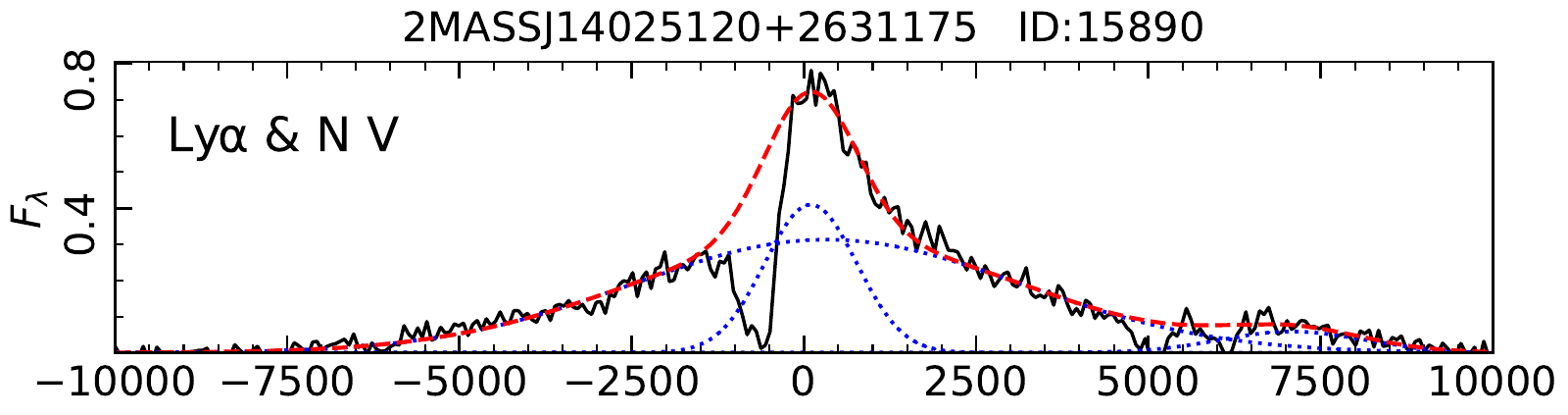}
}
\resizebox{0.95\hsize}{!}{
\includegraphics[angle=0]{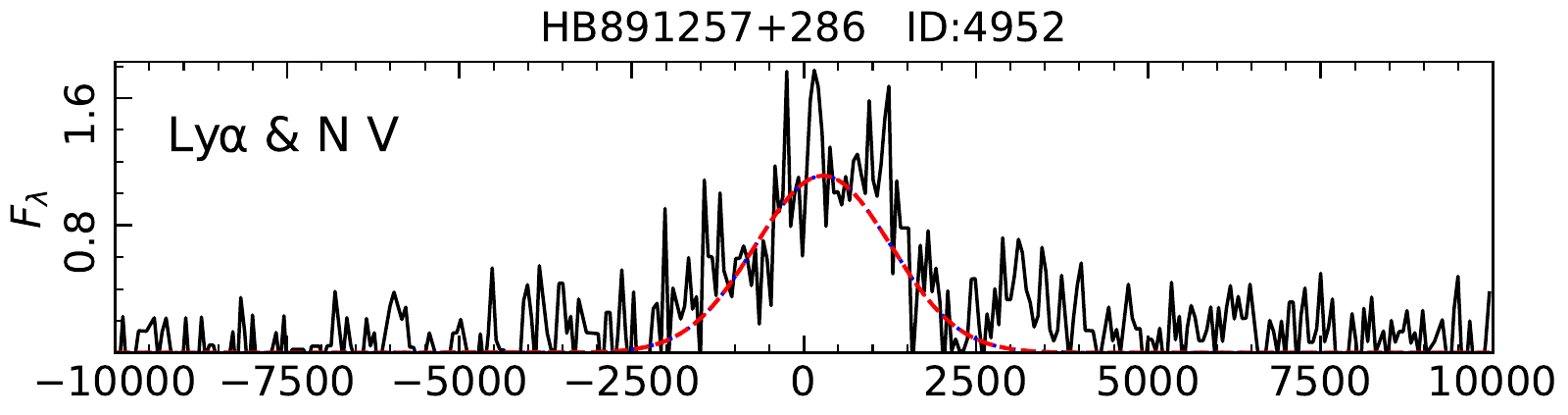}
\includegraphics[angle=0]{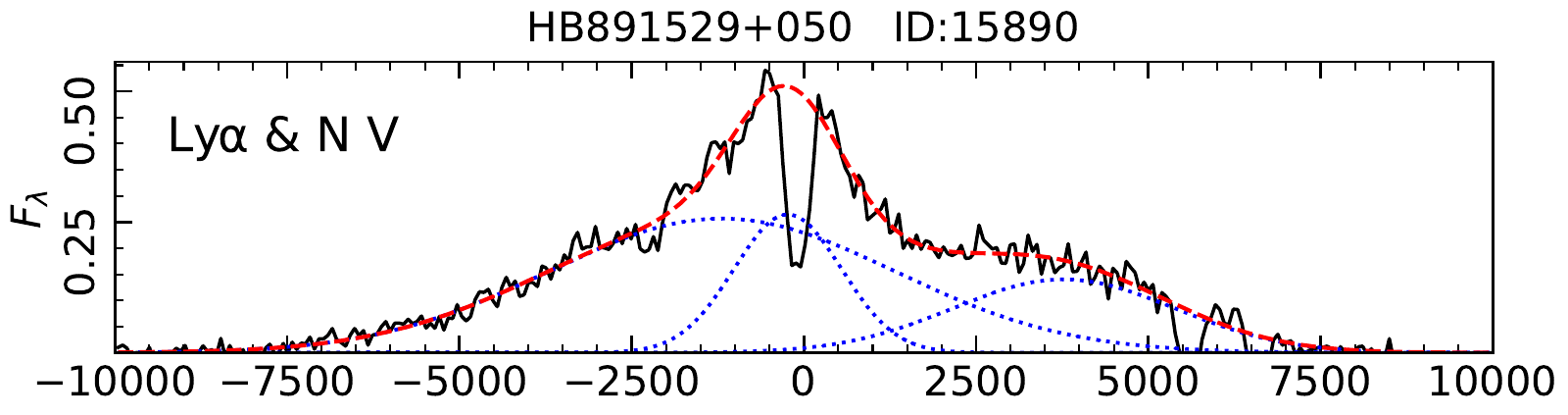}
}
\resizebox{0.95\hsize}{!}{
\includegraphics[angle=0]{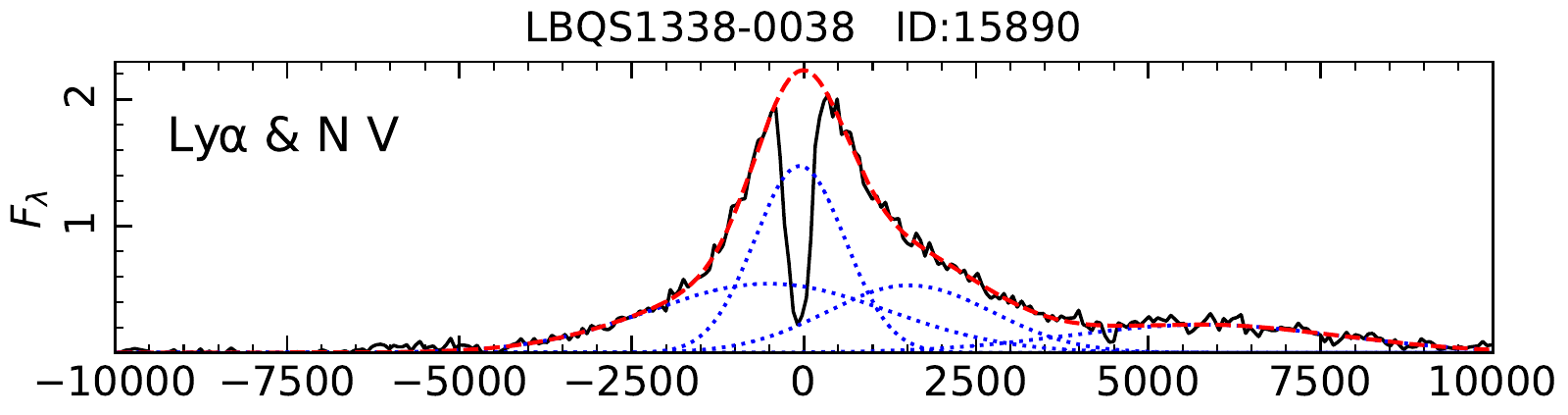}
\includegraphics[angle=0]{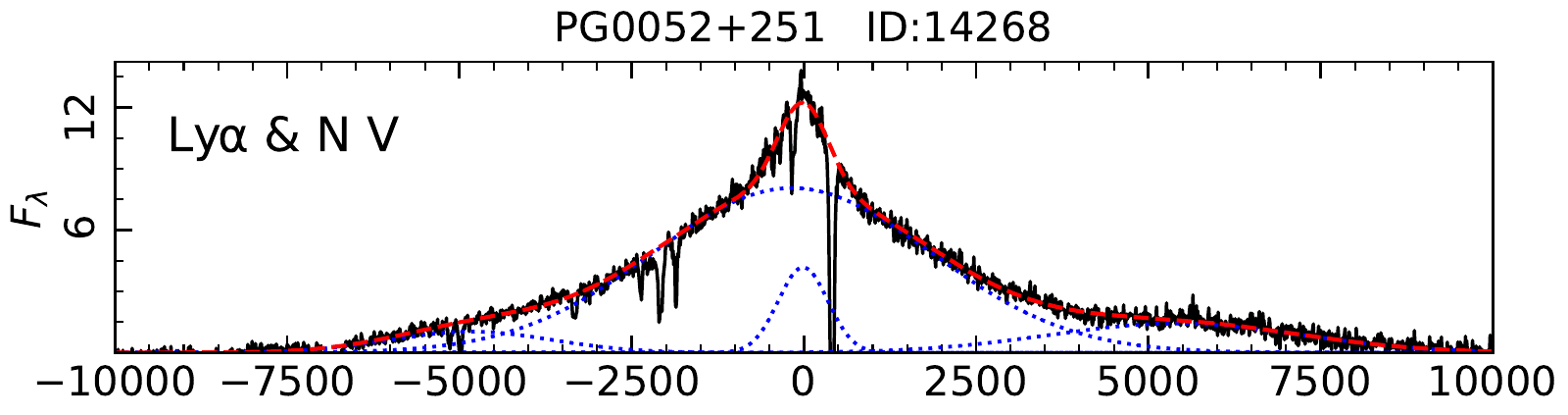}
}
\resizebox{0.95\hsize}{!}{
\includegraphics[angle=0]{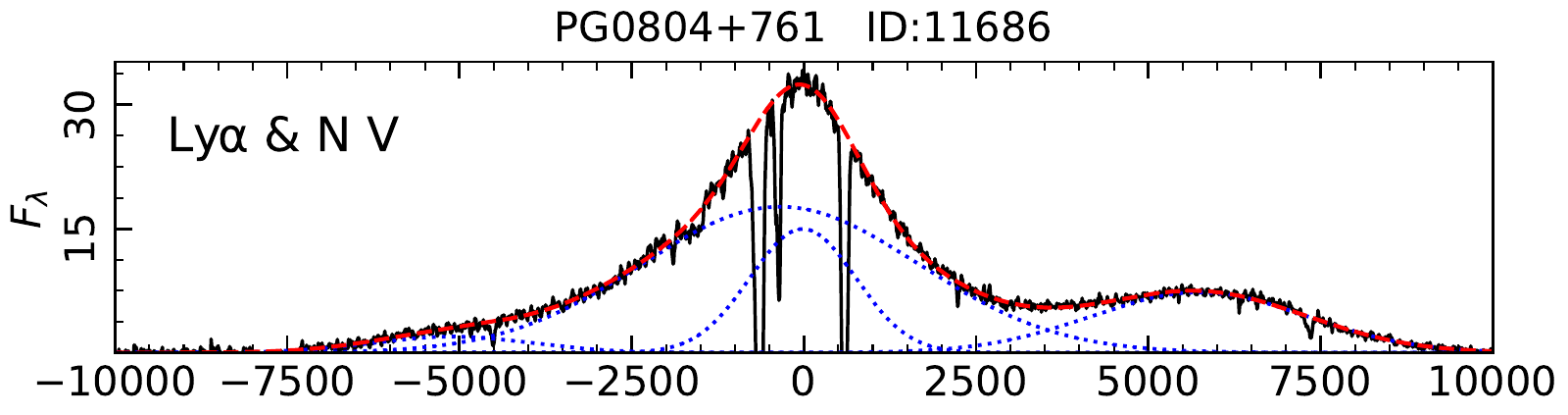}
\includegraphics[angle=0]{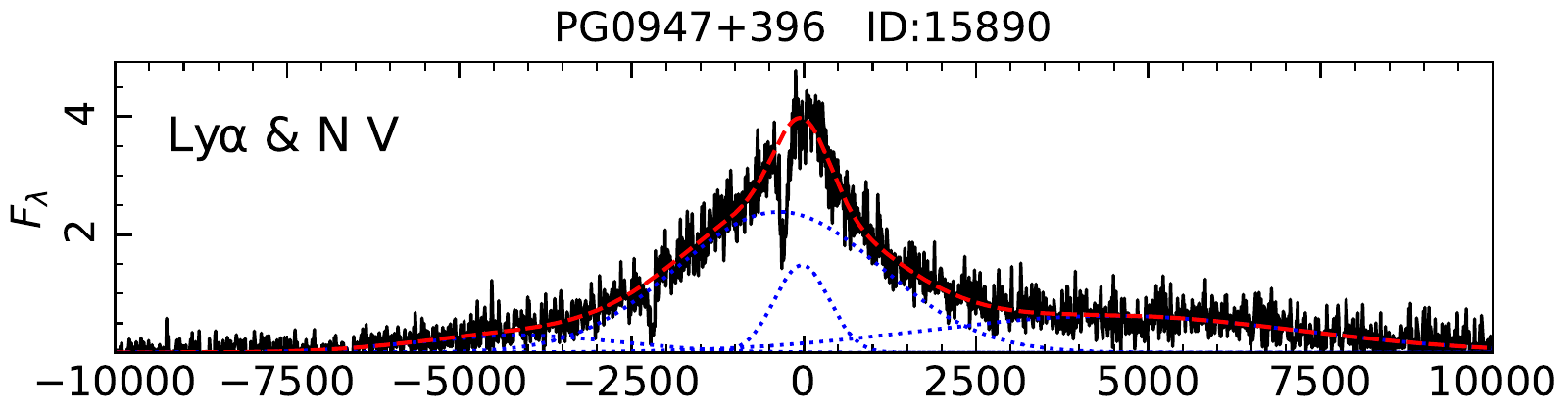}
}
\resizebox{0.95\hsize}{!}{
\includegraphics[angle=0]{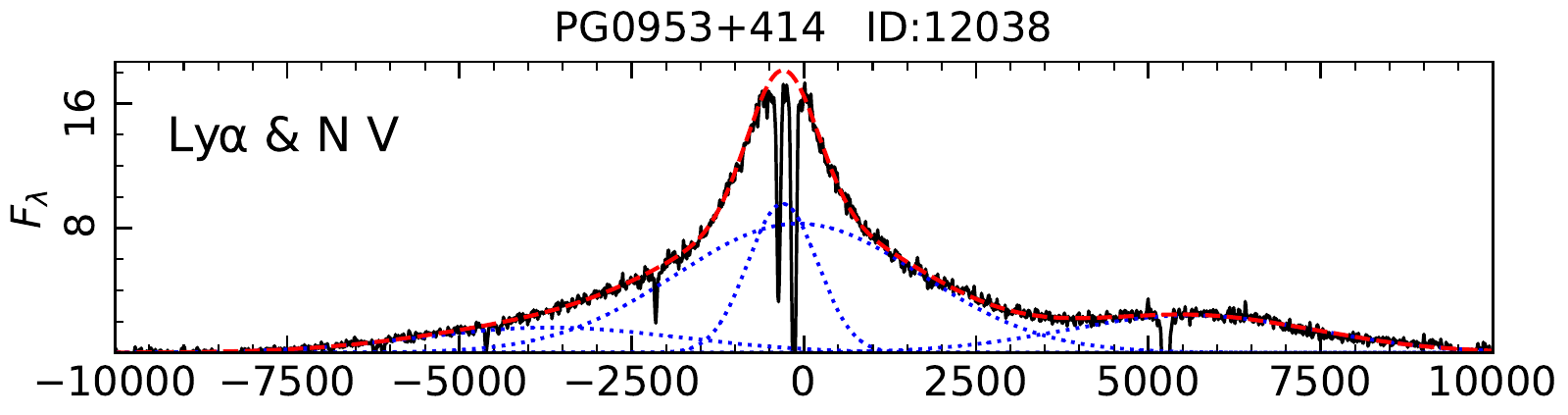}
\includegraphics[angle=0]{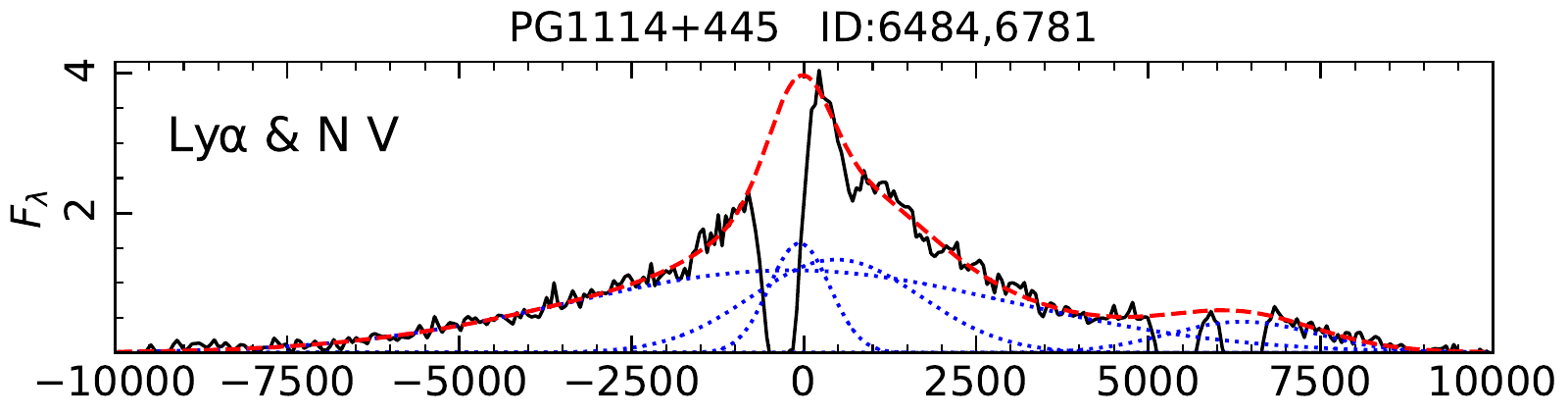}
}
\resizebox{0.95\hsize}{!}{
\includegraphics[angle=0]{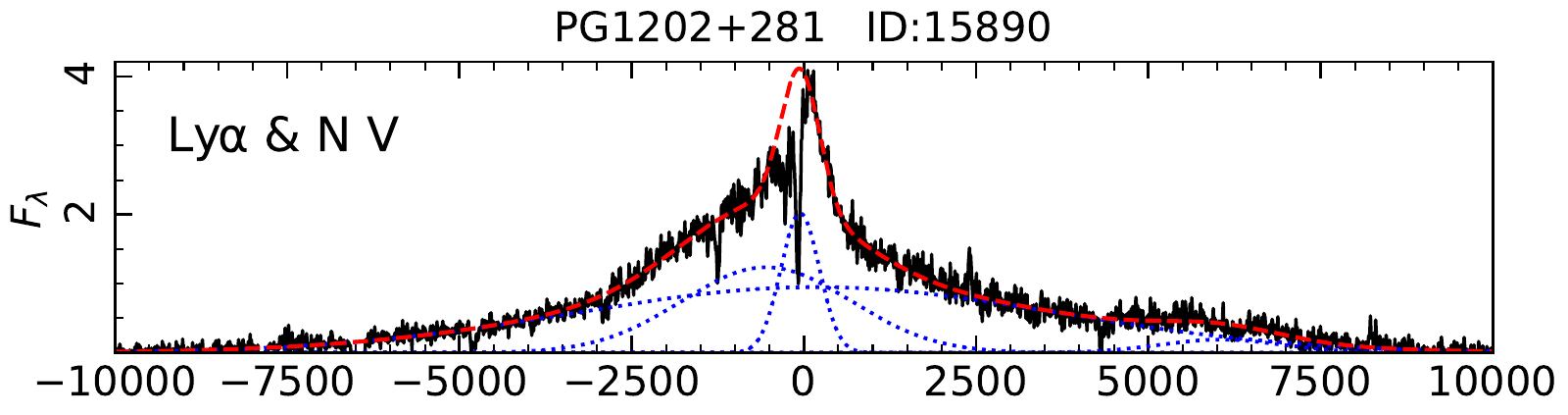}
\includegraphics[angle=0]{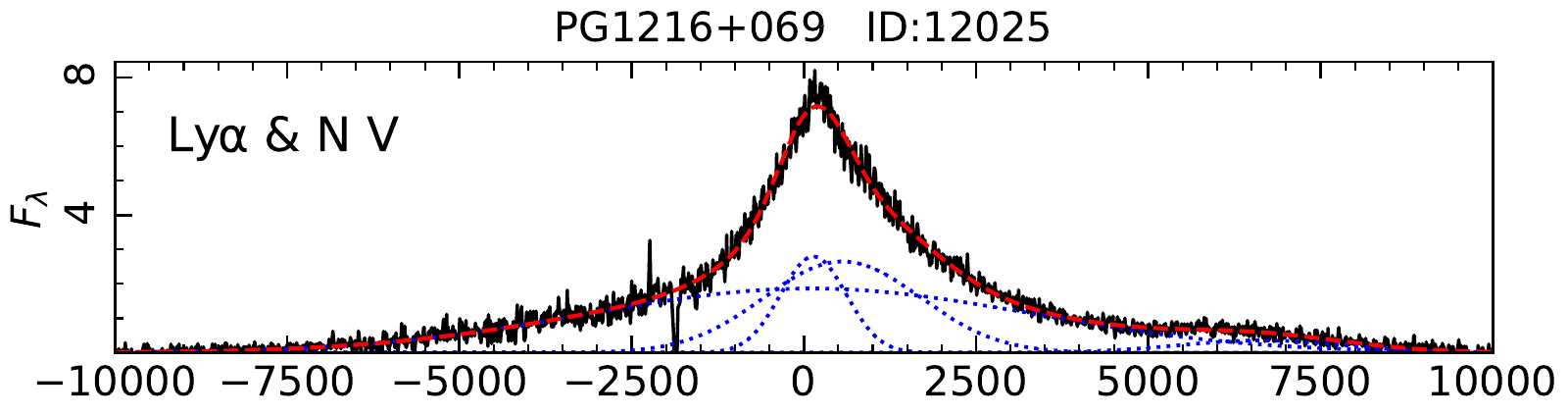}
}
\resizebox{0.95\hsize}{!}{
\includegraphics[angle=0]{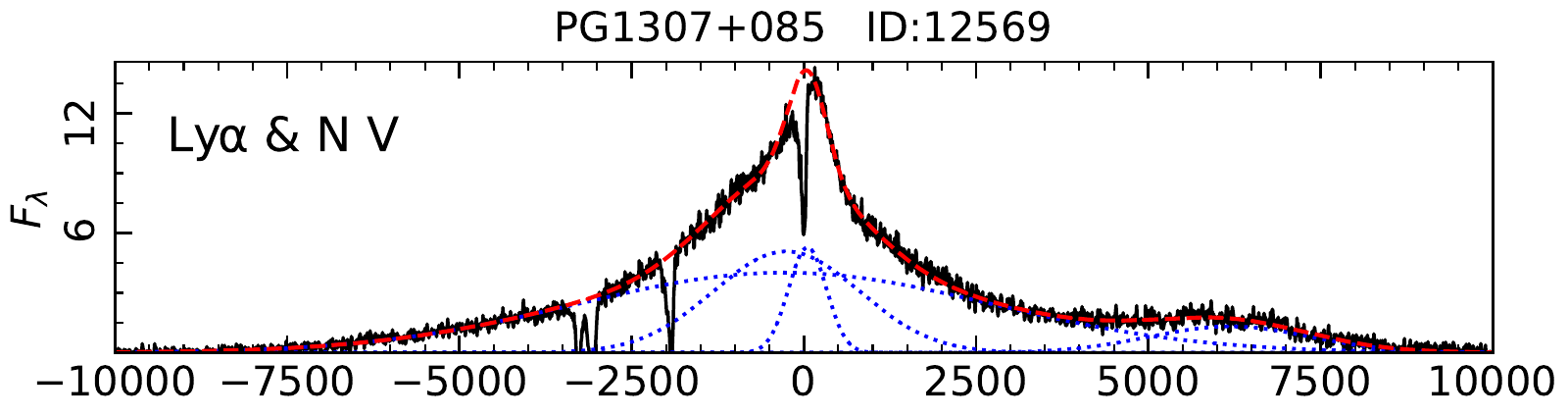}
\includegraphics[angle=0]{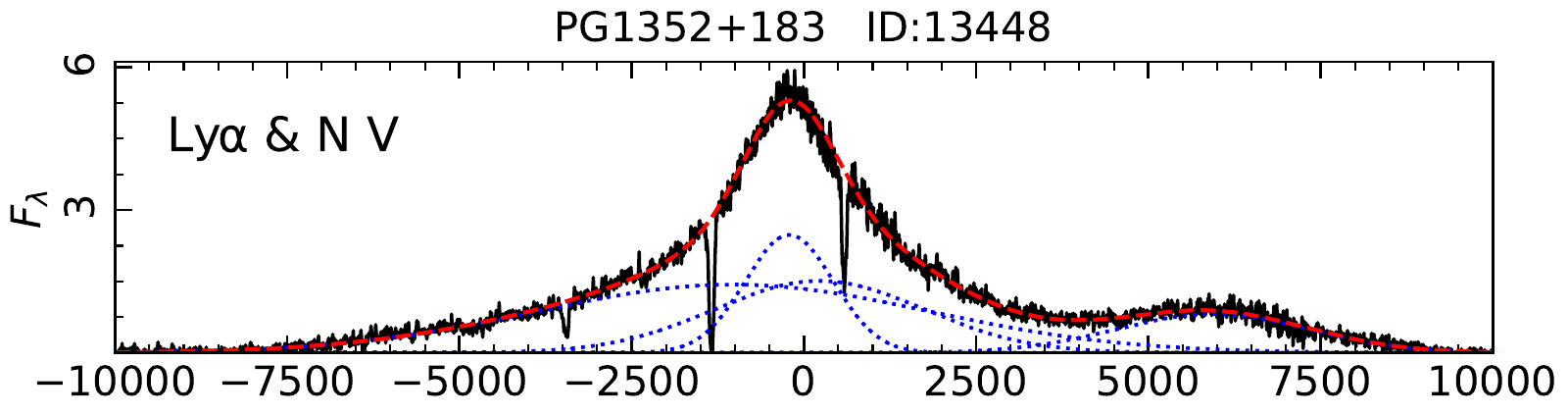}
}
\resizebox{0.95\hsize}{!}{
\includegraphics[angle=0]{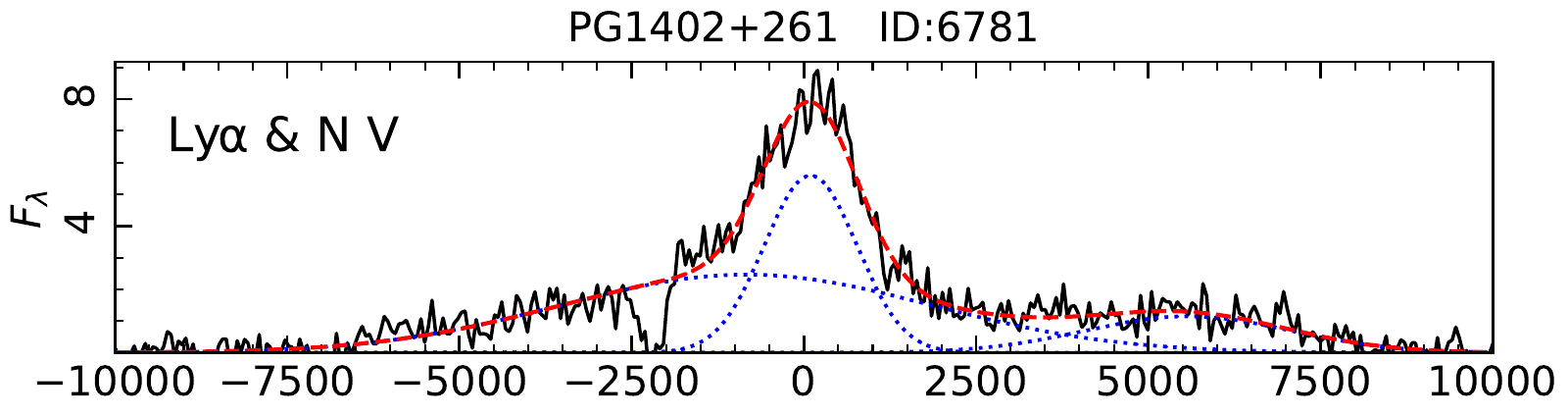}
\includegraphics[angle=0]{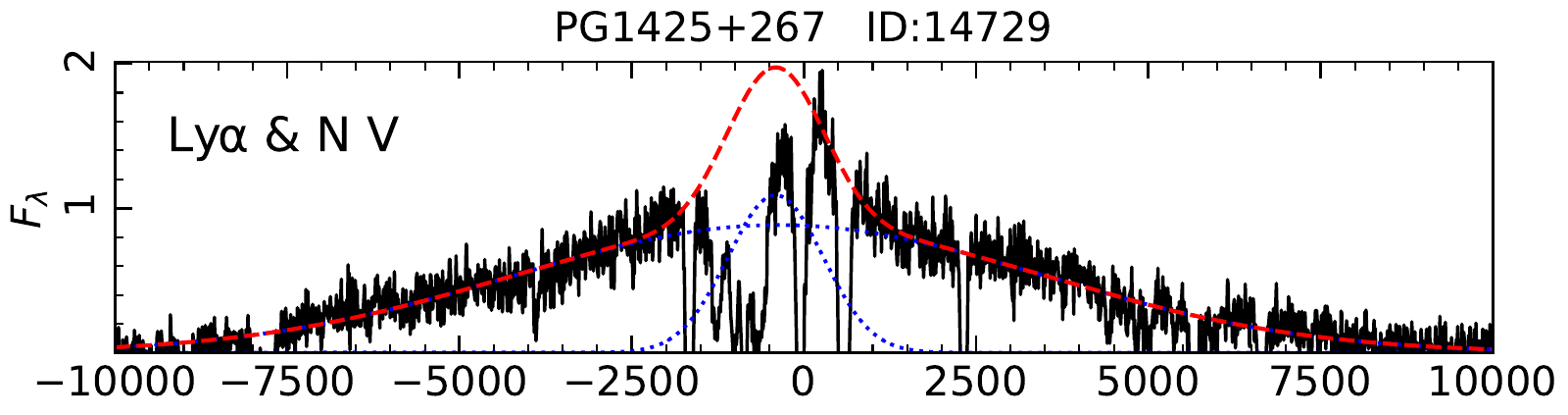}
}
\resizebox{0.95\hsize}{!}{
\includegraphics[angle=0]{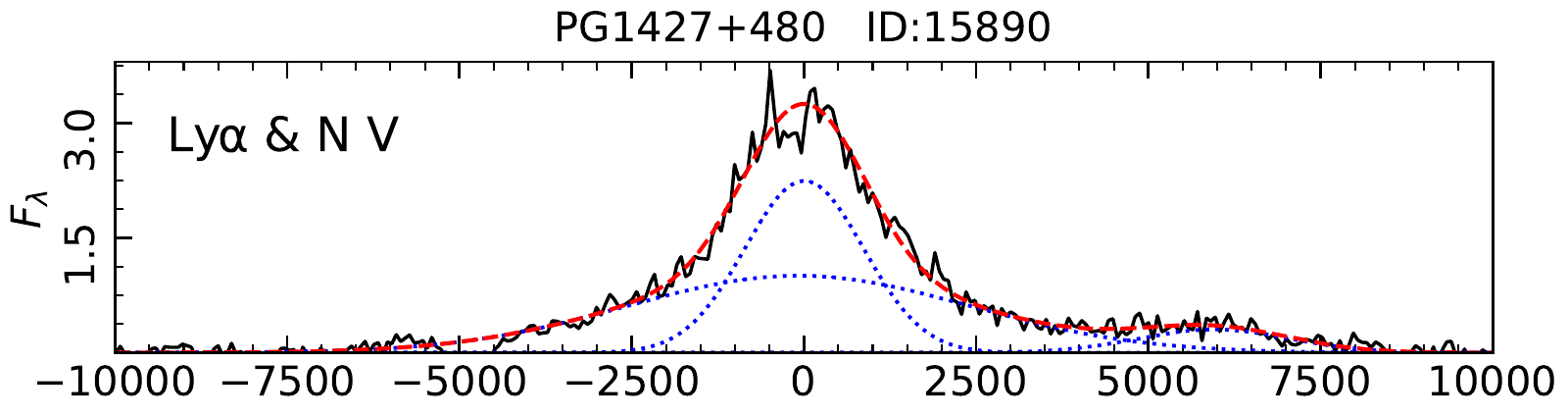}
\includegraphics[angle=0]{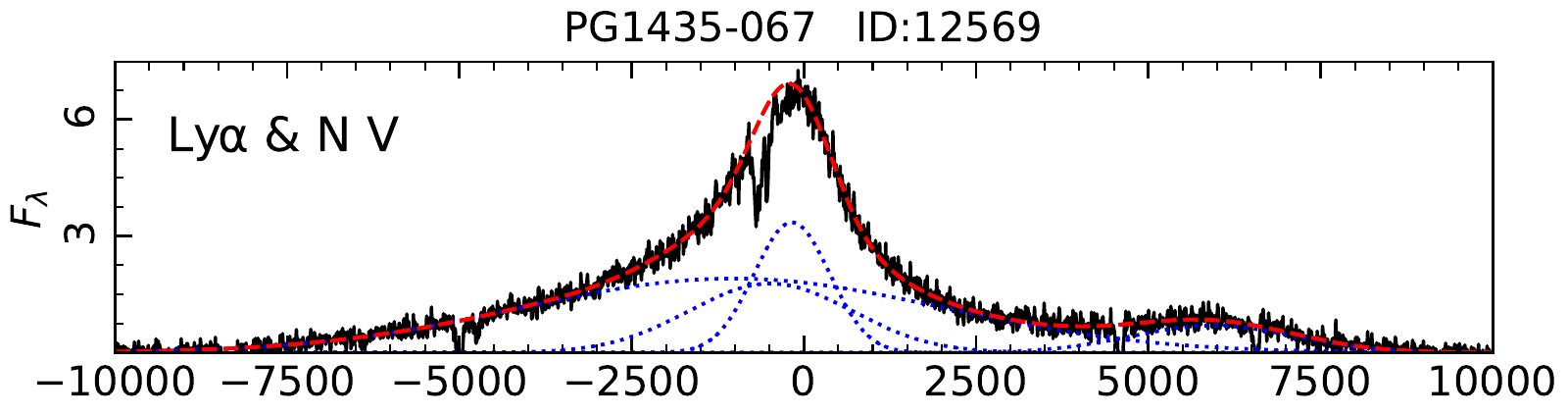}
}
\resizebox{0.95\hsize}{!}{
\includegraphics[angle=0]{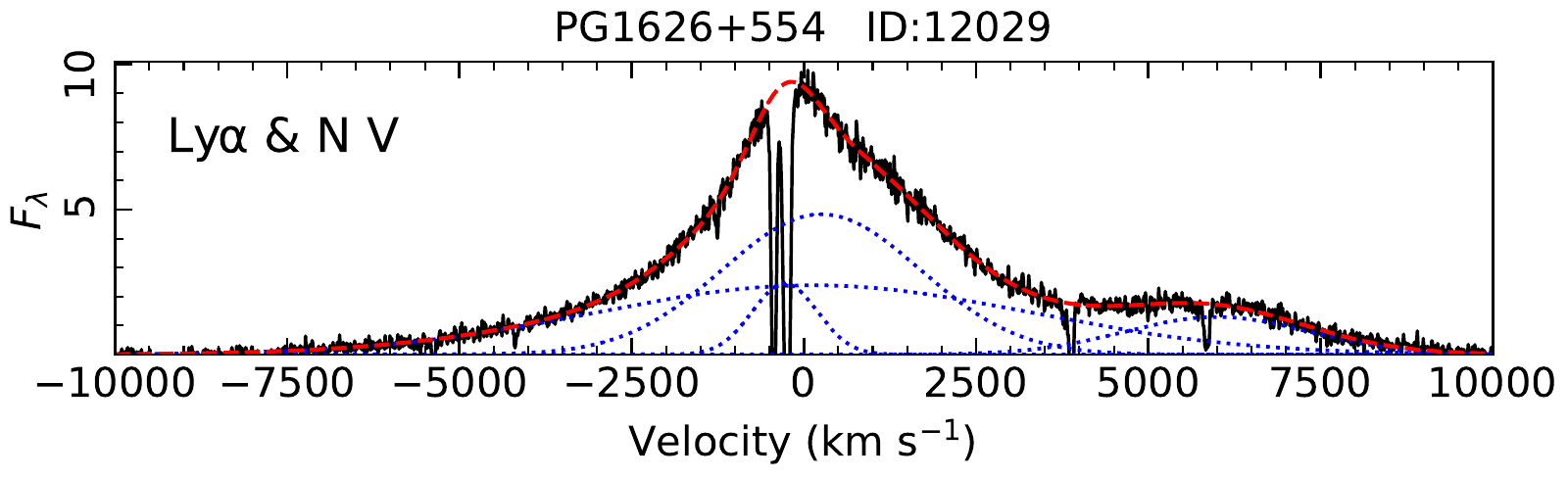}
\includegraphics[angle=0]{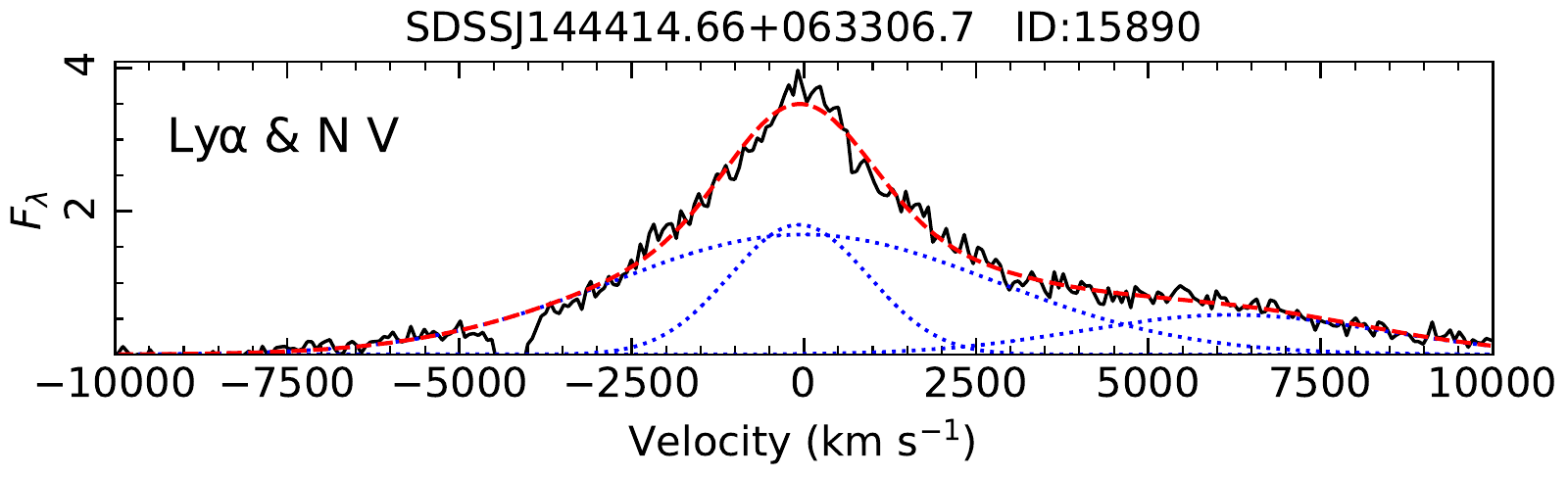}
}
\vspace{-0.15cm}
\caption{Ly $\alpha$ and \ion{N}{v} emission lines of the \sub sample. The HST UV spectra (in black) are continuum-subtracted. The total emission-line model is shown in dashed red line, and its individual emission components (Table \ref{table_emission}) in dotted blue line. The observed flux $F_{\lambda}$ is in ${10^{-14}}$ erg~s$^{-1}$~cm$^{-2}$~\AA$^{-1}$.
\label{fig_emission_lya}}
%\vspace{0.3cm}
\end{figure*}

\begin{figure*}
\centering
\resizebox{0.95\hsize}{!}{
\includegraphics[angle=0]{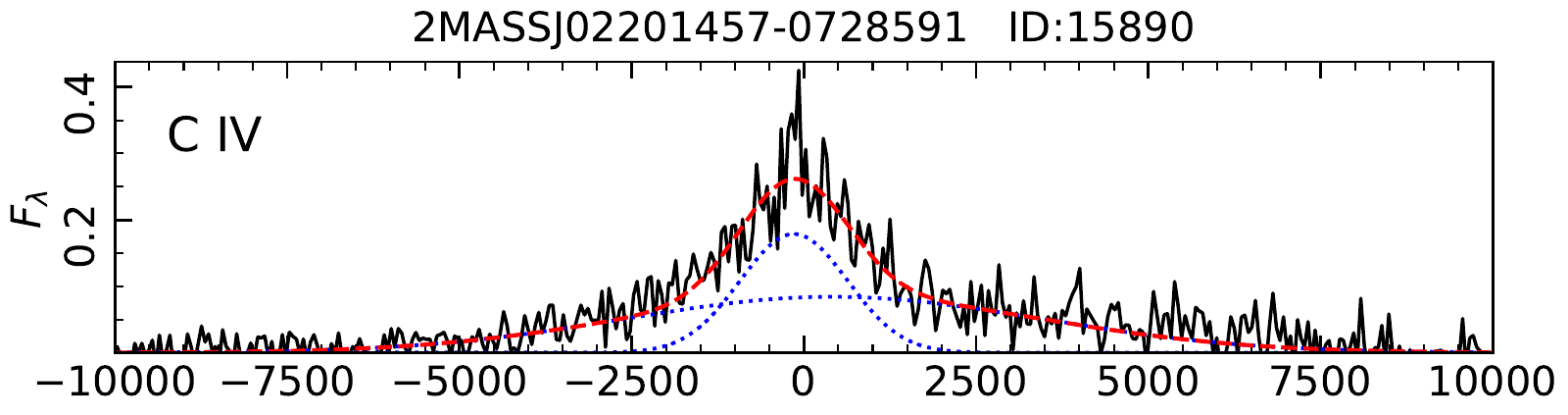}
\includegraphics[angle=0]{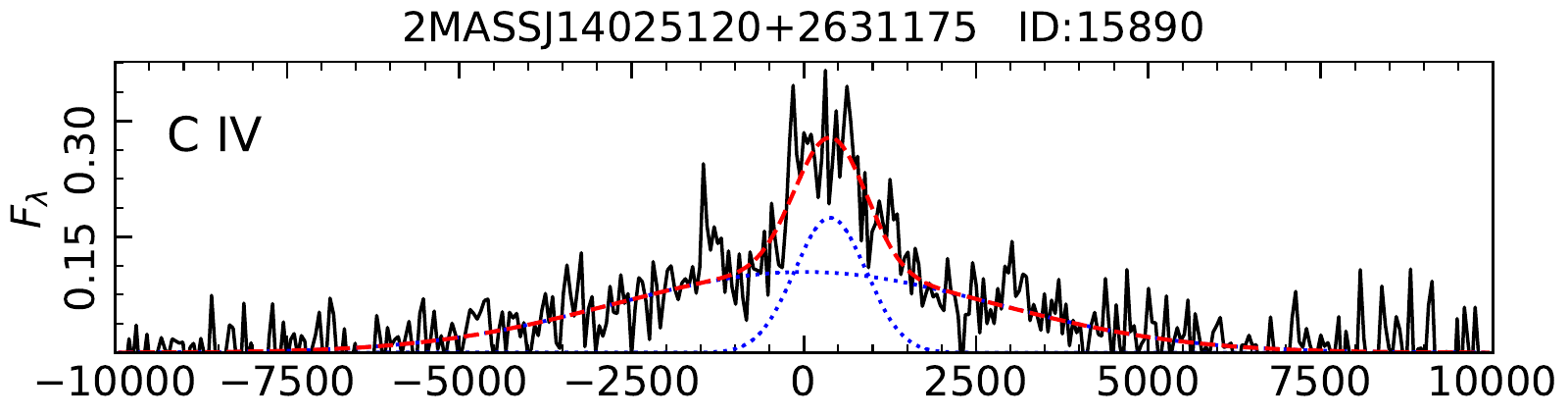}
}
\resizebox{0.95\hsize}{!}{
\includegraphics[angle=0]{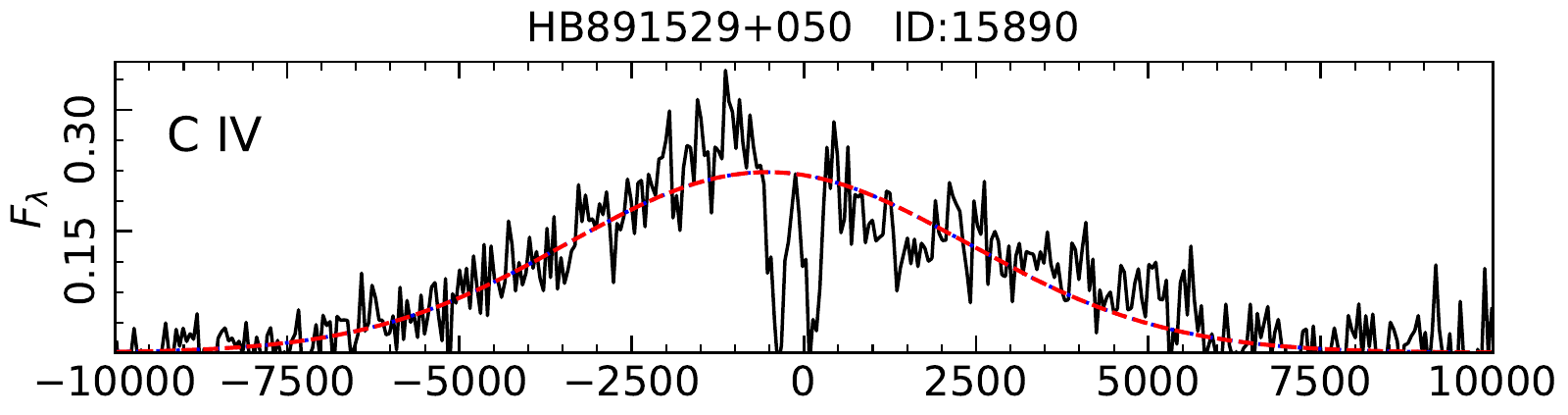}
\includegraphics[angle=0]{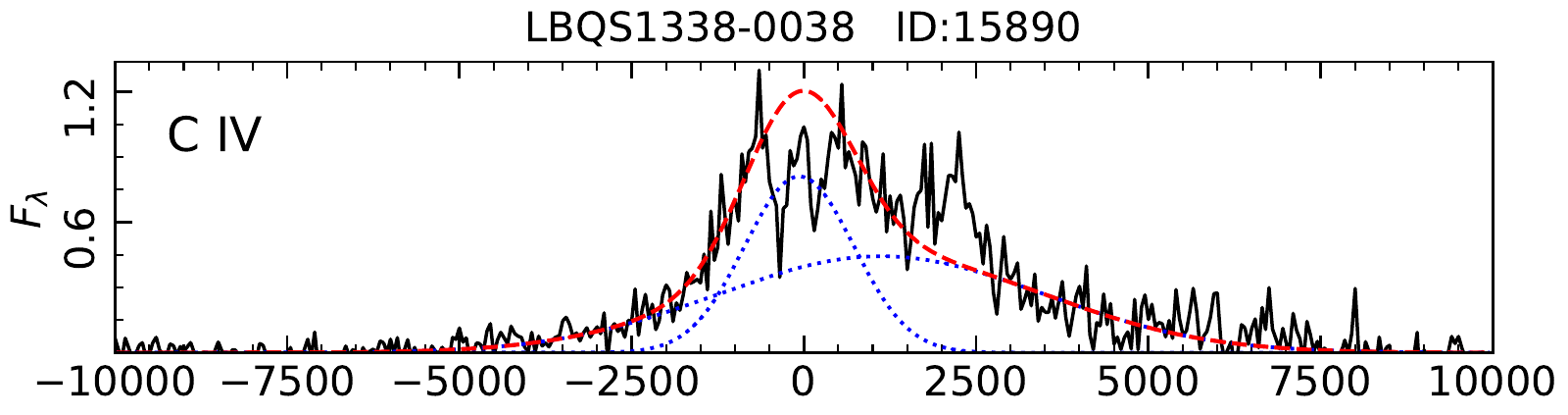}
}
\resizebox{0.95\hsize}{!}{
\includegraphics[angle=0]{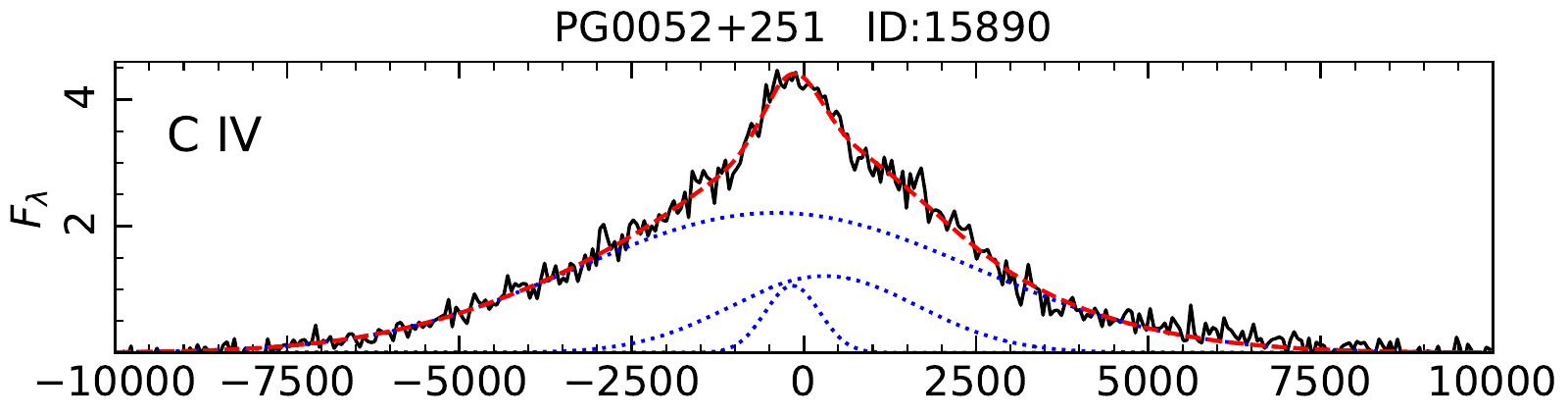}
\includegraphics[angle=0]{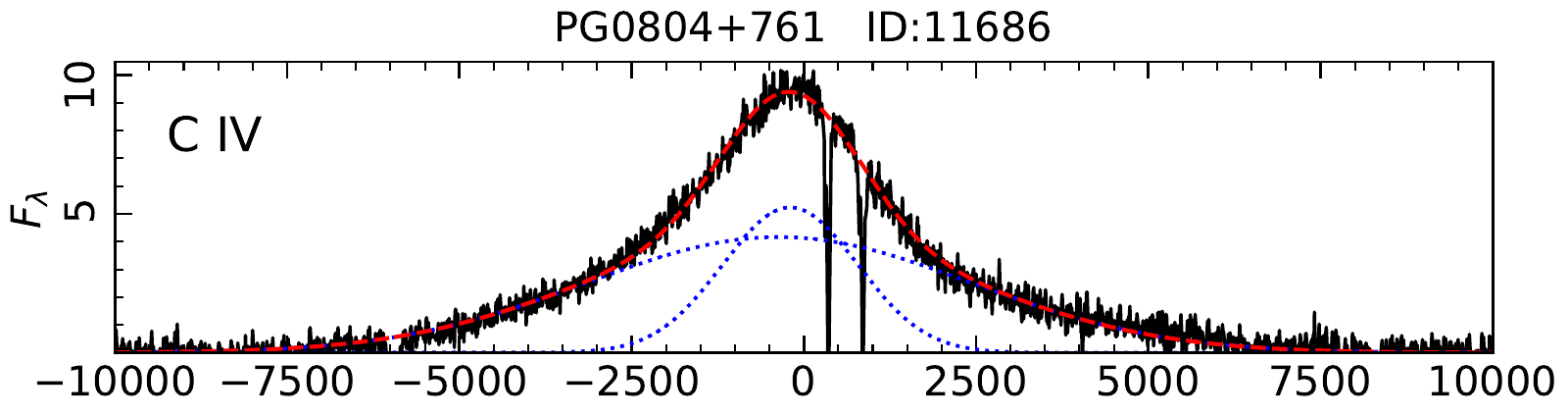}
}
\resizebox{0.95\hsize}{!}{
\includegraphics[angle=0]{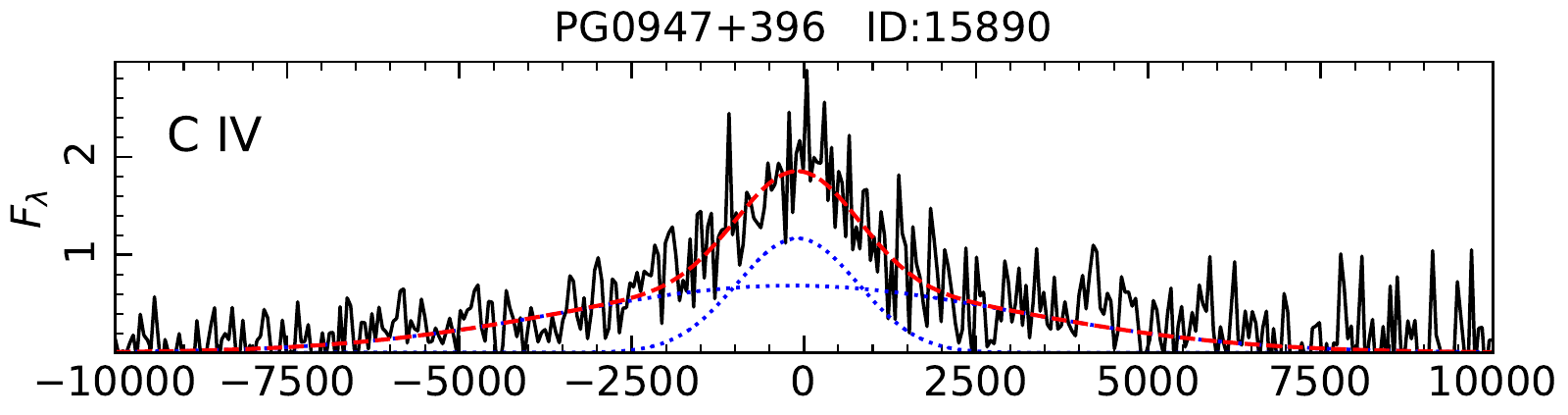}
\includegraphics[angle=0]{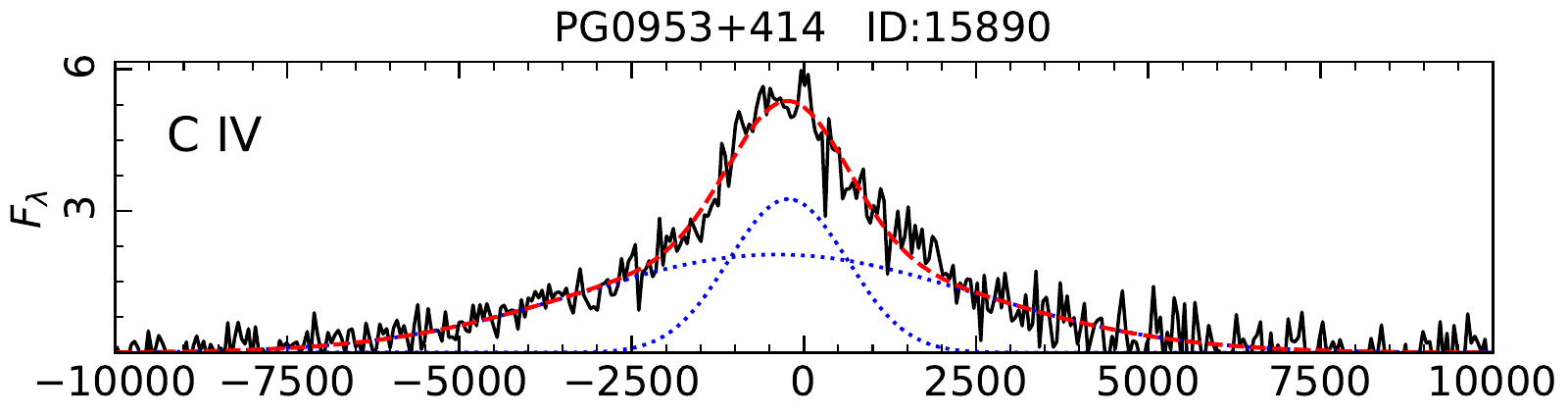}
}
\resizebox{0.95\hsize}{!}{
\includegraphics[angle=0]{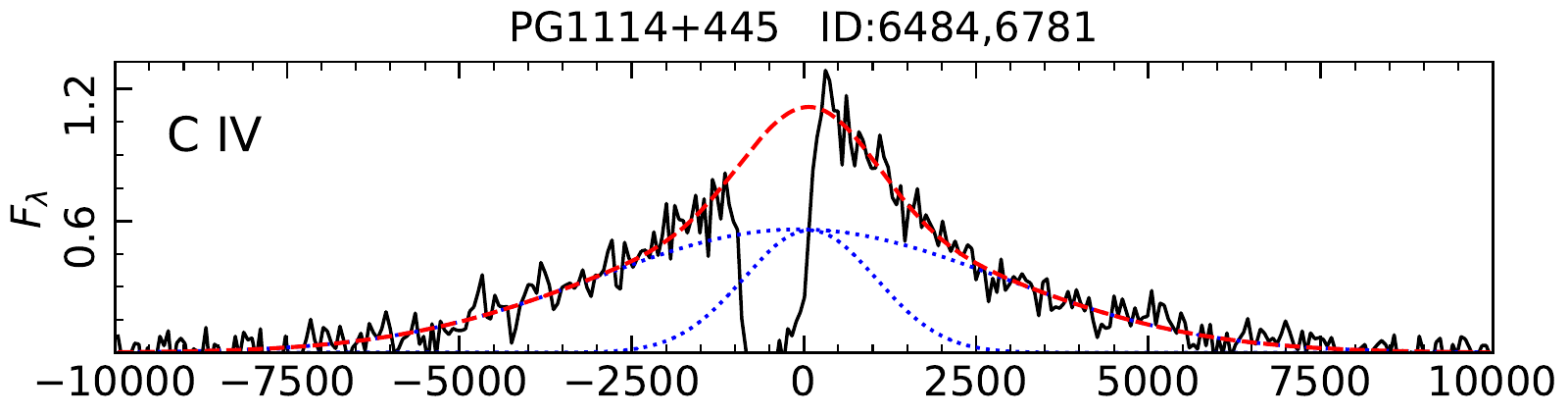}
\includegraphics[angle=0]{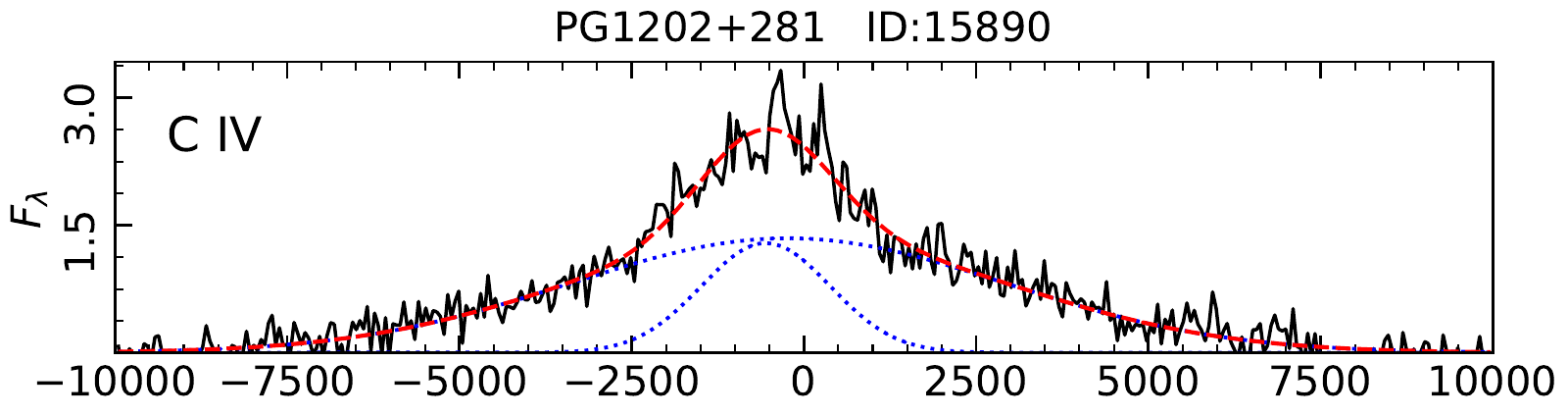}
}
\resizebox{0.95\hsize}{!}{
\includegraphics[angle=0]{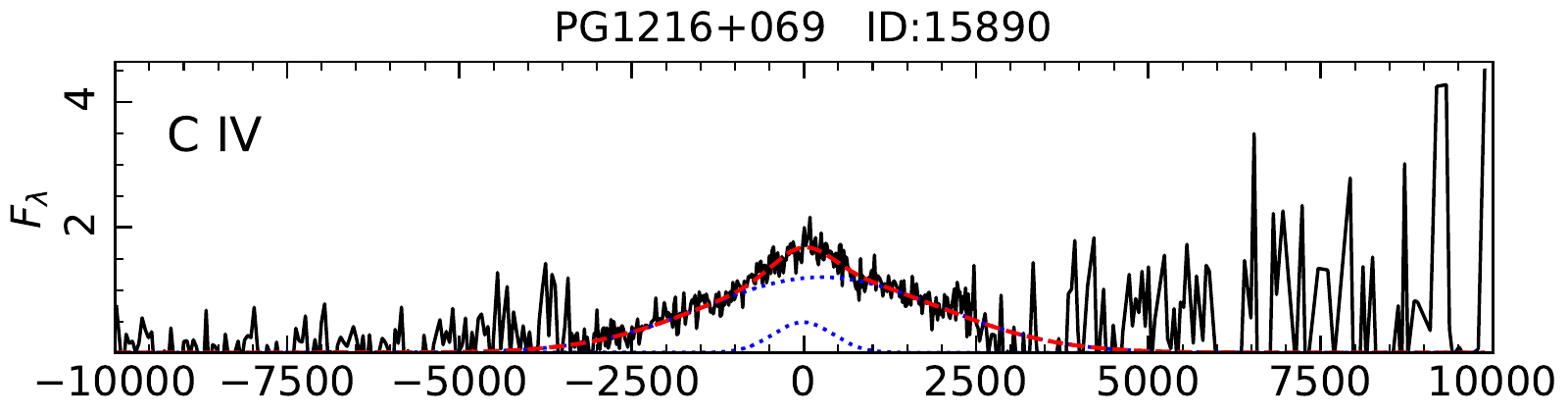}
\includegraphics[angle=0]{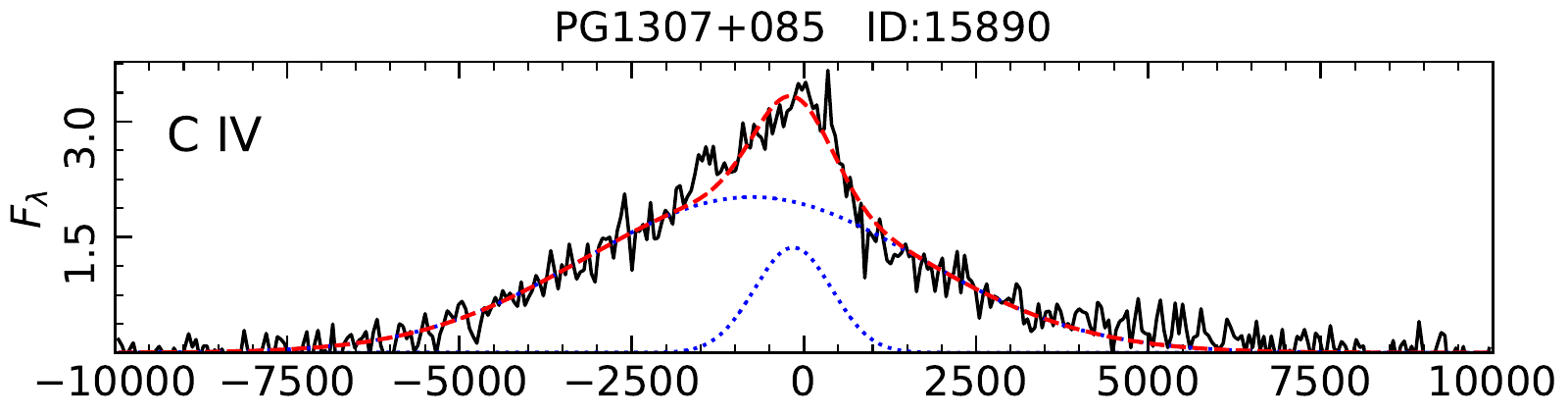}
}
\resizebox{0.95\hsize}{!}{
\includegraphics[angle=0]{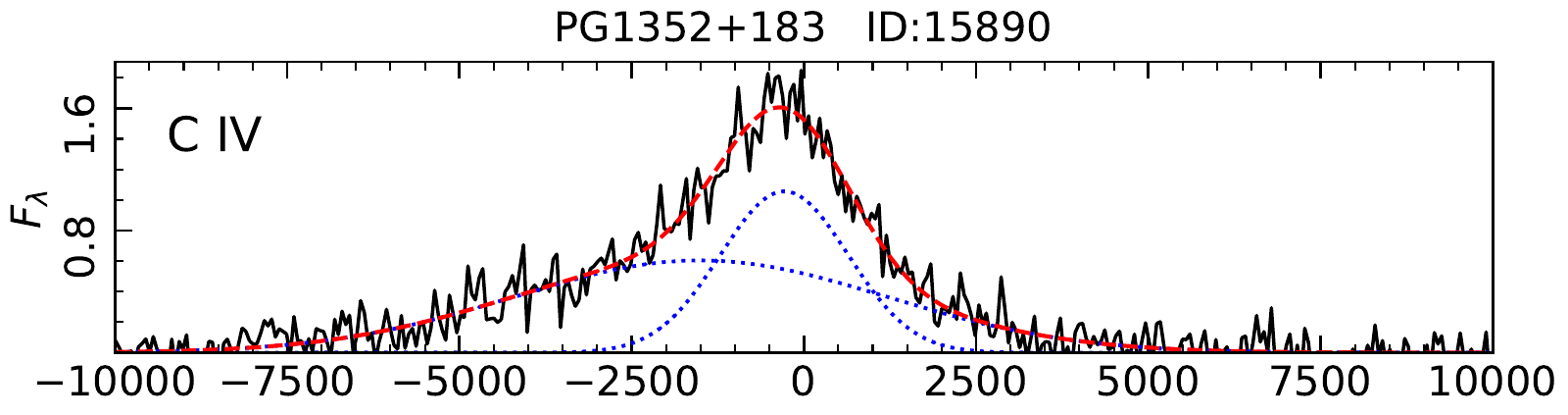}
\includegraphics[angle=0]{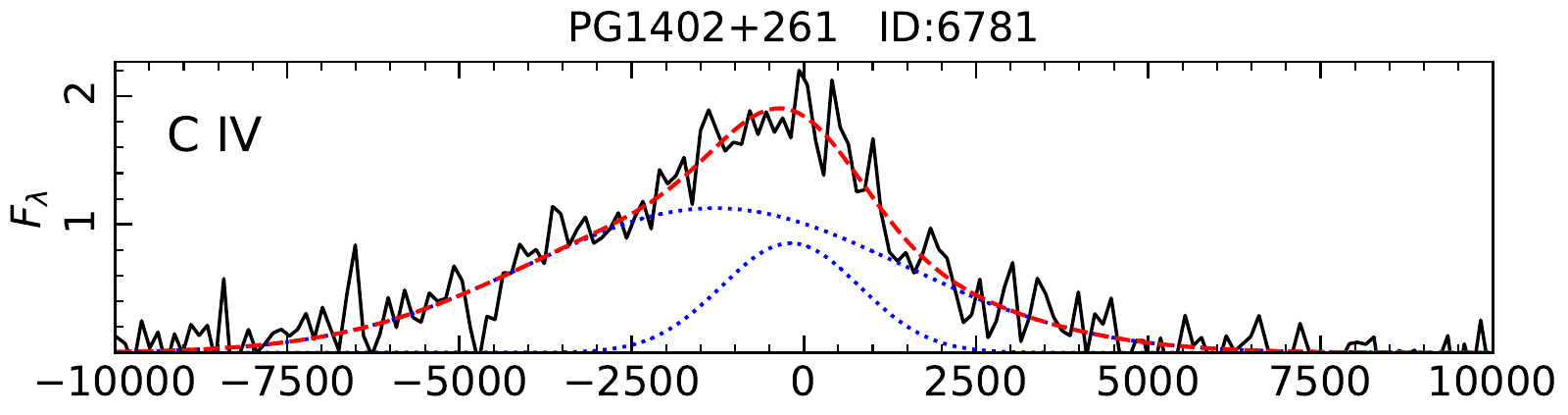}
}
\resizebox{0.95\hsize}{!}{
\includegraphics[angle=0]{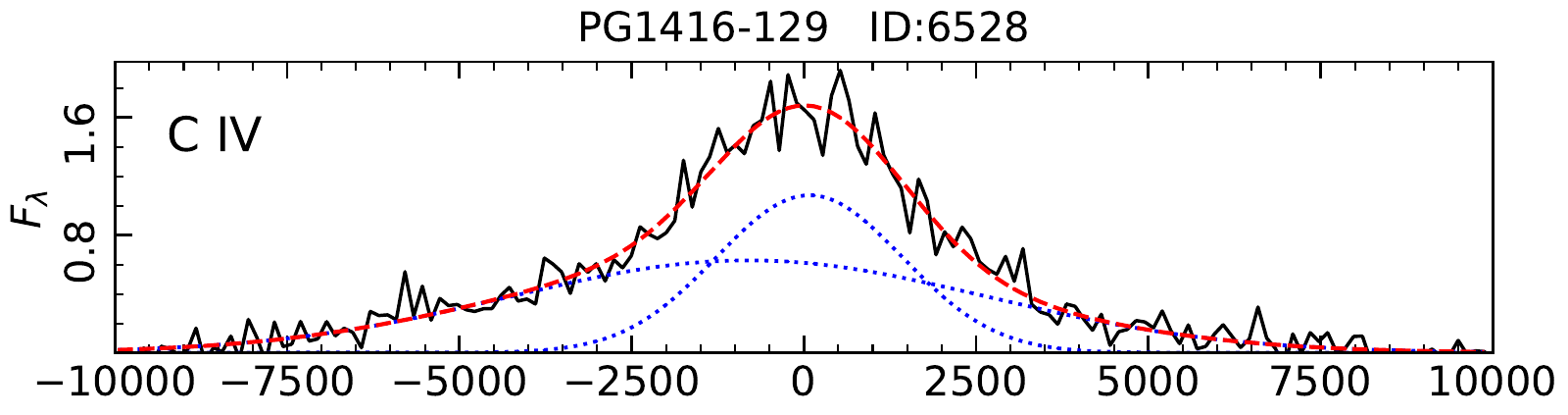}
\includegraphics[angle=0]{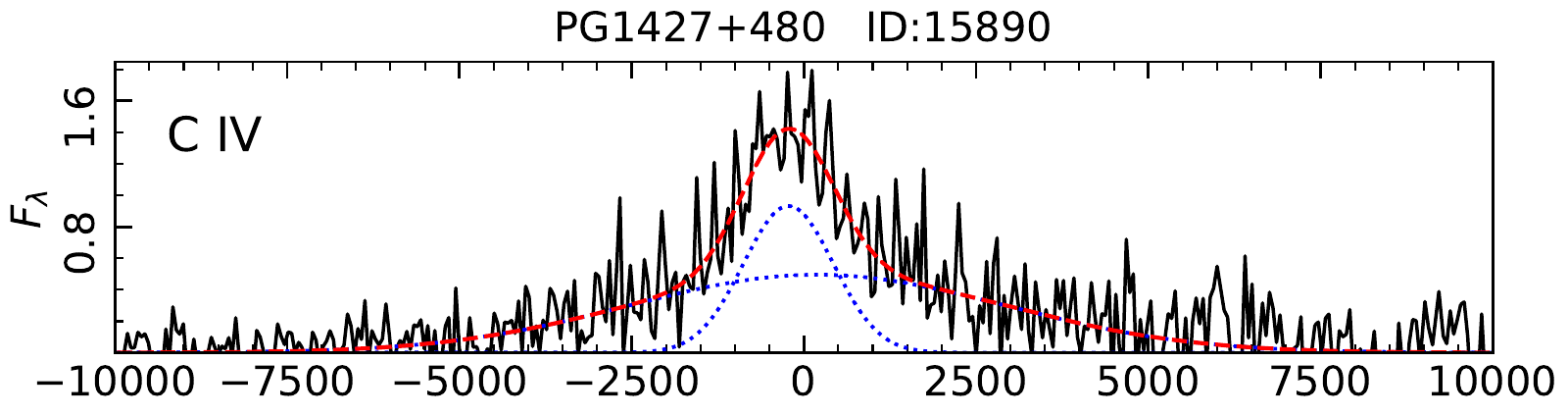}
}
\resizebox{0.95\hsize}{!}{
\includegraphics[angle=0]{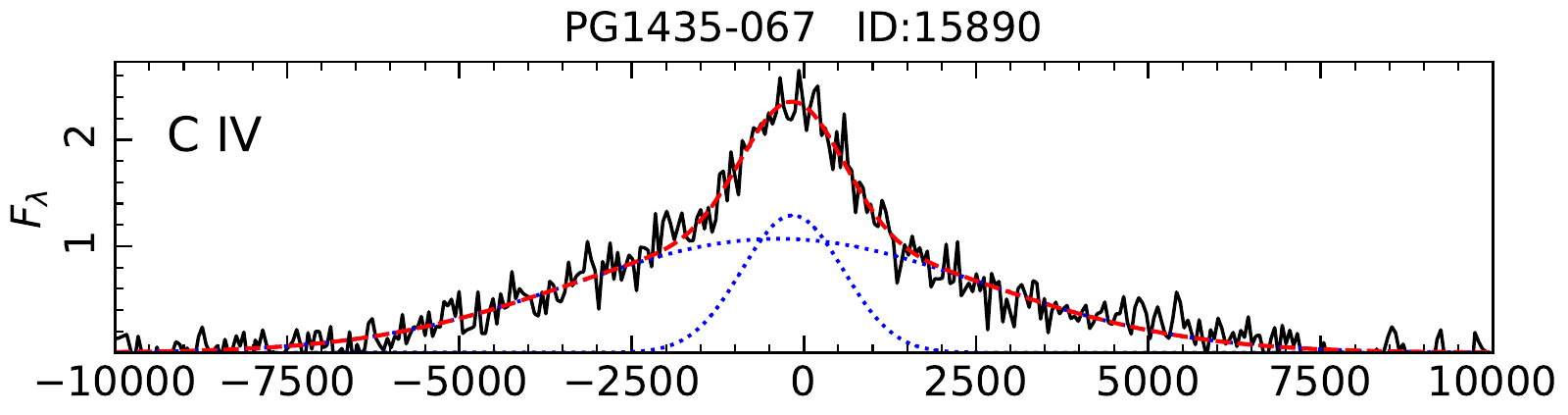}
\includegraphics[angle=0]{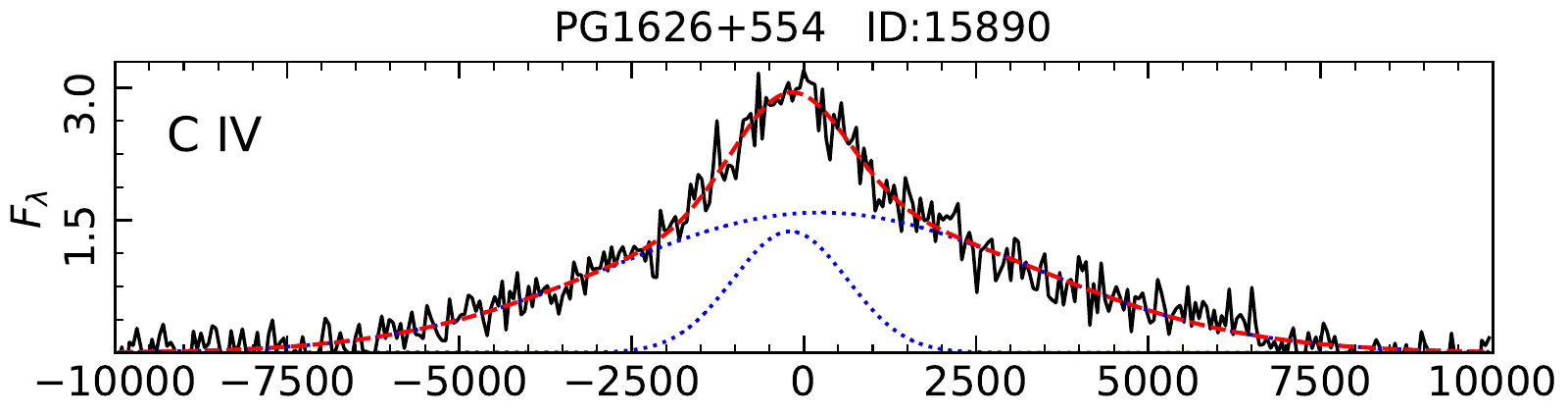}
}
\resizebox{0.49\hsize}{!}{
\includegraphics[angle=0]{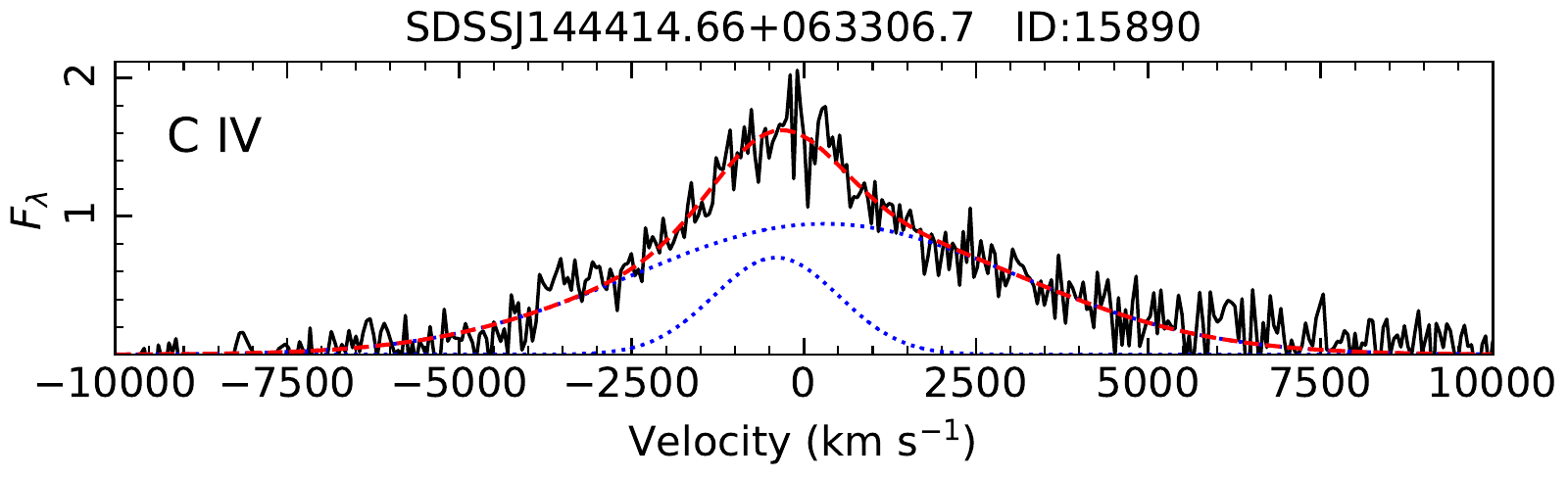}
}
\caption{\ion{C}{iv} emission lines of the \sub sample. The HST UV spectra (in black) are continuum-subtracted. The total emission-line model is shown in dashed red line, and its individual emission components (Table \ref{table_emission}) in dotted blue line. The observed flux $F_{\lambda}$ is in ${10^{-14}}$ erg~s$^{-1}$~cm$^{-2}$~\AA$^{-1}$.
\label{fig_emission_civ}}
%\vspace{0.3cm}
\end{figure*}

In Sect. \ref{sect_fit}, we described the modeling of the intrinsic UV spectral lines in the \sub HST sample. In this appendix, plots of the UV emission lines, and their best-fit models, are shown in Fig. \ref{fig_emission_lya} for \lya and \nv, and in Fig. \ref{fig_emission_civ} for \civ. The best-fit parameters of these emission lines are given in Table \ref{table_emission}. The best-fit parameters of all the intrinsic absorption lines are provided in Table \ref{table_abs} (spanning two pages).

%============================
% FIG: Lya & N V emission line profiles
%
%============================

%============================
% FIG: C IV emission line profiles
%
%============================

%============================
% TABLE: Emission line parameters
%
\begin{table*}[!tbp]
\begin{minipage}[t]{\hsize}
\setlength{\extrarowheight}{2pt}
\setlength{\tabcolsep}{1pt}
\caption{Best-fit parameters of the \lya, \nv, and \civ emission lines of the AGN in the \sub HST sample.}
\centering
\footnotesize
\renewcommand{\footnoterule}{}
\begin{tabular}{c | c c c | c c c | c c c}
\hline \hline
 & & \lya & & & \nv & & & \civ & \\
Object & $F$ & $v$ & FWHM & $F$ & $v$ & FWHM & $F$ & $v$ & FWHM \\
\hline
2MASS J02201457-0728591   & $0.35 \pm 0.07$ & $+60 \pm 30$ & $530 \pm 120$     & $1.28 \pm 0.07$ & $-510 \pm 130$ & $4010 \pm 280$   & $2.2 \pm 0.2$ & $-140 \pm 80$ & $2230 \pm 260$ \\
"                               & $2.9 \pm 0.2$  & $+140 \pm 20$ & $1760 \pm 120$    & - & - & -                                           & $4.0 \pm 0.6$ & $+420 \pm 230$ & $8590 \pm 790$ \\
"                               & $7.6 \pm 0.5$ & $+40 \pm 40$ & $6010 \pm 300$      & - & - & -                                           & - & - & -        \\
2MASS J14025120+2631175   & $3.3 \pm 0.1$   & $+100 \pm 50$ & $1860 \pm 120$   & $0.74 \pm 0.06$ & $+860 \pm 200$ & $2840 \pm 500$   & $1.4 \pm 0.1$ & $+380 \pm 60$ & $1450 \pm 180$ \\
"                         & $10.7 \pm 0.3$  & $+330 \pm 40$ & $7890 \pm 160$   & - & - & -                                           & $4.4 \pm 0.5$ & $+10 \pm 180$ & $7700 \pm 590$ \\
HB89 1257+286               & $12.1 \pm 0.8$ & $+280 \pm 60$ & $2520 \pm 140$    & - & - & -                                           & N/A & N/A & N/A  \\
HB89 1529+050               & $2.6 \pm 0.2$ & $-230 \pm 30$ & $2240 \pm 170$     & $2.8 \pm 0.4$ & $-2390 \pm 250$ & $4560 \pm 400$    & $10.2 \pm 0.3$ & $-530 \pm 80$ & $8320 \pm 240$ \\
"                           & $8.0 \pm 0.6$ & $-1190 \pm 320$ & $7240 \pm 380$   & - & - & -                                           & - & - & -   \\
LBQS 1338-0038            & $12 \pm 1$ & $-60 \pm 20$ & $1910 \pm 100$         & $5.5 \pm 0.2$ & $-410 \pm 180$ & $5660 \pm 390$     & $10 \pm 1$ & $-70 \pm 40$ & $2310 \pm 210$  \\
"                               & $8 \pm 7$ & $+1530 \pm 340$ & $3480 \pm 700$       & - & - & -                                           & $17 \pm 2$ & $+1120 \pm 140$ & $6870 \pm 380$ \\
"                               & $12 \pm 6$ & $-510 \pm 1330$ & $5070 \pm 1370$   & - & - & -                                             & - & - & -   \\
PG 0052+251                   & $17.7 \pm 0.3$ & $-10 \pm 10$ & $980 \pm 20$       & $30.8 \pm 0.5$ & $-800 \pm 70$ & $4900 \pm 130$     & $6.3 \pm 0.7$ & $-170 \pm 40$ & $1080 \pm 140$  \\        
"                             & $191 \pm 1$ & $-170 \pm 10$ & $5490 \pm 70$        & - & - & -                                           & $25 \pm 4$ & $+300 \pm 100$ & $3690 \pm 380$   \\
"                             & - & - & -                                          & - & - & -                                           & $96 \pm 7$ & $-410 \pm 70$ & $7860 \pm 200$   \\
PG 0804+761                   & $128 \pm 2$ & $-20 \pm 10$ & $1980 \pm 20$         & $128.0 \pm 0.3$ & $-460 \pm 10$ & $3970 \pm 20$     & $73.1 \pm 0.8$ & $-200 \pm 10$ & $2540 \pm 30$  \\         
"                             & $397 \pm 5$ & $-340 \pm 10$ & $5210 \pm 50$         & - & - & -                                          & $166 \pm 2$ & $-360 \pm 10$ & $7220 \pm 50$   \\ 
PG 0947+396                   & $7.2 \pm 0.6$ & $-30 \pm 30$ & $1130 \pm 90$       & $20.7 \pm 0.7$ & $-1710 \pm 310$ & $7550 \pm 840$   & $16 \pm 3$ & $-90 \pm 130$ & $2490 \pm 480$  \\
"                             & $43 \pm 2$ & $-360 \pm 60$ & $4210 \pm 250$        & - & - & -                                           & $35 \pm 8$ & $-180 \pm 330$ & $9310 \pm 1270$   \\
PG 0953+414                   & $60.5 \pm 0.5$ & $-310 \pm 10$ & $1470 \pm 20$     & $54.6 \pm 0.3$ & $-580 \pm 20$ & $5160 \pm 60$      & $44 \pm 2$ & $-230 \pm 40$ & $2490 \pm 140$   \\
"                             & $185 \pm 3$ & $-80 \pm 40$ & $5170 \pm 70$         & - & - & -                                           & $95 \pm 7$ & $-390 \pm 90$ & $8350 \pm 340$   \\
PG 1114+445                   & $8.0 \pm 0.9$ & $-60 \pm 50$ & $1180 \pm 120$      & $5.6 \pm 0.5$ & $+160 \pm 110$ & $2840 \pm 210$     & $8.0 \pm 0.7$ & $+90 \pm 70$ & $2590 \pm 190$  \\
"                             & $19 \pm 1$ & $+470 \pm 70$ & $3290 \pm 190$        & - & - & -                                           & $25 \pm 1$ & $-80 \pm 60$ & $7960 \pm 230$   \\
"                             & $44 \pm 2$ & $-190 \pm 100$ & $8700 \pm 190$       & - & - & -                                           & - & - & -   \\
PG 1202+281                   & $6.7 \pm 0.3$ & $-50 \pm 20$ & $770 \pm 30$        & $2.0 \pm 0.2$ & $-140 \pm 60$ & $2370 \pm 210$      & $18 \pm 1$ & $-550 \pm 60$ & $2580 \pm 210$   \\
"                             & $18.0 \pm 0.5$ & $-540 \pm 20$ & $3370 \pm 90$     & - & - & -                                           & $64 \pm 4$ & $-230 \pm 80$ & $8680 \pm 290$   \\
"                             & $40 \pm 1$ & $+190 \pm 60$ & $9620 \pm 190$        & - & - & -                                           & - & - & -   \\
PG 1216+069                   & $18.1 \pm 0.6$ & $+140 \pm 10$ & $1500 \pm 40$     & $5.9 \pm 0.3$ & $+450 \pm 60$ & $3720 \pm 200$      & $3.7 \pm 0.2$ & $-10 \pm 20$ & $1360 \pm 70$ \\
"                             & $41 \pm 1$ & $+560 \pm 20$ & $3540 \pm 70$         & - & - & -                                           & $35.5 \pm 0.6$ & $+260 \pm 20$ & $5350 \pm 80$ \\
"                             & $81 \pm 2$ & $+100 \pm 40$ & $10080 \pm 120$       & - & - & -                                           & - & - & -   \\      
PG 1307+085                   & $17.1 \pm 0.4$ & $+50 \pm 10$ & $750 \pm 20$       & $18.2 \pm 0.3$ & $-10 \pm 20$ & $3120 \pm 60$       & $11.5 \pm 0.7$ & $-150 \pm 30$ & $1530 \pm 110$ \\         
"                             & $63.2 \pm 0.9$ & $-250 \pm 10$ & $2890 \pm 40$     & - & - & -                                           & $74 \pm 2$ & $-750 \pm 40$ & $6620 \pm 140$   \\    
"                             & $142 \pm 2$ & $-280 \pm 20$ & $8210 \pm 80$        & - & - & -                                           & - & - & -   \\     
PG 1352+183                   & $18.6 \pm 0.4$ & $-210 \pm 10$ & $1740 \pm 20$     & $13.3 \pm 0.2$ & $-260 \pm 30$ & $3710 \pm 40$      & $14.6 \pm 0.8$ & $-280 \pm 40$ & $2520 \pm 140$ \\
"                             & $27 \pm 1$ & $+230 \pm 40$ & $4110 \pm 130$        & - & - & -                                           & $24 \pm 2$ & $-1510 \pm 140$ & $7370 \pm 310$  \\
"                             & $48 \pm 2$ & $-1030 \pm 100$ & $7760 \pm 90$       & - & - & -                                           & - & - & -   \\
PG 1402+261                   & $44 \pm 2$ & $+100 \pm 20$ & $1840 \pm 70$         & $20 \pm 1$ & $-610 \pm 140$ & $4020 \pm 350$        & $13 \pm 2$ & $-190 \pm 110$ & $2750 \pm 410$   \\
"                             & $79 \pm 5$ & $-830 \pm 130$ & $7410 \pm 330$       & - & - & -                                           & $46 \pm 5$ & $-1290 \pm 160$ & $7490 \pm 400$  \\
PG 1416-129                   & N/A & N/A & N/A                                    & N/A & N/A & N/A                                     & $21 \pm 2$ & $+80 \pm 70$ & $3590 \pm 250$   \\
"                             & -   & -   & -                                      & -   & -   & -                                       & $32 \pm 4$ & $-810 \pm 160$ & $9230 \pm 500$   \\
PG 1425+267                   & $11 \pm 1$ & $-410 \pm 50$ & $2230 \pm 120$        & - & - & -                                           & N/A & N/A & N/A \\
"                             & $47.0 \pm 0.7$ & $-360 \pm 40$ & $12330 \pm 140$   & - & - & -                                           & - & - & -   \\
PG 1427+480                   & $24 \pm 1$ & $+6 \pm 20$ & $2480 \pm 100$          & $4.2 \pm 0.3$ & $-240 \pm 110$ & $3090 \pm 330$     & $10 \pm 1$ & $-220 \pm 70$ & $1880 \pm 260$   \\
"                             & $31 \pm 3$ & $-80 \pm 60$ & $7140 \pm 380$         & - & - & -                                           & $21 \pm 3$ & $+190 \pm 250$ & $7820 \pm 830$  \\ 
PG 1435-067                   & $21 \pm 1$ & $-170 \pm 20$ & $1460 \pm 60$         & $9.6 \pm 0.2$ & $-260 \pm 40$ & $3090 \pm 100$      & $13.8 \pm 0.8$ & $-160 \pm 40$ & $1950 \pm 120$ \\
"                             & $24 \pm 2$ & $-490 \pm 40$ & $3140 \pm 180$        & - & - & -                                           & $47 \pm 2$ & $-370 \pm 70$ & $7940 \pm 230$   \\
"                             & $67 \pm 2$ & $-1030 \pm 40$ & $8140 \pm 120$       & - & - & -                                           & - & - & -   \\
PG 1626+554                   & $13.3 \pm 0.5$ & $-280 \pm 10$ & $1260 \pm 50$     & $20.1 \pm 0.8$ & $-200 \pm 50$ & $3510 \pm 100$     & $17 \pm 1$ & $-200 \pm 40$ & $2210 \pm 150$   \\
"                             & $80 \pm 2$ & $+250 \pm 30$ & $3820 \pm 80$         & - & - & -                                           & $72 \pm 3$ & $+240 \pm 60$ & $8200 \pm 200$   \\
"                             & $87 \pm 4$ & $+130 \pm 100$ & $8400 \pm 130$       & - & - & -                                           & - & - & -   \\
SDSS J144414.66+063306.7        & $22.4 \pm 0.8$ & $-70 \pm 20$ & $2870 \pm 90$      & $14.8 \pm 0.8$ & $+10 \pm 200$ & $6030 \pm 420$     & $10 \pm 1$ & $-400 \pm 80$ & $2580 \pm 320$   \\
"                         & $57 \pm 2$ & $-2 \pm 100$ & $7940 \pm 190$         & - & - & -                                           & $41 \pm 3$ & $+300 \pm 100$ & $7940 \pm 340$   \\
\hline
\end{tabular}
\end{minipage}
\tablefoot{
Each row in the table corresponds to one Gaussian line component. All line fluxes ($F$) are in units of $10^{-14}$~\ergflux. The line-centroid velocity shift $v$ and the FWHM are in units of \kms. Negative $v$ means systematic blueshift (outflow) and positive $v$ means systematic redshift (inflow) with respect to the local rest frame of the object.
}
\label{table_emission}
\end{table*}
%============================

%============================
% TABLE: Absorption line parameters
%
\begin{table*}[!tbp]
\begin{minipage}[t]{\hsize}
\setlength{\extrarowheight}{1.1pt}
\caption{Parameters of the intrinsic AGN absorption lines in the HST spectra of the \sub sample. Table continued next page.}
\centering
\footnotesize
\renewcommand{\footnoterule}{}
\begin{tabular}{c c c c c c c c}
\hline \hline
Object   & Comp. & Ion\,/  & $\lambda_{\rm obs}$ & EW    & $v$    & FWHM    & ${\log(N_{\rm ion})}$ \\
         &       & Line    & (\AA)               & (\AA) & (\kms) & (\kms)  & (cm$^{-2}$)           \\
\hline
2MASS J02201457-0728591   & -  & -      & -                 & -                & -              & -             & -      \\
2MASS J14025120+2631175   & 1  & \lya   & $1440.8 \pm 0.1$  & $1.59 \pm 0.02$  & $-610 \pm 10$  & $680 \pm 40$  & 14.47  \\
"                         & 1  & \nv(b) & $1468.4 \pm 0.1$  & $0.26 \pm 0.02$  & $-570 \pm 30$  & $500 \pm 60$  & 14.09  \\
"                         & 1  & \nv(r) & $1473.3 \pm 0.1$  & $0.13 \pm 0.02$  & $-550 \pm 20$  & $400 \pm 50$  & 14.09  \\
"                         & 1  & \ovi(r)& $1229.9 \pm 0.1$  & $0.51 \pm 0.01$  & $-590 \pm 20$  & $650 \pm 50$  & 14.91  \\
HB89 1257+286               & -  & -      & -                 & -                & -              & -             & -      \\
HB89 1529+050               & 1  & \lya   & $1480.4 \pm 0.1$  & $0.66 \pm 0.02$  & $-100 \pm 10$  & $420 \pm 20$  & 14.08  \\
"                         & 1  & \lyb   & $1249.2 \pm 0.1$  & $0.20 \pm 0.01$  & $-70 \pm 10$   & $480 \pm 20$  & 14.43  \\  
"                         & 1  & \nv(b) & $1508.6 \pm 0.1$  & $0.24 \pm 0.01$  & $-90 \pm 10$   & $340 \pm 30$  & 14.05  \\ 
"                         & 1  & \nv(r) & $1513.5 \pm 0.1$  & $0.14 \pm 0.01$  & $-90 \pm 10$   & $300 \pm 30$  & 14.12  \\
"                         & 1  & \ovi(b)& $1256.8 \pm 0.1$  & $0.23 \pm 0.01$  & $-80 \pm 10$   & $440 \pm 30$  & 14.26  \\ 
"                         & 1  & \ovi(r)& $1263.6 \pm 0.1$  & $0.23 \pm 0.01$  & $-90 \pm 10$   & $490 \pm 30$  & 14.56  \\
"                         & 1  & \civ(b)& $1885.2 \pm 0.1$  & $0.36 \pm 0.04$  & $-130 \pm 20$  & $290 \pm 40$  & 13.95  \\ 
"                         & 1  & \civ(r)& $1888.3 \pm 0.1$  & $0.34 \pm 0.04$  & $-130 \pm 20$  & $290 \pm 40$  & 14.22  \\
LBQS 1338-0038            & 1  & \lya   & $1504.0 \pm 0.1$  & $4.32 \pm 0.04$  & $-60 \pm 10$   & $490 \pm 20$  & 14.90  \\
"                         & 1  & \lyb   & $1269.1 \pm 0.1$  & $0.48 \pm 0.01$  & $-50 \pm 10$   & $530 \pm 30$  & 14.81  \\            
"                         & 1  & \ovi(b)& $1276.8 \pm 0.1$  & $0.45 \pm 0.01$  & $-40 \pm 10$   & $390 \pm 20$  & 14.56  \\    
"                         & 1  & \ovi(r)& $1283.8 \pm 0.1$  & $0.44 \pm 0.01$  & $-50 \pm 10$   & $390 \pm 20$  & 14.84  \\
"                         & 1  & \civ(b)& $1915.3 \pm 0.1$  & $1.20 \pm 0.15$  & $-80 \pm 20$   & $340 \pm 60$  & 14.47  \\    
"                         & 1  & \civ(r)& $1918.5 \pm 0.2$  & $1.10 \pm 0.16$  & $-80 \pm 20$   & $340 \pm 60$  & 14.73  \\
PG 0052+251                   & -  & -      & -                 & -                & -              & -             & -      \\
PG 0804+761               & 1  & \lya   & $1339.9 \pm 0.1$  & $18.01 \pm 0.02$ & $+560 \pm 10$  & $110 \pm 10$  & 15.52  \\
"                         & 1  & \nv(b) & $1365.4 \pm 0.1$  & $0.40 \pm 0.04$  & $+600 \pm 10$  & $50 \pm 10$   & 14.28  \\
"                         & 1  & \ovi(r)& $1143.1 \pm 0.1$  & $1.32 \pm 0.05$  & $+460 \pm 10$  & $70 \pm 10$   & 15.32  \\
"                         & 1  & \civ(b)& $1706.5 \pm 0.1$  & $2.97 \pm 0.03$  & $+600 \pm 10$  & $50 \pm 10$   & 14.87  \\    
"                         & 1  & \civ(r)& $1709.3 \pm 0.1$  & $2.18 \pm 0.04$  & $+600 \pm 10$  & $50 \pm 10$   & 15.03  \\
"                             & 2  & \lya   & $1330.2 \pm 0.2$  & $2.13 \pm 0.05$  & $-1570 \pm 50$ & $230 \pm 20$  & 14.59  \\
PG 0947+396                   & 1  & \lya   & $1464.1 \pm 0.1$  & $1.43 \pm 0.10$  & $-290 \pm 10$  & $170 \pm 20$  & 14.42  \\
"                               & 1  & \ovi(b)& $1242.8 \pm 0.1$  & $0.95 \pm 0.04$  & $-300 \pm 10$  & $180 \pm 20$  & 14.88  \\
"                               & 1  & \ovi(r)& $1249.6 \pm 0.1$  & $0.56 \pm 0.04$  & $-300 \pm 10$  & $180 \pm 20$  & 14.95  \\
"                               & 2  & \lya   & $1454.7 \pm 0.1$  & $0.52 \pm 0.05$  & $-2210 \pm 10$ & $140 \pm 20$  & 13.98  \\
PG 0953+414                   & 1  & \lya   & $1499.5 \pm 0.1$  & $9.06 \pm 0.02$  & $-150 \pm 10$  & $100 \pm 10$  & 15.22  \\      
"                             & 1  & \lyb   & $1265.2 \pm 0.1$  & $1.28 \pm 0.01$  & $-150 \pm 10$  & $70 \pm 20$   & 15.24  \\
"                             & 1  & \ovi(b)& $1272.9 \pm 0.1$  & $0.56 \pm 0.02$  & $-140 \pm 10$  & $80 \pm 20$   & 14.65  \\
"                             & 1  & \ovi(r)& $1280.0 \pm 0.1$  & $0.32 \pm 0.03$  & $-140 \pm 10$  & $100 \pm 20$  & 14.71  \\
"                             & 2  & \lya   & $1498.4 \pm 0.1$  & $4.93 \pm 0.04$  & $-370 \pm 10$  & $80 \pm 10$   & 14.96  \\
"                             & 2  & \lyb   & $1264.3 \pm 0.1$  & $0.25 \pm 0.02$  & $-360 \pm 10$  & $80 \pm 10$   & 14.53  \\
"                             & 3  & \lya   & $1495.7 \pm 0.1$  & $0.25 \pm 0.03$  & $-900 \pm 20$  & $60 \pm 10$   & 13.66  \\
"                             & 3  & \ovi(b)& $1270.5 \pm 0.1$  & $0.62 \pm 0.02$  & $-720 \pm 20$  & $90 \pm 10$   & 14.69  \\
"                             & 3  & \ovi(r)& $1277.2 \pm 0.1$  & $0.31 \pm 0.03$  & $-770 \pm 20$  & $90 \pm 11$   & 14.69  \\
"                             & 4  & \lya   & $1495.2 \pm 0.1$  & $0.10 \pm 0.02$  & $-1020 \pm 10$ & $40 \pm 20$   & 13.26  \\
"                             & 4  & \ovi(b)& $1268.9 \pm 0.1$  & $1.10 \pm 0.02$  & $-1080 \pm 10$ & $90 \pm 20$   & 14.94  \\
"                             & 4  & \ovi(r)& $1276.0 \pm 0.1$  & $0.55 \pm 0.02$  & $-1070 \pm 10$ & $70 \pm 20$   & 14.94  \\
PG 1114+445                   & 1  & \lya   & $1389.5 \pm 0.1$  & $3.75 \pm 0.09$  & $-200 \pm 10$  & $310 \pm 20$  & 14.84  \\
"                             & 1  & \lyb   & $1171.9 \pm 0.1$  & $0.87 \pm 0.06$  & $-322 \pm 20$  & $350 \pm 70$  & 15.07  \\
"                             & 1  & \nv(b) & $1415.6 \pm 0.1$  & $1.59 \pm 0.03$  & $-260 \pm 20$  & $350 \pm 30$  & 14.88  \\
"                             & 1  & \nv(r) & $1420.3 \pm 0.1$  & $1.05 \pm 0.04$  & $-250 \pm 20$  & $350 \pm 30$  & 14.99  \\
"                             & 1  & \ovi(b)& $1179.1 \pm 0.1$  & $1.11 \pm 0.03$  & $-280 \pm 20$  & $330 \pm 50$  & 14.95  \\
"                             & 1  & \ovi(r)& $1185.7 \pm 0.1$  & $0.57 \pm 0.08$  & $-270 \pm 20$  & $330 \pm 50$  & 14.96  \\
"                             & 1  & \civ(b)& $1769.1 \pm 0.2$  & $1.30 \pm 0.30$  & $-270 \pm 60$  & $250 \pm 110$ & 14.51  \\
"                             & 1  & \civ(r)& $1772.3 \pm 0.2$  & $1.30 \pm 0.30$  & $-240 \pm 60$  & $250 \pm 110$ & 14.81  \\
"                             & 2  & \lya   & $1388.0 \pm 0.1$  & $4.10 \pm 0.04$  & $-520 \pm 10$  & $350 \pm 20$  & 14.88  \\
"                             & 2  & \lyb   & $1170.9 \pm 0.1$  & $0.24 \pm 0.04$  & $-560 \pm 20$  & $250 \pm 50$  & 14.51  \\
"                             & 2  & \nv(b) & $1414.4 \pm 0.1$  & $1.15 \pm 0.06$  & $-520 \pm 20$  & $260 \pm 30$  & 14.73  \\
"                             & 2  & \nv(r) & $1419.0 \pm 0.1$  & $0.75 \pm 0.04$  & $-510 \pm 20$  & $260 \pm 30$  & 14.85  \\
"                             & 2  & \ovi(b)& $1178.1 \pm 0.1$  & $0.44 \pm 0.05$  & $-540 \pm 20$  & $280 \pm 50$  & 14.55  \\
"                             & 2  & \ovi(r)& $1184.8 \pm 0.1$  & $0.44 \pm 0.04$  & $-510 \pm 20$  & $280 \pm 50$  & 14.84  \\
"                             & 2  & \civ(b)& $1767.9 \pm 0.2$  & $1.20 \pm 0.25$  & $-470 \pm 60$  & $250 \pm 110$ & 14.47  \\
"                             & 2  & \civ(r)& $1770.9 \pm 0.2$  & $1.20 \pm 0.25$  & $-470 \pm 60$  & $250 \pm 110$ & 14.77  \\
\hline
\end{tabular}
\end{minipage}
\label{table_abs}
\end{table*}
%============================

\setcounter{table}{1}
%============================
% TABLE: Absorption line parameters (CONTINUED)
%
\begin{table*}[!tbp]
\begin{minipage}[t]{\hsize}
\setlength{\extrarowheight}{1.1pt}
\caption{Continued.}
\centering
\footnotesize
\renewcommand{\footnoterule}{}
\begin{tabular}{c c c c c c c c}
\hline \hline
Object   & Comp. & Ion\,/  & $\lambda_{\rm obs}$ & EW    & $v$    & FWHM    & ${\log(N_{\rm ion})}$ \\
         &       & Line    & (\AA)               & (\AA) & (\kms) & (\kms)  & (cm$^{-2}$)           \\
\hline
PG 1202+281                   & 1  & \lya   & $1415.9 \pm 0.1$  & $0.77 \pm 0.04$ & $-80 \pm 10$    & $80 \pm 10$   & 14.15  \\
"                             & 2  & \lya   & $1415.3 \pm 0.1$  & $1.94 \pm 0.08$ & $-210 \pm 10$   & $360 \pm 20$  & 14.55  \\
"                         & 3  & \lya   & $1410.4 \pm 0.1$  & $0.22 \pm 0.02$ & $-1250 \pm 10$  & $60 \pm 10$   & 13.61  \\
PG 1216+069                   & -  & -      & -                 & -               & -               & -             & -      \\
PG 1307+085                   & 1  & \lya   & $1388.3 \pm 0.1$  & $2.46 \pm 0.03$ & $-3090 \pm 10$  & $130 \pm 10$  & 14.65  \\
"                               & 1  & \ovi(b)& $1178.6 \pm 0.1$  & $0.34 \pm 0.02$ & $-3070 \pm 10$  & $50 \pm 10$   & 14.43  \\
"                               & 1  & \ovi(r)& $1185.0 \pm 0.1$  & $0.19 \pm 0.03$ & $-3080 \pm 10$  & $50 \pm 10$   & 14.48  \\
"                               & 2  & \lya   & $1387.4 \pm 0.1$  & $1.79 \pm 0.03$ & $-3280 \pm 10$  & $110 \pm 10$  & 14.52  \\
"                               & 2  & \ovi(b)& $1178.3 \pm 0.1$  & $0.20 \pm 0.03$ & $-3140 \pm 10$  & $80 \pm 30$   & 14.20  \\
"                               & 2  & \ovi(r)& $1184.7 \pm 0.1$  & $0.11 \pm 0.03$ & $-3150 \pm 10$  & $80 \pm 30$   & 14.24  \\
PG 1352+183                   & -  & -      & -                 & -               & -               & -             & -      \\
PG 1402+261                   & -  & -      & -                 & -               & -               & -             & -      \\
PG 1416-129                   & -  & -      & -                 & -               & -               & -             & -      \\
PG 1425+267                   & 1  & \lya   & $1660.9 \pm 0.1$  & $2.31 \pm 0.02$ & $+580 \pm 10$   & $250 \pm 10$  & 14.63  \\      
"                               & 1  & \lyb   & $1401.2 \pm 0.1$  & $0.40 \pm 0.09$ & $+540 \pm 20$   & $150 \pm 60$  & 14.73  \\
"                               & 1  & \ovi(b)& $1409.9 \pm 0.1$  & $0.62 \pm 0.06$ & $+580 \pm 20$   & $160 \pm 40$  & 14.69  \\
"                               & 1  & \ovi(r)& $1417.7 \pm 0.1$  & $0.31 \pm 0.03$ & $+580 \pm 20$   & $160 \pm 40$  & 14.69  \\
"                               & 2  & \lya   & $1657.4 \pm 0.1$  & $2.45 \pm 0.04$ & $-50 \pm 10$    & $250 \pm 10$  & 14.65  \\
"                               & 2  & \lyb   & $1398.4 \pm 0.1$  & $0.36 \pm 0.03$ & $-60 \pm 10$    & $90  \pm 10$  & 14.69  \\
"                               & 2  & \nv(b) & $1689.0 \pm 0.1$  & $0.55 \pm 0.02$ & $-50 \pm 10$    & $150 \pm 10$  & 14.41  \\
"                               & 2  & \nv(r) & $1694.4 \pm 0.1$  & $0.31 \pm 0.02$ & $-50 \pm 10$    & $150 \pm 10$  & 14.46  \\
"                               & 2  & \ovi(b)& $1407.0 \pm 0.1$  & $0.87 \pm 0.06$ & $-30 \pm 20$    & $190 \pm 40$  & 14.84  \\
"                               & 2  & \ovi(r)& $1414.8 \pm 0.1$  & $0.82 \pm 0.06$ & $-30 \pm 20$    & $190 \pm 40$  & 15.12  \\
"                               & 3  & \lya   & $1653.9 \pm 0.1$  & $5.68 \pm 0.08$ & $-690 \pm 10$   & $650 \pm 80$  & 15.02  \\
"                               & 3  & \lyb   & $1395.6 \pm 0.1$  & $0.62 \pm 0.12$ & $-670 \pm 20$   & $250 \pm 50$  & 14.93  \\
"                               & 3  & \nv(b) & $1685.0 \pm 0.1$  & $0.10 \pm 0.03$ & $-750 \pm 20$   & $40 \pm 10$   & 13.67  \\
"                               & 3  & \nv(r) & $1690.4 \pm 0.1$  & $0.08 \pm 0.03$ & $-750 \pm 20$   & $40 \pm 10$   & 13.88  \\
"                               & 3  & \ovi(b)& $1404.0 \pm 0.1$  & $0.62 \pm 0.05$ & $-670 \pm 20$   & $210 \pm 20$  & 14.69  \\
"                               & 3  & \ovi(r)& $1411.7 \pm 0.1$  & $0.50 \pm 0.03$ & $-670 \pm 20$   & $210 \pm 20$  & 14.90  \\
"                               & 4  & \lya   & $1652.9 \pm 0.1$  & $0.36 \pm 0.04$ & $-870 \pm 10$   & $100 \pm 10$  & 13.82  \\
"                               & 4  & \lyb   & $1394.6 \pm 0.1$  & $0.28 \pm 0.04$ & $-870 \pm 10$   & $90 \pm 20$   & 14.58  \\
"                               & 4  & \nv(b) & $1684.4 \pm 0.1$  & $0.15 \pm 0.02$ & $-870 \pm 10$   & $60 \pm 20$   & 13.85  \\
"                               & 4  & \nv(r) & $1689.8 \pm 0.1$  & $0.09 \pm 0.02$ & $-870 \pm 10$   & $60 \pm 20$   & 13.93  \\
"                               & 4  & \ovi(b)& $1403.2 \pm 0.1$  & $0.35 \pm 0.04$ & $-850 \pm 10$   & $110 \pm 30$  & 14.45  \\
"                               & 4  & \ovi(r)& $1410.9 \pm 0.1$  & $0.31 \pm 0.04$ & $-850 \pm 10$   & $110 \pm 30$  & 14.69  \\
"                               & 5  & \lya   & $1652.2 \pm 0.1$  & $0.59 \pm 0.06$ & $-1000 \pm 10$  & $140 \pm 20$  & 14.03  \\
"                               & 5  & \lyb   & $1394.1 \pm 0.1$  & $0.13 \pm 0.02$ & $-980 \pm 20$   & $120 \pm 30$  & 14.25  \\
"                               & 5  & \ovi(b)& $1402.8 \pm 0.1$  & $0.26 \pm 0.04$ & $-930 \pm 20$   & $160 \pm 40$  & 14.32  \\
"                               & 5  & \ovi(r)& $1410.5 \pm 0.1$  & $0.14 \pm 0.03$ & $-930 \pm 20$   & $160 \pm 40$  & 14.35  \\
"                               & 6  & \lya   & $1650.7 \pm 0.1$  & $1.60 \pm 0.07$ & $-1270 \pm 10$  & $280 \pm 20$  & 14.47  \\
"                               & 6  & \lyb   & $1392.7 \pm 0.1$  & $0.19 \pm 0.03$ & $-1280 \pm 20$  & $100 \pm 20$  & 14.41  \\
"                               & 6  & \nv(b) & $1682.1 \pm 0.1$  & $0.15 \pm 0.03$ & $-1270 \pm 20$  & $70 \pm 20$   & 13.85  \\
"                               & 6  & \nv(r) & $1687.5 \pm 0.1$  & $0.08 \pm 0.03$ & $-1280 \pm 20$  & $50 \pm 20$   & 13.88  \\
"                               & 6  & \ovi(b)& $1401.5 \pm 0.1$  & $0.61 \pm 0.11$ & $-1200 \pm 20$  & $200 \pm 40$  & 14.69  \\
"                               & 6  & \ovi(r)& $1409.3 \pm 0.1$  & $0.32 \pm 0.08$ & $-1200 \pm 20$  & $200 \pm 40$  & 14.71  \\
"                               & 7  & \lya   & $1648.5 \pm 0.1$  & $1.34 \pm 0.02$ & $-1660 \pm 10$  & $150 \pm 30$  & 14.39  \\
"                               & 7  & \lyb   & $1391.1 \pm 0.1$  & $0.06 \pm 0.02$ & $-1640 \pm 20$  & $50 \pm 20$   & 13.91  \\
PG 1427+480                   & 1  & \lya   & $1483.2 \pm 0.2$  & $0.72 \pm 0.18$ & $-130 \pm 40$   & $340 \pm 100$ & 14.12  \\
PG 1435-067                   & 1  & \lya   & $1370.3 \pm 0.2$  & $1.87 \pm 0.20$ & $-480 \pm 30$   & $610 \pm 60$  & 14.54  \\
"                               & 2  & \lya   & $1370.0 \pm 0.1$  & $0.35 \pm 0.03$ & $-540 \pm 10$   & $40 \pm 10$   & 13.81  \\
"                               & 3  & \lya   & $1369.4 \pm 0.1$  & $1.02 \pm 0.07$ & $-680 \pm 10$   & $130 \pm 10$  & 14.27  \\
PG 1626+554                   & 1  & \lya   & $1374.7 \pm 0.1$  & $7.35 \pm 0.02$ & $-240 \pm 10$   & $140 \pm 10$  & 15.13  \\
"                             & 1  & \lyb   & $1159.9 \pm 0.1$  & $0.61 \pm 0.03$ & $-240 \pm 10$   & $110 \pm 10$  & 14.92  \\
"                                     & 2  & \lya   & $1373.8 \pm 0.1$  & $4.70 \pm 0.02$ & $-440 \pm 10$   & $100 \pm 10$  & 14.94  \\
"                                     & 2  & \lyb   & $1159.1 \pm 0.1$  & $0.47 \pm 0.02$ & $-440 \pm 10$   & $80 \pm 10$   & 14.80  \\
"                                     & 3  & \lya   & $1370.0 \pm 0.1$  & $0.22 \pm 0.02$ & $-1260 \pm 10$  & $50 \pm 10$   & 13.61  \\
SDSS J144414.66+063306.7        & -  & -      & -                 & -               & -               & -             & -      \\ 
\hline
\end{tabular}
\end{minipage}
\tablefoot{
Negative $v$ means outflow and positive $v$ inflow velocity. The (b) and (r) labels denote the "blue" (shorter wavelength) and "red" (longer wavelength) lines of a doublet.
}
\label{table_abs}
\end{table*}
%============================

\end{appendix}

%%%%%%%%%%%%%%%%%%%%%%%%%%%%%%%%%%%%%%%%%%%%%%%%%%%%%%%%%%%%%%%%%%%%%%%%%%%%%%%%%%%%%%%%%%%%%%%%%%%%%%%
%%%%%%%%%%%%%%%%%%%%%%%%%%%%%%%%%%%%%%%%%%%%%%%%%%%%%%%%%%%%%%%%%%%%%%%%%%%%%%%%%%%%%%%%%%%%%%%%%%%%%%%
%%%%%%%%%%%%%%%%%%%%%%%%%%%%%%%%%%%%%%%%%%%%%%%%%%%%%%%%%%%%%%%%%%%%%%%%%%%%%%%%%%%%%%%%%%%%%%%%%%%%%%%
\end{document}